\documentclass[twocolumn]{aastex631}
\usepackage[dvipsnames]{xcolor}
\usepackage{amsmath}
\usepackage{booktabs}
\usepackage{float}
\usepackage{ragged2e}
\usepackage{CJK}
\usepackage{soul}
\usepackage{pifont}

\newcommand{\astrid}{\textsc{ASTRID}}
\newcommand{\Msun}{$M_{\odot}$}
\newcommand{\SCnR}{\texttt{SC-noR}}
\newcommand{\SCR}{\texttt{SC-R}}
\newcommand{\Gas}{\texttt{G-noR}}
\newcommand{\Halo}{\texttt{H-noR}}

\shorttitle{Short Title}
\shortauthors{Y. Zhou et al.}

\begin{document}
\begin{CJK*}{UTF8}{bsmi}

\title{
ACORN I. Massive Black Hole Seeding and Tidal Disruption Events from Star Clusters in Cosmological Simulations}

\author[0000-0002-8828-8461]{Yihao Zhou (周亦豪)}
\affiliation{McWilliams Center for Cosmology, Department of Physics, Carnegie Mellon University, Pittsburgh, PA 15213, USA}

\author[0000-0002-6462-5734]{Tiziana Di Matteo}
\affiliation{McWilliams Center for Cosmology, Department of Physics, Carnegie Mellon University, Pittsburgh, PA 15213, USA}

\author[0000-0003-2369-2911]{Claire E. Williams}
\affiliation{Department of Physics and Astronomy, UCLA, Los Angeles, CA 90095, USA}
\affiliation{Mani L. Bhaumik Institute for Theoretical Physics, Department of Physics and Astronomy, UCLA, Los Angeles, CA 90095, USA}

\author[0000-0002-7080-2864]{Aklant Kumar Bhowmick}
\affiliation{University of Virginia, 530 McCormick Rd, Charlottesville, VA 22904, USA}
\affiliation{Virginia Institute for Theoretical Astronomy, University of Virginia, Charlottesville, VA 22904, USA}
\affiliation{The NSF-Simons AI Institute for Cosmic Origins, USA}

\author[0000-0002-8556-4280]{Marta Reina-Campos}
\affiliation{Instituto Galego de F\'isica de Altas Enerx\'ias, Universidade de Santiago de Compostela, 15782 Santiago de Compostela, Galicia, Spain}

\author[0000-0003-4175-8881]{Carl L. Rodriguez}
\affiliation{Department of Physics and Astronomy, University of North Carolina at Chapel Hill, 120 E. Cameron Ave, Chapel Hill, NC, 27514, USA}

\author[0000-0002-0933-6438]{Elena Gonz\'{a}lez Prieto}
\affiliation{Center for Interdisciplinary Exploration \& Research in Astrophysics (CIERA) and Department of Physics \& Astronomy, Northwestern University, Evanston, IL
60208, USA}

\author[0000-0002-2183-1087]{Laura Blecha}
\affiliation{Department of Physics, University of Florida, Gainesville, FL 32611, USA}


\author[0000-0003-0697-2583]{Rupert Croft}
\affiliation{McWilliams Center for Cosmology, Department of Physics, Carnegie Mellon University, Pittsburgh, PA 15213, USA}



\correspondingauthor{Yihao Zhou}
\email{yihaoz@andrew.cmu.edu}
\shorttitle{Star-Cluster-Based MBH seeding model}
\shortauthors{Y.Zhou et al.}

\begin{abstract}
    JWST has revealed massive star clusters at $z\gtrsim 6$, which provide promising environments for massive black hole (MBH) seed formation. 
    We introduce a new MBH seeding model for cosmological simulations, physically motivated by runaway collisions in star clusters. Subgrid star clusters are generated according to the local ISM and evolved through stellar evolution and two-body relaxation. The masses of MBH seeds formed through runaway collisions are determined by the properties of their host clusters, including the effects of metallicity-dependent stellar winds. We further follow tidal disruption events (TDEs) with the evolving clusters, which can be enhanced in MBH binaries, and account for gravitational-wave (GW) recoil following MBH mergers. 
    We implement this model in a constrained simulation of a $5\sigma$ overdense region, where seeds of $10^{3}\!-\!10^{5.6}\,M_\odot$ form in clusters with $10^{4}\leq m_{\rm SC}\leq 10^{8}\,M_\odot$. 
    Because both the seed abundance and mass are directly connected to the local environment, we have a more physical $M_{\rm BH}\!-\!M_{\rm galaxy}$ scaling for low-mass galaxies ($M_{\rm galaxy}\sim10^{6-9}\,M_\odot$), a regime poorly described by previous seeding models.
    Multiple MBHs are naturally allowed to form \textit{in situ} within individual halos, producing a spatially clustered population. This increases the MBH merger rate by $\sim100$ relative to other seeding models, significantly enhancing the expected LISA event rate. TDEs can dominate the growth of $10^{3-4}\,M_\odot$ MBHs and produce luminous flares of $L_{\rm peak}\sim 10^{42-43}\,{\rm erg\,s^{-1}}$, increasing the observability of low-mass seeds. 
    This seeding prescription accelerates early MBH growth at $z\gtrsim 8$, while subsequent self-regulated gas accretion drives the central MBH toward similar masses $\sim10^{9}\, M_\odot$ by $z=6$ across different seeding models.
\end{abstract}


\section{Introduction} 

The James Webb Space Telescope (JWST) has revealed a group of massive black holes (MBHs) at $z\gtrsim 6$, including GN-z11, CEERS-1019, and UHZ1 \citep{Larson2023_ceers_obs, Maiolino2024_gnz11, Bogdan2024_uhz1_obs, Goulding2023_uhz1_obs}. 
The inferred black hole (BH) masses and luminosities require efficient growth within the first billion years of cosmic history, which challenges our understanding of BH formation and growth theories \citep{Pacucci2023}.

To explain these high-redshift MBHs, several MBH seeding channels have been proposed \citep{Volonteri2012, Inayoshi2020, Greene2020}. Dense stellar clusters provide a promising pathway for forming high-redshift MBHs. 
In compact young clusters with central stellar densities $\rho_{\rm c}\gtrsim 10^{7}\ M_{\odot}\,{\rm pc}^{-3}$, massive stars rapidly segregate toward the cluster center and undergo repeated collisions \citep{Gurkan2004, PortegiesZwart2004}. 
In low-metallicity environments, mass loss through stellar winds is limited, allowing the formation of a massive remnant  \citep{Heger2003, Mapelli2009, Vink2018}. 
The resulting very massive star (VMS), with masses potentially reaching $10^{3}-10^{5}$~\Msun, can subsequently collapse into an intermediate-mass black hole (IMBH) seed \citep{Devecchi2009, Katz2015}. 
This formation pathway operates rapidly, typically producing the seed within the first few Myr after cluster formation \citep{Freitag2006, Sakurai2017}. 
Gas inflows can further facilitate MBH seed formation in dense stellar clusters. \citet{Davies2011} showed that rapid gas accretion can drive cluster contraction and core collapse, which makes the subsequent MBH seed formation possible. 

In addition, dense stellar systems play an important role in the early MBH evolution. 
Extreme stellar densities in star clusters significantly increase the probability of close encounters between IMBHs and surrounding stars, triggering tidal disruption events (TDEs, \citealt{Rees1988,Strubbe2009}). 
TDEs provide an additional channel for feeding young MBHs, especially when sustained gas accretion is inefficient \citep{Stone2017, Lee2023}. 
Their luminous flares may also make otherwise faint, low-mass MBHs detectable \citep{Fialkov2017, Gezari2021}. 
\citet{Kritos2026_NSC_LRD} combined TDEs, BH captures, and gas accretion in environments similar to observed LRDs and found that they can rapidly evolve the central BH seeds into the supermassive regime. 
\citet{Kritos2025_NSCmerger} further demonstrated that hierarchical assembly of nuclear star clusters can enhance MBH growth by repeatedly bringing together dense stellar environments and their embedded BHs. 
Large gravitational wave (GW) recoil kicks can eject the MBH from its host system, while the deep gravitational potentials of compact nuclear star clusters help retain the merger remnant.

The importance of this scenario has been reinforced by recent JWST observations of compact stellar systems at $z\gtrsim 5$. 
A population of young star cluster systems is being uncovered through gravitational lensing in the high-redshift Universe, such as the Cosmic Gems \citep{Adamo2024_cosmicGems}, the Sunrise Arc \citep{Vanzella2023}, and the Firefly Sparkle \citep{Mowla2024_firefly}. 
These systems contain clusters with masses of $\approx 10^{5}-10^{6}$~\Msun\ and sizes of only a few parsecs. 
Recent observations have also provided possible evidence for very massive stars at high redshift \citep{Marques-Chaves2026}. 
Some works have also proposed a connection between star clusters and little red dots (LRDs). 
\citet{Chisholm2026} suggested that LRDs may represent globular clusters in formation hosting short-lived supermassive stars, which can form through runaway stellar mergers in sufficiently dense young clusters. The predicted evolution of the LRD population was shown to be broadly consistent with the mass function and abundance of present-day globular clusters. 
With the observed high stellar densities, \citet{Pacucci2025} and \citet{Escala2025} showed that 
LRDs unavoidably host MBH seeds that form from runaway stellar collisions.
These works together indicate that the dense environments required for runaway collisions and MBH seed formation may already be observable in the early Universe.


Significant progress has been made in modeling star cluster formation and evolution in galactic and cosmological environments. To bridge the wide range of scales and physics involved, different numerical approaches have been used: from subgrid modeling techniques in cosmological zoom-in simulations \citep[e.g. the E-MOSAICS project,][]{Pfeffer2018,Kruijssen2019}, 
cosmological hydrodynamical simulations of dense star cluster formation at cosmic dawn \citep[e.g.][]{Williams2025_simulation, Williams2026_PopIII},
single-star simulations of clustered star formation in low-metallicity dwarf galaxies and merger-driven starbursts \citep[e.g.~the GRIFFIN project,][]{Lahen2020_griffin,Lahen2023,Lahen2025}, to dynamical modeling of star clusters spawned from star-forming giant molecular clouds (GMCs) from FIRE-2 \citep[the Great Balls of FIRE series,][]{Grudic2023, Rodriguez2023, Bruel2024}. These efforts demonstrate that massive, compact star clusters can arise naturally in gas-rich galactic environments and provide increasingly detailed predictions for their masses, sizes, metallicities, and subsequent evolution. However, these approaches generally do not translate the resulting cluster populations into MBH seeds formed through runaway stellar collisions. 

Meanwhile, dynamical studies have provided important insights into IMBH formation in star clusters. 
Direct $N$-body and hydrodynamic simulations have demonstrated that both the formation environment and subsequent cluster assembly can strongly influence runaway stellar collisions. 
\citet{Chung2026} self-consistently followed gas fragmentation, star cluster formation, and the subsequent runaway collisions. They found that dense, metal-poor gas clouds can naturally form compact clusters with VMSs of $\sim 3\times 10^{2}\!-\!5\times 10^{3}\ M_{\odot}$. 
The FROST-CLUSTERS simulations \citep{Rantala2024_Frost_clusterI, Rantala2025_Frost_clustersII, Rantala2026_Frost_clusterIII}, which combine the direct $N$-body code BIFROST with stellar evolution from SEVN, show that hierarchical cluster assembly can 
promote the formation and the subsequent growth of VMSs and IMBHs.
Complementary Monte Carlo and analytic approaches have explored a much broader range of cluster parameter space. 
Using an extended grid of  Cluster Monte Carlo (\texttt{CMC}) simulations, \citet{GonzalezPrieto2024} showed how VMS formation depends on cluster density, the stellar mass function, and binary properties. 
Building on these studies, \citet{Williams2026} developed an analytic framework that predicts the  VMS mass based on the properties of their host clusters, and showed that MBH seeds produced by runaway collisions are widespread in dense high-redshift stellar systems. 
An alternative pathway to MBH seed formation in star clusters has been explored by \citet{Kritos2023} and \citet{Kritos2024}, who showed that sufficiently massive and compact nuclear clusters can produce an MBH seed through successive BH mergers, with residual gas accretion further accelerating its growth. 
These works demonstrate that MBH seed formation is highly sensitive to cluster structure, stellar population, and assembly history. 
However, they generally model individual clusters or prescribed cluster populations rather than coupling these formation channels self-consistently to the evolved cluster population within a cosmological context. 
Therefore, they cannot directly predict the cosmological demographics of MBH seeds.


A key challenge for cosmological simulations is therefore developing a subgrid model that connects galaxy-scale environments to MBH seeds formed in star clusters. 
Several previous works have explored novel MBH seeding subgrid models. 
\citet{Habouzit2017} developed an implementation of BH formation in dense, low-metallicity environments, as advocated by models invoking the collapse of the first generation of stars, or dense nuclear star clusters. 
The BRAHMA simulation suite \citep{2024MNRAS.531.4311B,2024MNRAS.533.1907B,2025MNRAS.538..518B,2026ApJ..1008...30B} also employed gas-based seeding models~\citep{2021MNRAS.507.2012B} designed to identify and initialize seeds in pockets of dense, low-metallicity gas that could be candidate sites for seed formation through Pop III remnants or stellar-cluster pathways. More recently, the AMBRA simulation \citep{Zhou2026_ambra} combines this gas-based seeding prescription with the galaxy-formation model and large cosmological volume of ASTRID to statistically study the high-redshift MBH population. However, the \texttt{BRAHMA} gas-based seeding model does not include an explicit subgrid treatment of star cluster formation and evolution, and therefore cannot distinguish between the stellar-cluster and Pop III remnant seeding pathways. In this work, we address this limitation by developing a new subgrid model for the stellar-cluster seeding pathway in cosmological hydrodynamic simulations, in which the formation and evolution of unresolved dense star clusters are explicitly modeled.
By coupling the subgrid treatment of star clusters with an analytical model for VMS formation, our model establishes a physically motivated connection between galaxy formation and MBH seed formation through the stellar-cluster pathway.
We investigate the resulting high-redshift MBH population, providing a new framework for assessing the role of stellar clusters in the origin of the first massive black holes. 
This work is the first paper of the ACORN\footnote{\textbf{A}STRID \textbf{C}osmological simulations \textbf{O}f the massive-black-hole seeds in star cluste\textbf{R} populatio\textbf{N}s} project.

This paper is organized as follows. Section~\ref{sec:method} describes the star-cluster-based seeding model along with the galaxy formation model used in our simulations.  
Section~\ref{sec:setup} shows the setup of the simulations. 
We present our results in Section~\ref{sec:results} and conclude in Section~\ref{sec:conclusion}.

\section{Method}\label{sec:method}

\subsection{ASTRID galaxy formation model}

In this work we use the same galaxy formation model as \astrid\
 \citep{Bird2022, Ni2022_astrid, Ni2025,Zhou2026_astrid}. 
Hosting $0.33$ trillion particles in a simulation box of $250\,{\rm Mpc}/h$ per side, \astrid\ is one of the largest hydrodynamical simulations evolved to $z=0$. It is performed using the massively parallel simulation code MP-GADGET \citep{Feng2018_mpgadget} and includes a full-physics sub-grid treatment for modeling galaxy formation, black holes, stellar and AGN feedback, and inhomogeneous reionization. 
In the following, we briefly summarize the key features of the \astrid\ galaxy formation model and refer readers to \citet{Bird2022} and \citet{Ni2022_astrid} for more details.

In \astrid, gas is allowed to cool through primordial gas cooling following \citet{Katz1996}, as well as through metal-line cooling. 
The metal cooling rate is estimated by scaling a solar metallicity template according to the gas metallicity, following \citet{Wiersma2009}. 
Star formation is implemented based on the multi-phase star formation model in \citet{Springel2003}, where gas above a density threshold of $0.13\ {\rm cm}^{-3}$ is allowed to form stars.
Following \citet{Krumholz2011_h2cooling}, a correction is included to account for the effects of molecular hydrogen on star formation at low metallicity, which suppresses star formation at high redshifts. 
Stars are formed with one-fourth of the mass of gas particles; the remaining gas mass is kept in the gas particle for subsequent star formation. 
Type II supernova wind feedback is implemented following \citet{Okamoto2010}, assuming wind speeds proportional to the local dark matter (DM) velocity dispersion. 
\astrid\ models metal return by treating each star particle as a single stellar population with a Chabrier initial mass function (IMF) \citep{Chabrier2003}. 
We follow the general approach of \citet{Vogelsberger2013} and \citet{Pillepich2018}, while using our own mass and metal tables for AGB stars, Type Ia and Type II supernovae \citep{Karakas2010,Doherty2014_metallII,Doherty2014_metallIII, Kobayashi2006, Nomoto1997}.

MBHs are evolved with subgrid prescriptions for seeding, accretion, feedback, and dynamics. 
We will describe the BH seeding prescriptions in detail in Section~\ref{sec:SC_seeding_method}, and here we briefly summarize the other aspects of the BH model. 
The BH accretion rate $\dot{M}_{\mathrm{BH}}$ is estimated using the Bondi-Hoyle formalism \citep{BondiHoyle1944, DiMatteo2005_BH_model}, which is based on the local properties of nearby gas particles:
\begin{equation}    
\label{equ:Mdot}
\dot{M}_{\mathrm{BH}}=4\pi\alpha \,G^{2}\,M_{\mathrm{BH}}^{2}\,\rho_{\mathrm{BH}}\left(c_{\mathrm{s}}^{2}+v^{2}_{\mathrm{rel}}\right)^{-3/2},
\end{equation}
where $c_{\mathrm{s}}$ is the local sound speed, $\rho_{\mathrm{BH}}$ is the gas density around the BH, and $v_{\mathrm{rel}}$ is the velocity of the BH relative to the surrounding gas. 
The dimensionless boost $\alpha=100$ is adopted to account for the underestimation of the accretion rate due to the unresolved interstellar medium. 
Super-Eddington accretion is allowed with an upper limit of twice the Eddington accretion rate $\dot{M}_{\mathrm{Edd}}$. Therefore, the BH accretion rate $\dot{M}_{\mathrm{BH}}$ is determined by $\dot{M}_{\mathrm{BH}} = \min\left(\dot{M}_{\mathrm{B}}, 2\dot{M}_{\mathrm{Edd}}\right)$.

\astrid\ adopts two-mode AGN feedback: 
high accretion mode (or thermal feedback) and low accretion mode (or kinetic feedback). 
As the low-accretion mode only comes into play at low redshifts \citep[$z\lesssim 2$;][]{Weinberger2017_tng}, while not affecting the early growth of BHs, we describe only the high-accretion mode here.
In this feedback mode, 5\% of the radiated energy $\Delta \dot{E}_{\rm high}=0.05\,\eta\, \dot{M}_{\rm BH}c^{2}$ is thermally injected into the gas within twice the radius of the SPH smoothing kernel of the BH \footnote{
The remaining 95\% of the radiated energy is emitted as AGN luminosity, which is not explicitly tracked in the simulation. 
}. 
The energy injection is carried out according to the SPH kernel weight, without any preferred direction.

Following \citet{Tremmel2015} and \citet{Chen2022_DF}, a subgrid model is adopted 
to account for the unresolved dynamical friction for BHs from surrounding gas, stars, and dark matter. 
This model not only gives a better estimation of the BH hardening timescale, but also provides well-defined BH trajectories and velocities.
This allows us to impose a more physical criterion for BH mergers. Following \citet{Bellovary2011}  and \citet{Tremmel2017}, we merge two BHs based on their separation and whether they are gravitationally bound:
\begin{equation}
\text{\textit{bound if}}\quad
\left\{
\begin{aligned}
|\Delta r| &< 2\,\epsilon_g,\\
\tfrac12\,|\Delta v|^{2} &< \Delta a \cdot \Delta r ,
\end{aligned}
\right.
\label{equ:merger}
\end{equation}
where $\Delta a$, $\Delta v$, and $\Delta r$ are the relative acceleration, velocity, and position of the two BHs. $\epsilon_{\rm g}$ is the gravitational softening length. 
To alleviate artificial heating and stabilize the BH motion in the early growth phase, we assign a dynamical mass tracer, $M_{\rm dyn}$, which is only used for the gravity calculation for small-mass BH with $M_{\rm BH}\leq M_{\rm dyn}$. After the BH grows to the regime $M_{\rm BH}\geq M_{\rm dyn}$, $M_{\rm BH}$ is used to calculate the dynamical friction force applied to the BHs.
The adopted value of $M_{\rm dyn}$ differs among the simulation runs and is specified in Section~\ref{sec:setup}.

The original ASTRID has a resolution of $m_{\rm DM} = 6.7 \times 10^6$~$h^{-1}M_\odot$  and $m_{\rm gas} = 1.3 \times 10^6$~$h^{-1}M_{\odot}$. The gravitational softening length is $\epsilon_{\rm g} = 1.5\ h^{-1}{\rm ckpc}$ for both DM and gas particles. 
In this work, we adopted a higher resolution: $m_{\rm DM} = 1.1 \times 10^6$~$h^{-1}M_\odot$, $m_{\rm gas} = 2.0 \times 10^5$~$h^{-1}M_{\odot}$, and a gravitational softening length $\epsilon_{\rm g} = 0.8\ h^{-1}{\rm ckpc}$.

\begin{figure*}[!t]
    \centering
    \includegraphics[width=1\linewidth]{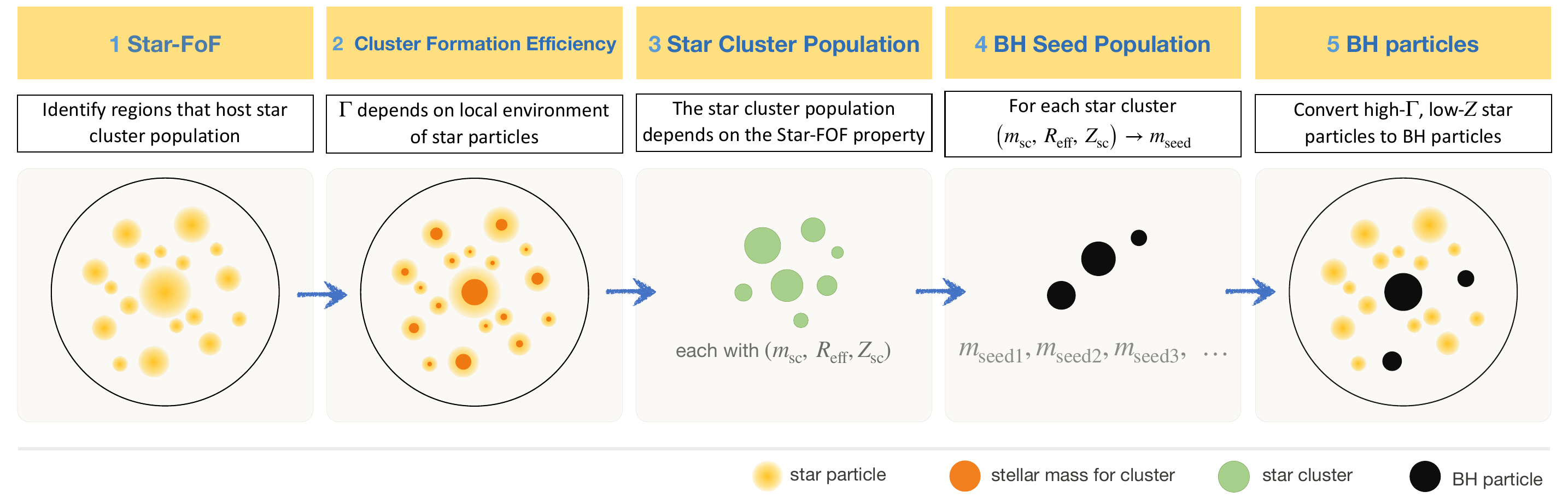}
    \caption{A schematic illustration of our star-cluster-based MBH seeding model, described in full in Section~\ref{sec:method_StarFOF} to \ref{sec:method_seed}. In our cosmological hydrodynamic simulations, we first identify StarFOF structures using a FoF algorithm with star particles as the primary linking particles. 
    We then use the star particles younger than 100 Myr and bound to the local potential within each StarFOF to determine the total mass budget of the star clusters.
    We use the model developed in \citet{Kruijssen2012} and calculate a cluster formation efficiency $\Gamma$ based on the local environment of each eligible star particle.
    The star cluster population is then sampled from a power-law ICMF and assigned sizes and metallicities. 
    We use the model from \citet{Williams2026} to predict the BH seed mass formed in each star cluster. 
    Finally, we seed the MBH in the simulation by converting high-$\Gamma$ and low-metallicity star particles into BH particles.
    }\label{fig:pipeline}
\end{figure*}

\subsection{Star-Cluster-based MBH seeding model}
\label{sec:SC_seeding_method}

Our star-cluster-based MBH seeding model follows five steps: (i) identify the stellar structures that host the star cluster populations; 
(ii) estimate the mass budget for the star clusters; 
(iii) generate the star cluster population; 
(iv) estimate the MBH seed mass within each star cluster; 
and (v) seed the MBH particle in the simulations. 
We visualize this procedure in Fig.~\ref{fig:pipeline} and describe each step in turn below.

\subsubsection{StarFOF} 
\label{sec:method_StarFOF}
In this work, our typical star particle mass is a quarter of the gas particle mass: $m_{\rm star}=5\times 10^{4}\ M_{\odot}\,h^{-1}$. 
At this resolution, individual star clusters cannot be resolved.
We therefore aim to identify larger stellar complexes that host populations of star clusters, analogous to the observed cluster systems revealed by JWST at high redshift, such as Cosmic Gems and FireFly systems \citep{Adamo2024_cosmicGems, Mowla2024_firefly}.
To search for these structures on the fly at a feasible computational cost, we apply a Friends-of-Friends (FoF) algorithm with star particles as the primary linking particles and the linking length $\ell_{\star}=0.1\,\overline{d}_{\rm DM}$, where $\overline{d}_{\rm DM}$ is the mean separation between DM particles. 
Similar FoF-based methods to identify the star clusters have been adopted in previous works \citet{Bhowmick2024_gasbased_subgrid},
\citet{Williams2025_simulation, Deng2026, Chon2026}. 

We refer to these stellar FOF structures as ``StarFOF'', and the conventional dark-matter FOF halos are ``DM-FOF''.
The selected $\ell_{\star}$  is half of the conventional $\ell=0.2\,\overline{d}_{\rm DM}$ used to identify DM-FOF and close to the size of our gravity kernel ($2.8\,\epsilon_{g}$).
With our resolution ($L_{\rm box}=12.5\,{\rm cMpc}/h$, $N_{\rm DM}=N_{\rm gas}=512^3$, $m_{\rm DM} = 1.1 \times 10^6$~$h^{-1}M_\odot$), $\ell=0.1\,\overline{d}_{\rm DM}$ corresponds to a physical scale of $\sim 0.3\!-\!0.5\ {\rm kpc}$ over $z\sim 6-10$, which is comparable to the size of observed star cluster complexes. 
In Appendix~\ref{app:linkstar} we test the sensitivity of the StarFOF catalog to the adopted $\ell_{\star}$
and we find that adopting a different $\ell_{\star}$ does not significantly affect our main conclusions. 

We include only StarFOFs that host at least 16 star particles and apply a boundedness criterion to retain only gravitationally bound stars. Following \citet{Williams2025_simulation}, we include the DM particles within the size of the StarFOF in addition to the member stars when calculating the potential. 
This is implemented as a correction when the star is not bound to the stellar structures alone. 
Gas particles are not included, which may underestimate the star mass in gas-rich regions. 
We confirm the effect is minor: including the gas would raise the bound stellar mass by $\lesssim 0.5\%$. 
Beyond that, we exclude star particles older than 100 Myr to ensure star clusters consist of young stellar populations. 
In practice, this $100$ Myr age limit is rarely reached: 95\% of the star particles contributing to seed formation do so within $\sim 36$Myr of their formation at $z=10$, and within $\sim 10$ Myr at $z=6$. 

\subsubsection{Cluster Formation Efficiency} 
\label{sec:Gamma}

To estimate the mass budget of the star clusters within each StarFOF, we follow the E-MOSAICS framework \citep{Pfeffer2018} and assign a cluster formation efficiency, $\Gamma$, to every star particle. 
$\Gamma$ is defined as the fraction of newly formed stellar mass that is born in bound clusters. 
The total mass initially assigned to the subgrid cluster population with a star particle of mass $m_{\star}$ is therefore $\Gamma\, m_{\star}$. 
We calculate $\Gamma$ using the local formulation of the analytical model developed by \citet{Kruijssen2012}. 
This model assumes that the hierarchically structured ISM follows a lognormal density distribution, and predicts that star cluster formation preferentially occurs in the high-density tail of this distribution, where the free-fall time is short enough to convert gas into stars efficiently before feedback terminates star formation.
We exclude the `cruel cradle effect', which describes the early tidal disruption of the forming clusters by their natal environment. 
$\Gamma$ is therefore a function of local gas density $\rho$, gas velocity dispersion $\sigma_{\rm loc}$, and thermal sound speed $c_{\rm s}$. 

Because our simulations do not explicitly resolve the multiphase ISM, we use the approach taken by E-MOSAICS and adopt a fixed sound speed of $c_{\rm s}=0.3\ {\rm km}\,{\rm s}^{-1}$ for the cold ISM. 
However, unlike E-MOSAICS, which approximates the local turbulent velocity dispersion as $\sigma_{\rm loc} \simeq (P_{\rm gas}/\rho_{\rm gas})^{1/2}$, where $P_{\rm gas}$ is the local gas pressure, we estimate $\sigma_{\rm loc}$ directly from the velocities of star-forming gas within the SPH kernel of the parent gas particle: $\sigma_{\rm loc} = \sigma_{\rm SF}$. 
This choice is motivated by the physical interpretation of the \citet{Kruijssen2012} model, in which the turbulent density distribution represents cold, GMC-associated gas where star formation occurs. 
Restricting the calculation to star-forming gas hence reduces contamination from diffuse warm gas. 
If the kernel contains fewer than eight star-forming gas particles, we set $\sigma_{\rm loc}=0$. 
These local environment properties and the resultant $\Gamma$ are measured at the time the parent gas particle is converted into a star particle.

\subsubsection{Star Cluster Population}
\label{sec:method_starcluster}

We populate each StarFOF with a subgrid star cluster population based on the $\Gamma$ of its member star particles.
The total cluster mass budget in a StarFOF is computed as $M_{\rm SC,sFOF}=\sum_{i}\Gamma_{i}\,m_{\star,i}$,
where the sum is over the member star particles.
To avoid double counting, we include only star particles that have not contributed to a previous MBH seeding.
More specifically, if the star cluster population generated based on the mass budget $M_{\rm SC,sFOF}$ produces at least one MBH seed following the prescription in Section~\ref{sec:method_seed}, we flag all particles that contribute to $M_{\rm SC,sFOF}$ as ``seeded'' and exclude them from future star cluster formation. If no MBH seed is produced, the star particles remain eligible at later timesteps.

We assume that cluster masses follow a power-law initial cluster mass function (ICMF) with a slope of $-2$ within the mass range $10^{2}-10^{8}\ M_{\odot}$, consistent with observations of young star clusters in the local Universe \citep{Zhang1999, Messa2018, Cook2019} and high-resolution simulations \citep[e.g.][]{Lahen2020_griffin}:
\begin{equation}
    N\, dM \propto M^{-2} \, dM, \qquad 10^{2}\leqslant M \leqslant 10^{8}\,M_{\odot}.
    \label{equ:ICMF}
\end{equation}
The mean cluster mass for this distribution is about $\bar{m}_{\rm SC}=1.38\times 10^{3}\ M_{\odot}$.
We then draw the number of star clusters in a StarFOF from a Poisson distribution 
\begin{equation}
    N_{\rm SC} = \mathcal{P}(M_{\rm SC, \rm sFOF} / \bar{m}_{\rm SC}).
\end{equation}
and independently sample $N_{\rm SC}$ cluster masses from Equation~\ref{equ:ICMF}.
This construction ensures that the total sampled cluster mass is statistically consistent with the available mass budget $M_{\rm SC, \rm sFOF}$.
As a physical upper limit, we cap the total sampled cluster mass by the stellar mass contained in StarFOF $M_{\star,\rm sFOF}=\sum_{i}m_{\star,i}$.
If this limit is exceeded, we randomly remove sampled clusters until the total cluster mass falls below $M_{\star,\rm sFOF}$.
In practice, this correction is rarely applied because the cluster mass budget $M_{\rm SC, \rm sFOF}$ is typically much smaller than $M_{\star, \rm sFOF}$.

For each sampled cluster, we assign an effective radius $R_{\rm eff}$ using the mass-radius relation from \citet{Brown2021}: 
\begin{equation}
    R_{\rm eff} = 2.365\,{\rm pc}\ \left(\frac{m_{\rm SC}}{10^{4}\,M_{\odot}}\right)^{0.18}
\end{equation}
with a lognormal scatter of 0.318 dex at fixed $m_{\rm SC}$. 
The effective radius is related to the half-mass radius $r_{\rm h}$ by $R_{\rm eff}=1.3\,r_{\rm h}$.
This relation is fitted for the young star clusters with age $\leqslant 10\ {\rm Myr}$ in the nearby galaxies from the LEGUS survey. 
However, at high redshift, more compact star clusters may form with
the high gas pressures and surface densities in the early galaxies \citep{Marks2012, Choksi2021, Lahen2025}, consistent with recent JWST observations of extremely dense clusters at $z\gtrsim 6$ \citep[e.g.,][]{Adamo2024_cosmicGems, Mowla2024_firefly}.
Therefore, adopting the local $R_{\rm eff}\!-\! m_{\rm SC}$ relation may provide a conservative estimate of cluster densities and therefore the resulting MBH seed mass. 
We assign each cluster a metallicity by randomly drawing from the metallicity of the member star particles in the host StarFOF.
This ensures that the cluster metallicity follows the metallicity distribution of the stellar population that contributes to star cluster formation. 
Finally, we obtain a subgrid population of star clusters, each with mass $m_{\rm SC}$, size $R_{\rm eff}$, and metallicity $Z_{\rm SC}$.

\subsubsection{MBH Seeds in Star Clusters}
\label{sec:method_seed}

With the star cluster population within StarFOFs, we apply the analytic model developed by \citet{Williams2026} to predict the mass of the MBH seed formed in individual star clusters. 
This model developed a radially resolved analytic framework for predicting the mass of a VMS formed through runaway stellar collisions.
An alternative pathway to MBH seed formation is hierarchical growth through repeated mergers of stellar-mass black holes \citep[e.g.,][]{Antonini2019, Gerosa2021}. 
However, this channel requires successive merger remnants to remain bound to the cluster after GW recoil, which can be challenging because during the first few generations, merger remnants typically have high spins that can enhance the recoil in subsequent mergers, particularly for misaligned spin configurations \citep[e.g.,][]{Campanelli2007}. 
Therefore, runaway stellar collisions provide a more favorable pathway for forming MBH seeds in young, dense clusters \citep[e.g.,][]{PortegiesZwart2004, Sakurai2017}. 
In the following, we briefly summarize the main features of this model and refer readers to \citet{Williams2026} for more details. 

Each cluster is represented by a spherical power-law stellar density profile
\begin{equation}
    \rho_{\star} \propto r^{-\alpha},
\end{equation}
whose normalization is set by the cluster mass $m_{\rm SC}$ and the effective radius $R_{\rm eff}$.
The velocity dispersion is calculated from the enclosed stellar mass under the assumption of virial equilibrium. 
We adopt the fiducial parameters of the no-initial-BH branch in \citet{Williams2026}, including a density power-law index $\alpha=1.2$, a representative collider mass of $1\,M_{\odot}$, a stellar binary fraction of $0.2$, and a characteristic orbital eccentricity of $0.5$. 
Rather than the Salpeter IMF used by \citet{Williams2026}, we evaluate the IMF mass fraction that sets the inflow of collision products with the Chabrier IMF, consistent with the ASTRID galaxy formation model \citep{Bird2022}.

We first calculate the mass of the central VMS formed through runaway stellar collisions. 
Its growth rate is obtained by integrating the inward mass flux of collision products over the disruption timescale $t_{\rm d}$. $t_{\rm d}$ is limited by the timescale of stellar main-sequence evolution, cluster core collapse, and the age of the Universe 
\footnote{
We remove the limitation of the halo encounter timescale $t_{\rm merger}$ in \citet{Williams2026}
when determining $t_{\rm d}$, as $t_{\rm merger}$ is estimated statistically using halo abundance and velocity calibrated to a separate cosmological simulation \citep{Williams2025_simulation}, which may not accurately represent the environments in our constrained overdense regions. 
Moreover, as shown in Appendix~B of \citet{Williams2026}, $t_{\rm merger}$ primarily limits low-mass, extended clusters, which are in general less favorable for MBH seed formation.}.
We also confirm through a post-processing test that including $t_{\rm merger}$ has a negligible effect on the predicted MBH seed masses and abundance. 
The mass supply is reduced by depletion of the local stellar population, mass loss during collisions, and energy injection from binary heating. The final VMS is then determined from the balance 
\begin{equation}
    \begin{split}
    \frac{\mathrm{d}M_{\mathrm{VMS}}}{\mathrm{d}t}
    &= \left(1-f_{\mathrm{ml}}\right)
       \left(
       \dot{M}_{\mathrm{df}}
       - \dot{M}_{\mathrm{dep}}
       - \dot{M}_{\mathrm{bin}}
       \right) \\
    &\quad - \dot{M}_{\mathrm{wind}},
    \end{split}
    \label{eq:vms_growth}
\end{equation}
where $\dot{M}_{\rm df}$ is the dynamical-friction-driven inflow rate, $\dot{M}_{\rm dep}$ is the depletion of collision, and $\dot{M}_{\rm bin}$ represents the effect of binary heating. 
The mass loss fraction during the collisions $f_{\rm ml}$ depends on the energy ratio of the colliding stars (see \citealt{Rose2023} for details). 
The metallicity-dependent stellar wind rate follows \citet{Vink2018}:
\begin{equation}
    \dot{M}_{\rm wind} \propto M_{\rm VMS}^{2.1}\,Z_{\rm SC}^{0.74}.
\end{equation}
We integrate Equation~\ref{eq:vms_growth} over the disruption timescale $t_{\rm d}$ with the equilibrium condition $d\,M_{\rm VMS}/dt=0$ to obtain the VMS mass $M_{\rm VMS}$.
If $M_{\rm VMS}\geqslant 200\,M_{\odot}$, we assume it collapses directly into an MBH seed and 100\% of the VMS mass is retained in the MBH seed.
With the mass resolution of our fiducial run, accurately following the dynamics of MBH seeds near the minimal collapse mass of $200\ M_{\odot}$ is numerically challenging. Although we adopt an effective dynamical mass $M_{\rm dyn}$ to stabilize MBH motion (see Section~\ref{sec:setup}), this remains an approximate treatment, particularly for very low-mass MBHs. 
We therefore impose a minimum MBH seed mass $M_{\rm seed,min}=10^{3}\,h^{-1}M_{\odot}$ to reduce these numerical uncertainties: 
\begin{equation}
    M_{\rm seed}=M_{\rm VMS}\qquad (M_{\rm VMS}\geqslant M_{\rm seed,min}).
\end{equation}
In Appendix~\ref{app:M_seed_min}, we investigate the sensitivity of the resultant MBH population to the choice of $M_{\rm seed,min}$. We find that lowering $M_{\rm seed,min}$ from $10^{3}$ to $300\ h^{-1}M_{\odot}$ has little effect on MBH growth histories and mass function at $z=9$.

\begin{figure*}[!t]
    \centering
    \includegraphics[width=0.9\linewidth]{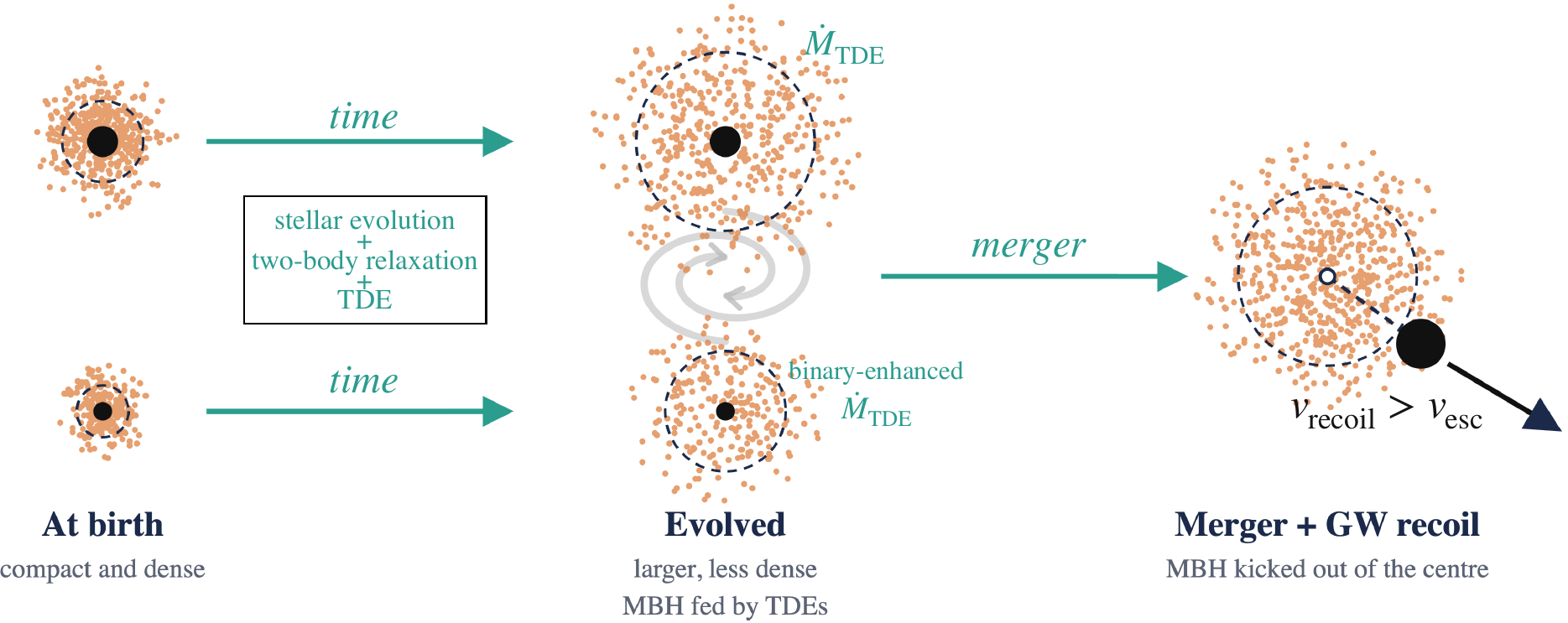}
    \caption{ A schematic illustration of the star cluster and MBH coevolution model, which is described in full in Sections~\ref {sec:method_SC_evolution} to \ref{sec:GW_recoil}. 
    After a cluster forms, we trace its evolution caused by stellar evolution, two-body relaxation, and TDEs. This evolution makes the star cluster less compact and less massive. 
    The MBH is fed by TDEs, whose rate is determined by the evolved star cluster properties, and can be boosted within binary systems. After the merger, we compare the escape velocity $v_{\rm esc}$ of the cluster and the GW recoil velocity $v_{\rm recoil}$ assuming zero spin for MBHs to determine whether the cluster retains the merger remnant. 
    }\label{fig:evolution_pipeline}
\end{figure*}

For each cluster with $M_{\rm seed}\geqslant M_{\rm seed,min}$, we convert a parent star particle into an MBH particle of mass $M_{\rm seed}$. 
MBH seeds are expected to form in high-$\Gamma$ and metal-poor regions. Hence, to select the parent particle, we consider both the metallicity $Z_{\star}$ and the cluster mass budget $\Gamma m_\star$ of each star particle. 

We first identify all available star particles in the host StarFOF with metallicity $Z_{\star}\leqslant Z_{\rm crit}$. 
Rather than imposing a universal metallicity threshold, we calculate a physically motivated $Z_{\rm crit}$ based on the properties of individual clusters.
It is defined as the metallicity at which a cluster with the same mass and size ($m_{\rm SC}$, $R_{\rm eff}$) would produce the smallest MBH seed allowed in our simulation $M_{\rm seed,min}$.\footnote{Using the predicted seed mass $M_{\rm seed}$ instead would recover the cluster-averaged metallicity $Z_{\rm SC}$, rather than the maximum metallicity for the threshold.} 
This threshold is estimated from the equilibrium condition in Eq.~\ref{eq:vms_growth}. For fixed ($m_{\rm sc}, R_{\rm eff}$), the metallicity dependence of the resulting $M_{\rm VMS}$ enters through the stellar wind term, which gives a scaling relation 
\begin{equation}
    \frac{Z_{\rm crit}}{Z_{\rm SC}}=\left(\frac{M_{\rm seed,min}}{M_{\rm seed}}\right)^{-2.8}.
\end{equation}
Among eligible star particles, we select the one with the largest $\Gamma m_{\star}$ as the parent particle, corresponding to the largest contribution to the star cluster mass budget.
In this way, metallicity serves as a threshold for selecting eligible star particles, while $\Gamma m_{\star}$ determines the parent particle for each MBH seed.
We stop seeding when all eligible star particles are converted to MBH seeds or all clusters with $M_{\rm seed}\geqslant M_{\rm seed,min}$ are seeded. In practice, the number of eligible star particles is always larger than the number of seeding MBHs, so the availability of parent particles does not limit MBH seed formation.


In this work, we track only star clusters hosting MBHs, using the associated MBH particles as tracers. 
Clusters that do not produce MBH seeds, or whose MBHs are ejected by GW recoil (see Section~\ref{sec:GW_recoil}), are not included in the subsequent cluster evolution.

\begin{figure*}[!t]
    \centering
    \includegraphics[width=1\linewidth]{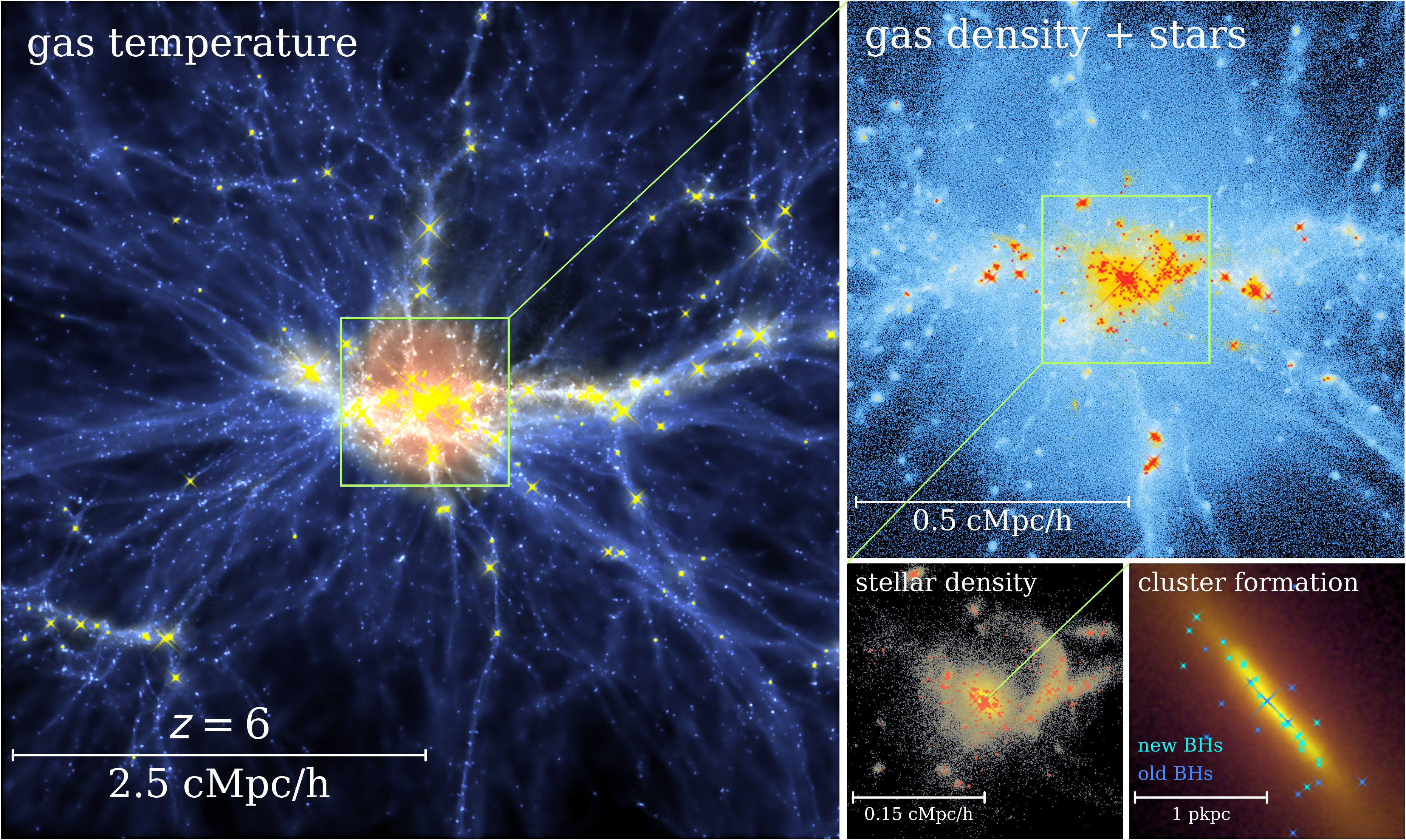}
    \caption{Multi-scale visualizations of the most massive halo in the \SCnR\ run at $z=6$, from cosmological environment to the central galaxy. 
    MBHs are shown as crosses, with marker size scaling with MBH mass. We change the color of MBH markers in each panel for visual clarity.  
    Green boxes indicate the regions enlarged in the next panel, with the green lines connecting the next zoom levels.  
    \textit{Left:} Gas surface density colored by temperature in a $5\ {\rm cMpc}\,h^{-1}$ region centered on the simulation volume, with redder colors indicating hotter gas. 
    \textit{Upper right}: Gas surface density in a $1\ {\rm cMpc}\,h^{-1}$ region centered on the most massive halo, overlaid with the star particles (yellow dots). 
    \textit{Lower middle:} Stellar surface density in a $300\ {\rm ckpc}\,h^{-1}$-per-side region. 
    \textit{Lower right:} Edge-on view of the central $2\ {\rm pkpc}\,h^{-1}$ around the host galaxy of the most massive MBH ($M_{\rm BH}\approx 10^{9}\, M_{\odot}$). We plot the stellar density field colored by the cluster formation efficiency ($\Gamma$). Cyan crosses show MBHs newly seeded within the previous $20\, {\rm Myr}$ (23 of the 37 BHs in the panel); the dark blue crosses mark other old BHs. 
    }\label{fig:large_vis}
\end{figure*}

\subsubsection{Star Cluster Evolution}
\label{sec:method_SC_evolution}

With the star cluster population generated in each StarFOF, we include a subgrid model to track their subsequent evolution and interaction with their host MBHs.
In Fig.~\ref{fig:evolution_pipeline} we show a schematic pipeline of our star cluster and MBH coevolution model.

The star clusters experience mass loss due to stellar evolution, two-body relaxation, and TDEs: 
\begin{equation}
    \frac{dm}{dt} = \left(\frac{dm}{dt}\right)_{\rm sev} + \left(\frac{dm}{dt}\right)_{\rm rlx} + \left(\frac{dm}{dt}\right)_{\rm TDE}. 
\end{equation}
$\left(dm/dt\right)_{\rm TDE}$ accounts for the star mass lost from the cluster through TDEs and accreted by the central MBH, which we will discuss in more detail in Section~\ref{sec:method_TDE}. 
The mass loss caused by stellar evolution is directly inherited from the mass return model in ASTRID, which is described in detail in \citet{Bird2022}. 
A simple stellar population (SSP) of mass $m_{\rm SC}$ and metallicity $Z_{\rm SC}$ is assumed to form within each cluster.  
We assume a Chabrier initial mass function (IMF) \citep{Chabrier2003}, and massive stars in the population return mass and metals at the end of their finite lifetime. 
Because our cluster mass is represented by a subgrid budget rather than by explicit star particles, we reduce $m_{\rm SC}$ without returning mass to the surrounding gas. 
For the mass loss caused by two-body relaxation, we adopt the compact object enhanced evaporation model from \citet{Gieles2023} \citep[also implemented in][]{ContrerasGuerra2026}:
\begin{equation}
    \left(\frac{dm}{dt}\right)_{\rm rlx} = -a\,\xi\,\frac{m}{t_{\rm rh}}\left(\frac{m}{m_{\rm i}}\right)^{-2/3},
\end{equation}
where $a=1.5$, $m_{\rm i}$ is the cluster mass after stellar-evolution mass loss, and $t_{\rm rh}$ is the half-mass relaxation time \citep{Spitzer1987} computed with a mean stellar mass $0.62\ M_{\odot}$ (see Eq.~14 of \citealt{ContrerasGuerra2026}).  
The escape rate $\xi$, i.e., the fraction of stars escaping per relaxation time, follows the 
fit to $N$-body models of \citet{Alexander2012} with the parameters of \citet{Gieles2014} (see Eqs.~55-56 of \citealt{Reina-Campos2022}).
$\xi$ depends on the cluster size relative to its tidal radius $r_{\rm t}=\left(Gm_{\rm SC}/T\right)^{1/3}$, where $T$ is the tidal field strength. 
Following \citet{Pfeffer2018} and \citet{Reina-Campos2022}, we measure $T$ from the tidal tensor $T_{\rm ij} = -\partial^{2} \Phi/\partial x_{i}\partial x_{j}$ at the position of each MBH particle, where $\Phi$ is the local gravitational potential.
With $\lambda_i$ the eigenvalues of $T_{\rm ij}$:
\begin{equation}
    T=\max(\lambda_{i})+1/3\left|\sum_{i} \lambda_{i}\right|
\end{equation}
Because we adopted the TreePM and hierarchical time integration algorithm to calculate the gravity \citep{Springel2021}, we can only access the whole gravity tree and therefore the tidal field at the Particle-Mesh (PM) step. 
We compute $T$ at each PM step and use this value to estimate 
$\xi$ until $T$ is updated at the next PM step. The typical timescale between the PM steps is $\lesssim 3$ Myr. 
We have verified that this approximation agrees very well with that calculated by updating $T$ at every individual timestep.

Besides the cluster mass, we also track the cluster size evolution due to stellar evolution and two-body relaxation. 
We follow the prescription of \citet{ContrerasGuerra2026} and \citet{Reina-Campos2022} but neglect the contribution of tidal shocks. 
\begin{equation}
    \frac{dr_{\rm h}}{r_{\rm h}} = \left(\frac{dr_{\rm h}}{r_{\rm h}}\right)_{\rm sev} + \left(\frac{dr_{\rm h}}{r_{\rm h}}\right)_{\rm rlx} 
\end{equation}
The star cluster experiences an adiabatic expansion due to stellar evolution, which is given by $\left(dr_{\rm h}/r_{\rm h}\right)_{\rm sev} = -dm_{\rm sev}/m_{\rm SC}$, where $dm_{\rm sev}$ is the mass loss due to stellar evolution at this timestep. 
The dynamical evolution of the cluster size is given by 
\begin{equation}
    \left(\frac{dr_{\rm h}}{r_{\rm h}}\right)_{\rm rlx} = \left(2-\frac{\zeta}{\xi}\right) \frac{dm_{\rm rlx}}{m_{\rm SC}},
\end{equation}
where $dm_{\rm rlx}$ is the mass loss due to the two-body relaxation at this timestep, $\zeta=0.1$ is a dimensionless constant for the fraction of energy change per relaxation time-scale \citep{Henon1961,Gieles2011}.

We do not include the mass loss or size expansion from tidal shocks because this cannot be traced with the hierarchical time integration, which means including these effects would make the simulation more computationally expensive. 
Tidal shocks are expected to be less effective for massive and compact, i.e., high-density, star clusters \citep{Reina-Campos2022, Rodriguez2023, ContrerasGuerra2026}. 
In our simulation, the average density within the halo-mass radius of 
MBH-hosting star clusters $\rho_{\rm h}$ has a median value of
$9\times 10^{3}\ M_{\odot}/{\rm pc}^{3}$ at formation, with a $1\sigma$ range of $210\!-\!4\times 10^{4}\ M_{\odot}/{\rm pc}^{3}$. These high densities suggest that our clusters are relatively resistant to tidal fields. 
Hence, we do not expect tidal shocks to significantly affect our results. 

When two MBHs merge according to the criteria in Eq.~\ref{equ:merger}, the remnant retains only the more massive of the two host clusters with its full evolution state (e.g., mass, size, metallicity, and age). 
In our simulation, MBHs numerically merge on the scales of $\sim 1\ {\rm kpc}$, so the subsequent evolution on small scales is unresolved. 
Direct $N$-body simulations show that clusters spiraling into a nucleus hosting MBHs are strongly tidally disrupted at parsec-scale distances, with the stripped stars forming a diffuse component rather than a compact bound cluster \citep{Antonini2013, Gnedin2014}. In addition, the hardening MBH binary ejects stellar mass through three-body scattering \citep{Merritt2006}.
Because the mass loss driven by these unresolved processes cannot be followed, simply summing the two cluster masses would let clusters grow without limit through repeated mergers, producing objects well above the upper truncation of the ICMF, i.e., $10^{8} M_{\odot}$, beyond the regime where the cluster evolution model is calibrated. 


\begin{figure}[!t]
    \centering
    \includegraphics[width=1\linewidth]{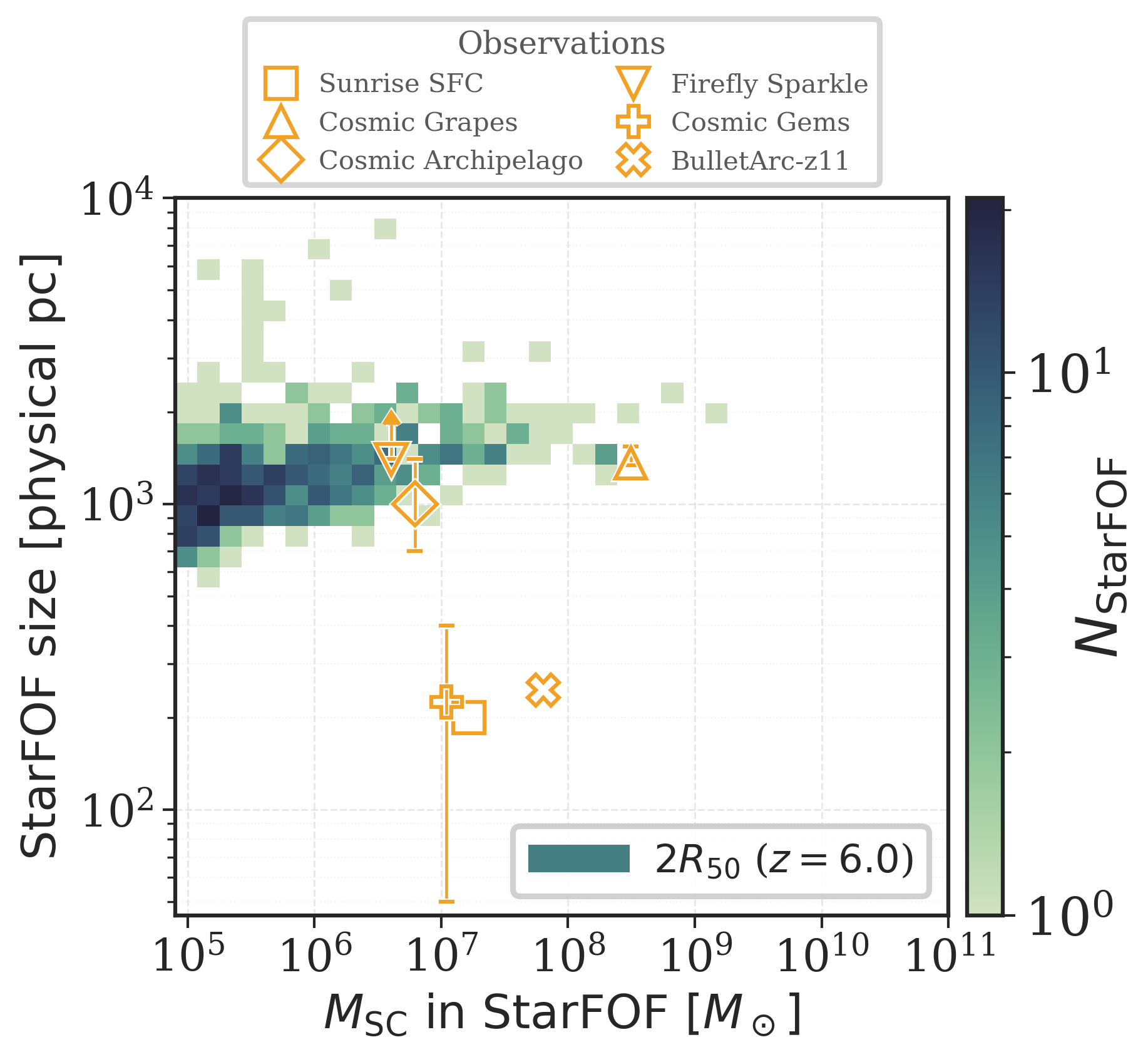}
    \caption{The size and total star cluster mass of StarFOFs in \SCnR\ at $z=6$. For each StarFOF, we plot its half-mass radius $R_{\rm 50}$ and the total star cluster mass $M_{\rm SC}$. 
    The observed star cluster systems are shown by yellow markers, including Cosmic Gems \citep{Adamo2024_cosmicGems, Messa2026_CosmicGems_new}, 
Firefly Sparkle \citep{Mowla2024_firefly}, Sunrise Arc \citep{Vanzella2023}, Bullet Arc \citep{Bradac2025_BulletArc}, Cosmic Grapes  \citep{Fujimoto2025_CosmicGrapes}, and Cosmic Archipelago \citep{Messa2025_CosmicArchipelago}.
    }\label{fig:StarFOF_vs_obs}
\end{figure}

\begin{figure*}[!t]
    \centering
\includegraphics[width=1\linewidth]{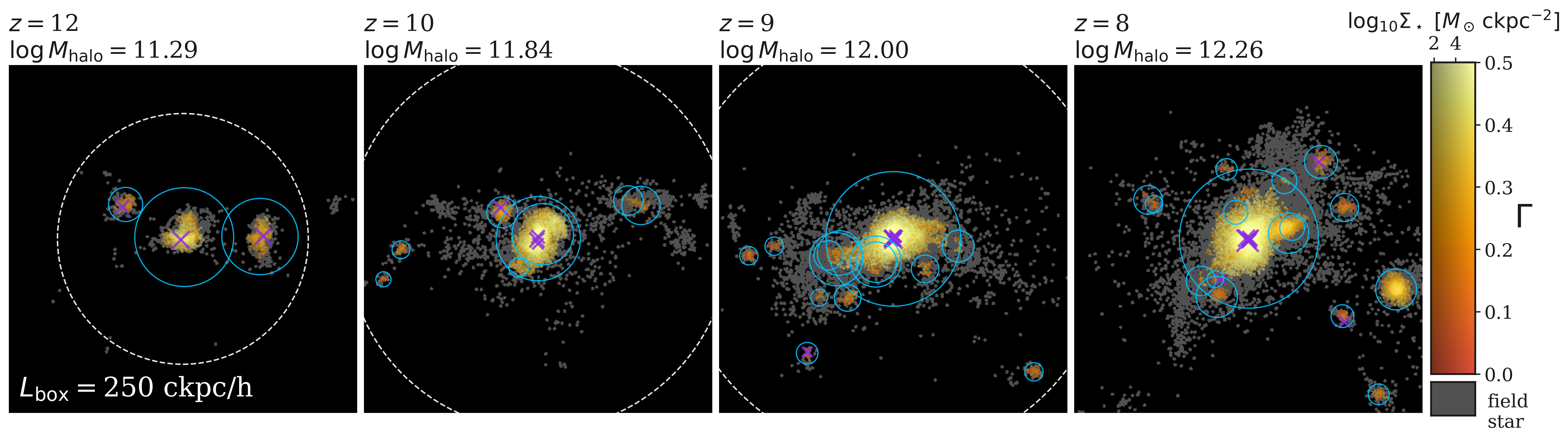}
    \caption{Evolution of the stellar density field within the most massive DM-FOF in the run \texttt{SC-noR} from $z=12$ to $z=8$.
    Each panel shows a region of side length $L_{\rm box}=250\ {\rm ckpc}\, h^{-1}$ around the halo center. 
    The redshift and halo mass are labeled above each panel. The white dashed circles mark the $R_{\rm 500c}$ of the DM-FOF halo. 
    Blue circles show the individual StarFOFs identified in this region, and the radius corresponds to $R_{\rm max}$ of each StarFOF.
    Stellar density of the StarFOFs is colored by their cluster formation efficiency $\Gamma$, while star particles not belonging to any StarFOF (field stars) are plotted in gray.
    Purple crosses indicate MBH seeds formed within the 10 Myr before the snapshot, and the marker size is proportional to the MBH seed mass.
    }\label{fig:Sfof_evolve}
\end{figure*}

\subsubsection{TDEs in Star Clusters} 
\label{sec:method_TDE}

Dense star clusters harboring MBHs provide favorable environments for TDEs, in which stars are torn apart by the strong tidal forces of the MBH. The star cluster populations tracked within our simulation therefore provide a framework to model TDEs and their contributions to MBH growth.
Beyond TDEs, the mergers from stellar-mass BHs in star clusters may provide an additional growth channel for the central MBH \citep{GonzalezPrieto2025}, which we do not include in this work. 

For an isolated MBH, we adopt the TDE rate of \citet{Rizzuto2023}, who assume that the bound stars within the influence radius of the BH $R_{\rm inf}$ are scattered into the loss cone on the local two-body relaxation timescale:
\begin{equation}
    \begin{aligned}
    \dot{N}_{\rm TDE, single} = {}& 1.1 F f_{\rm b} \ln\left(0.22\,\frac{M_{\rm BH}}{m_{\star}}\right)\left(\frac{M_{\rm BH}}{10^{3}M_{\odot}}\right) \\  &\times \left(\frac{\rho}{10^{7}\,M_{\odot}{\rm pc}^{-3}}\right)\left(\frac{100\ {\rm km\,s^{-1}}}{\sigma}\right)^{3}\ {\rm Myr}^{-1}.
    \end{aligned}
\end{equation}
Here $F=0.8$ is a factor calibrated against the high-resolution $N$-body simulations, $f_{\rm b}=0.2$ is the fraction of stars within the influence that are bound to the BH, $m_{\star}=1\ M_{\odot}$ is the mass of the disrupted star \citep{Rizzuto2023}.
The $\rho$ and $\sigma$ are the stellar density and velocity dispersion of the host star cluster
evaluated at $R_{\rm inf}$. When the star cluster is entirely embedded within the BH influence radius, i.e., $R_{\rm inf} > R_{\rm max}$, we use $\rho$ and $\sigma$ at $R_{\rm max}$ instead.
To estimate $\sigma$, we assume virial equilibrium. 
For $\rho$, we approximate the stellar distribution around the MBH with a Bahcall-Wolf cusp \citep{Bahcall1976}, $\rho_{\star}\propto r^{-7/4}$, truncated at $R_{\rm max}=1.4\,R_{\rm eff}$. 
The Bahcall-Wolf cusp is the classical density profile for a dynamically relaxed stellar population around a central MBH \citep{Alexander2017}, and has been reproduced in direct $N$-body simulations of star clusters hosting central MBHs \citep{Baumgardt2004, Preto2004}. 
We then normalize this profile to the cluster mass. 

Within MBH binary systems, the TDE rate can be enhanced by the eccentric Kozai-Lidov (EKL) mechanism, through which the MBH companion excites the eccentricities of the stellar orbits \citep{Naoz2016, Melchor2024, Mockler2023}.
As we will show later, the star-cluster-seeding model produces more MBH merger events in the early Universe, which may lead to a significant enhancement of TDEs. 
We include this contribution for the star clusters around the secondary MBH, where the primary MBH can efficiently perturb the stellar population. For the primary MBH, secular perturbations from the weaker companion are easy to overcome by relativistic precession \citep{Chen2011, Li2015, Naoz2016}. 
We estimate the available stellar reservoir using the number of stars within the hierarchical limit 
\begin{equation}
a_{\rm hier} = 0.1\,a_{\rm b}\frac{1-e_{\rm b}^{2}}{e_{\rm b}},
\end{equation}
where $a_{\rm b}$ and $e_{\rm b}$ are the binary semimajor axis and eccentricity measured when the two MBHs are gravitationally bound. Due to our resolution limit, we cannot resolve the binary evolution to this scale before the numerical merger, and we therefore adopt the same value as \citet{Mockler2023}: $a_{\rm b}=0.5 R_{\rm inf}$ and $e_{\rm b}=0.5$ for all MBH binaries, which produces $a_{\rm hier}=0.075\,R_{\rm inf}$. For the stellar density profile $\rho_{\star}\propto r^{-7/4}$, the number of stars within $a_{\rm hier}$ is 
\begin{equation}
        N_{\star,\rm hier} = \frac{M_{\star}(<a_{\rm hier})}{m_{\star}} \approx  0.039 \frac{2\,M_{\rm BH}}{m_{\star}}. 
\end{equation}
Here we adopt the definition of $R_{\rm inf}$ as the radius enclosing a stellar mass of $2M_{\rm BH}$, 
and require $m_{\rm SC}\geqslant 2M_{\rm BH}$. For small clusters with $m_{\rm SC}<2M_{\rm BH}$, we instead set $N_{\star,\rm hier}=m_{\rm SC}/m_{\star}$. This occurs only when the secondary MBH is massive and the host cluster goes through a significant mass loss. In this case, the TDE rate is limited by the available stellar reservoir.
Based on the Monte Carlo simulations in \citet{Mockler2023} and \citet{Melchor2024}, we assume a fraction $\epsilon_{\rm dis}=0.2$ of these stars will be disrupted by the secondary MBH, and these EKL-enhanced TDEs occur within a timescale of $\tau_{\rm EKL}=10\ {\rm Myr}$. 
Hence, for each numerical MBH merger in our simulation, the cumulative number of enhanced TDEs is 
\begin{equation}
    N_{\rm TDE,binary} = \epsilon_{\rm dis}\,N_{\star,\rm hier},
\end{equation}
and the corresponding TDE rate is $\dot{N}_{\rm TDE,binary} = N_{\rm TDE,binary}/\tau_{\rm EKL}$.

The total TDE rate for each MBH is then given by
\begin{equation}
    \dot{N}_{\rm TDE} = \dot{N}_{\rm TDE, single} + \dot{N}_{\rm TDE,binary},
\end{equation}
where the first term $\dot{N}_{\rm TDE, single}$ is above zero as long as the host star cluster has $m_{\rm SC}>0$, and $\dot{N}_{\rm TDE,binary}$ only becomes non-zero within the $\tau_{\rm EKL}$ time window of the merger event.
For each TDE event, we assume 50\% of the stellar mass is accreted by the MBH following \citet{Rizzuto2023}. This gives an MBH mass growth rate
\begin{equation}
    \dot{M}_{\rm BH,TDE} = 0.5\,m_{\star}\,\dot{N}_{\rm TDE},
\end{equation}
where the disrupted stellar mass is $m_{\star}=1\ M_{\odot}$.
Accordingly, we remove the mass from the host star cluster at a rate of 
\begin{equation}
    \left(\frac{dm}{dt}\right)_{\rm TDE} = -m_{\star}\,\dot{N}_{\rm TDE}.
\end{equation}

\subsubsection{Gravitational Wave Recoil}
\label{sec:GW_recoil}  

As we will show later, MBH mergers play an important role in the early growth of MBH seeds.
However, the anisotropic emission of gravitational waves during MBH mergers carries away linear momentum, which can cause a recoil kick to the remnant MBH \citep{Peres1962, Bekenstein1973, Fitchett1983, Campanelli2007, Lousto2011}.
Depending on the escape velocity of the host galaxy, the remnant may be displaced from the galactic center or ejected from the galaxy entirely, therefore suppressing the further growth of the MBH. 

To account for this effect, we implement a GW recoil model in which both progenitor MBHs are assumed to be non-spinning. 
We discuss the influence of this assumption in Section~\ref{sec:results_GWrecoil}.
Under zero spin, the recoil is generated entirely by the mass asymmetry of the merging MBHs, and its magnitude is determined by the mass ratio:
\begin{equation}
    v_{\rm recoil} = A\,\eta^{2}\sqrt{1-4\eta}\left(1+B\eta\right),
    \label{equ:GW_recoil}
\end{equation}
where $A = 12000\ \rm{km/s}$, $B=-0.93$, $\eta = q/\left(1+q\right)^2$, and $q=m_2/m_1$ $\left(0\leqslant q\leqslant 1\right)$ is the mass ratio of the two progenitor MBHs \citep{Fitchett1983, Campanelli2007}. 
Whenever an MBH merger occurs, we add the recoil velocity kick to the merger remnant, whose direction is randomly selected within the orbital plane of the merging MBHs.
Assuming the Bahcall-Wolf cusp profile for the star cluster, the escape velocity $v_{\rm esc}$ is given by 
\begin{equation}
    v_{\rm esc} = 42\ {\rm km\,s^{-1}}\, \frac{m_{\rm SC}}{10^{5}\,M_{\odot}}^{1/2} \left(\frac{R_{\rm eff}}{1\ {\rm pc}}\right)^{-1/2}, 
\end{equation}
which depends on the evolved cluster mass and size.
If the recoil velocity exceeds the escape velocity of the host star cluster, the remnant MBH is thought to be ejected from the cluster, and we label its host cluster mass as zero. 

For non-spinning MBHs, the recoil velocity in Equation~\ref{equ:GW_recoil} reaches a maximum of $\sim 175\ {\rm km}\,{\rm s}^{-1}$ at $q\approx 0.36$ and vanishes for equal-mass mergers ($q=1$).
Hence, the large non-spinning recoil occurs for moderately unequal-mass mergers, which are common during the seeding-dominated regime of MBH assembly because all MBH masses are initially small and comparable to each other.
The GW recoil can therefore have a strong impact on the early growth of MBHs.

\subsection{Halo-based and Gas-based MBH Seeding Models}
To investigate how different seeding prescriptions affect the early growth of MBHs, in this work we run two comparison simulations with the same initial conditions (ICs) while implementing the halo-based and gas-based MBH seeding models, respectively.
In this section, we briefly introduce these two seeding models. 

The halo-based MBH seeding model is widely used in many large-volume cosmological simulations, like ASTRID \citep{Zhou2026_astrid}, IllustrisTNG \citep{Springel2018_illustrisTNGI,Naiman2018_illustrisTNGII, Marinacci2018_illustrisTNGIII, 
Nelson2018_illustrisTNGIV, Pillepich2018_illustrisTNGV}, and EAGLE \citep{Schaye2015_eagle}.  
In this model, a DM-FOF halo is eligible for seeding if it is massive enough. This is a phenomenological treatment which is partly motivated by the observation that almost all massive galaxies in the local Universe host a central MBH.
In \astrid, to seed an MBH, a DM-FOF must satisfy both the total mass $M_{\rm halo,FOF}\geqslant 5\times 10^{9}\, h^{-1}M_{\odot}$ and the stellar mass $M_{\star,\rm FOF}\geqslant 2\times 10^{6}\, h^{-1}M_{\odot}$. 
The stellar-mass threshold ensures that BHs are seeded only in halos that contain sufficient cold dense gas to form stars. 

The BRAHMA simulation suite~\citep{2024MNRAS.531.4311B,2024MNRAS.533.1907B,2025MNRAS.538..518B,2026ApJ..1008...30B} introduced a family of gas-based seeding models~\citep{2021MNRAS.507.2012B} in which MBHs form in dense, metal-poor gas pockets. This model is motivated by the heavy seeds from stellar collisions in star clusters or from the rapid growth of Pop III remnants in dense gas clouds.
The AMBRA simulation \citep{Zhou2026_ambra} applies the most lenient gas-based seeding model among the BRAHMA to the large cosmological volume of ASTRID, allowing statistical study of the high-redshift MBH population.
In AMBRA, a DM-FOF becomes eligible for seeding if it contains sufficient star-forming, metal-poor gas, whose mass is above a threshold of $M_{\rm gas,sfmp}=1.26\times 10^{6}\,h^{-1}M_{\odot}$.

In both halo-based and gas-based seeding models, seeding is performed only for the DM-FOF halos hosting no BH, i.e., only one MBH is seeded in each DM-FOF, and DM-FOFs can only gather multiple MBHs through hierarchical mergers.
The newly formed MBH is placed at the potential minimum of the DM-FOF halo. 
Different from the star-cluster-based seeding model, in which the MBH seed mass is determined self-consistently from its host star cluster, both halo-based and gas-based seeding models treat the MBH seed mass as a free parameter. 
In ASTRID and AMBRA, the BH seed mass is drawn from a power-law distribution with a slope of $-1$ in the range of $3\times 10^{4}-3\times 10^{5}\,h^{-1}M_{\odot}$. 
This makes the seed mass independent of the local environment, which is in contrast to the star-cluster-based seeding model.

\begin{table}[!ht]
\centering
\caption{Summary of the simulation suite.}
\begin{tabular}{ccc} 
\hline
\hline
Simulation & Seeding Model & GW recoil \\
\hline
\texttt{SC-noR} & star-cluster-based &  \ding{56} \\
\texttt{SC-R} & star-cluster-based & \ding{52} \\
\texttt{H-noR} & halo-based & \ding{56} \\
\texttt{G-noR} & gas-based & \ding{56} \\
\hline
\end{tabular}\label{tab:simulationsetup}
\end{table}

\section{Simulation Setup} 
\label{sec:setup}

In this work, we have a suite of simulations to investigate how different MBH seeding prescriptions affect the early formation and growth of MBHs within the same cosmological environment. 
We focus on a rare, highly overdense region that is expected to evolve into a massive halo and host rapidly growing MBHs at high redshift, similar to the systems observed by JWST.
To construct this environment, we use constrained ICs generated with \texttt{Gaussian-CR} \citep{Ni2025} to reproduce the most massive halo at $z=8$ in the ASTRID simulation.  
This halo originates from a $5\sigma$ density peak smoothed over $R_{\rm G}=1\,h^{-1}\,{\rm Mpc}$ in the ICs and reaches a halo mass of $M_{\rm halo}\approx 10^{12}\,M_{\odot}$ by $z=8$. 
We measure the properties of this density peak from the ASTRID ICs and impose the corresponding constraints at the center of our simulation volume.
In Appendix~\ref{app:astrid_compare}, we show that the constrained simulation successfully reproduces the halo assembly history of the targeted ASTRID halo. 
Because we deliberately select a rare overdense region, our simulations are intended to provide a controlled comparison between seeding models rather than to provide a statistical prediction for the cosmological MBH population.

The simulation has a periodic box of side length $12.5\,{\rm cMpc}/h$, and contains $512^3$ DM and $512^3$ gas particles in the IC. The 
masses of DM and gas particles are $1.1\times 10^{6}\,h^{-1}\,M_{\odot}$ and $2.0\times 10^{5}\,h^{-1}\,M_{\odot}$, respectively. The typical star particle mass is $\approx 5\times 10^{4}\,h^{-1}\,M_{\odot}$.
The gravitational softening length is $0.8\,{\rm ckpc}/h$. 
The ICs are set at $z=99$ and the cosmological parameters are from \citet{Planck}: $\Omega_{\rm 0}=0.3089$, $\Omega_{\rm b}=0.0486$, $\Omega_{\Lambda}=0.6911$, $h=0.6774$, $\sigma_8=0.816$, and $n_s=0.9667$.
Except for MBH seeding prescriptions and the treatment of GW recoil, all simulations use the same galaxy formation model as ASTRID \citep{Zhou2026_astrid}, including star formation, stellar feedback, chemical enrichment, BH accretion, and AGN feedback.

We perform four simulations, which are summarized in Table~\ref{tab:simulationsetup}. The runs \SCR\ and \SCnR\ implement the star-cluster-based seeding model with and without the GW recoil, respectively.
The comparison runs \Halo\ and \Gas\ apply the halo-based and gas-based seeding models, respectively,  without the GW recoil. All simulations are evolved down to $z=6$. 
The \Halo\ run follows the same halo-based seeding criteria in ASTRID. The DM-FOF halo is eligible for seeding when it satisfies both the total halo mass $M_{\rm halo,FOF}\geqslant 5\times 10^{9}\, h^{-1}M_{\odot}$ and the stellar mass $M_{\star,\rm FOF}\geqslant 2\times 10^{6}\, h^{-1}M_{\odot}$. 
The \Gas\ run adopts the gas-based seeding criteria in AMBRA. In this model, a DM-FOF halo is eligible for seeding if it contains at least $M_{\rm gas, sfmp}=1.26\times 10^{6}\,h^{-1}M_{}\odot$ of star-forming, metal-poor gas.
Such gas is defined as satisfying $Z_{\rm gas}\leqslant 10^{-4}\,Z_{\odot}$ and $n_{\rm H}\geqslant 0.13\,{\rm cm}^{-3}$.
In both comparison runs, seeding occurs only if the DM-FOF does not already host an MBH, and the newly formed MBH is placed at the potential minimum of the DM-FOF halo. 
To make a direct comparison and reduce the stochastic variations associated with small volume, we adopt a fixed seed mass of $M_{\rm seed}=10^{5}\ M_{\odot}/h$ in both \Halo\ and \Gas, rather than drawing seed masses from the power-law distribution as in ASTRID and AMBRA.

We also adopt an effective dynamical mass, $M_{\rm dyn}$, for newly formed MBHs to stabilize their movement. 
For \Halo\ and \Gas\ runs, we use the same value as in ASTRID and AMBRA: $M_{\rm dyn}=10^{7}\,h^{-1}M_{\odot}$. 
For the star-cluster-seeding runs, \SCnR\ and \SCR, we instead set $M_{\rm dyn}=m_{\rm SC}$, where $m_{\rm SC}$ is the mass of the host star cluster. 
In this case, $M_{\rm dyn}$ approximately accounts for the unresolved gravitational contribution of the host star cluster surrounding the MBH seed.
We impose a lower limit set by the gas particle mass, $M_{\rm dyn}=\max(m_{\rm SC}, m_{\rm gas})$, to ensure that with the star cluster evolution, the effective dynamical mass does not fall below the mass resolution of the surrounding gas particles. 
Allowing smaller MBH seeds would also introduce more MBHs hosted by low-mass star clusters, for which the effective dynamical mass is increasingly dominated by the imposed numerical floor $m_{\rm gas}$ rather than the physical cluster mass. This further motivates our choice of $M_{\rm seed,min}=10^{3}\ M_{\odot}h^{-1}$. 
This $M_{\rm dyn}$ is only used when the physical MBH mass $M_{\rm BH}$ is smaller than $M_{\rm dyn}$. Once the MBH grows beyond $M_{\rm dyn}$, its dynamics are determined by its physical mass.

\begin{figure}[htbp]
    \centering
    \includegraphics[width=0.9\linewidth]{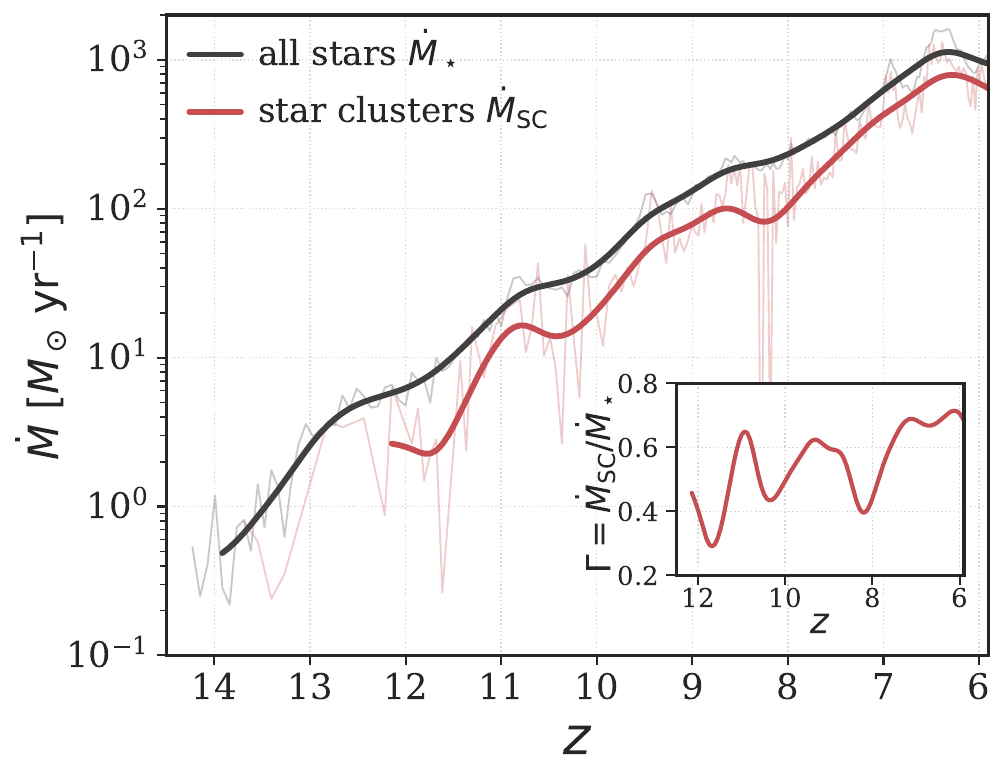}
    \caption{Redshift evolution of star formation within the most massive DM-FOF in the \SCnR\ simulation, which forms from a $5\sigma$ density peak and reaches a halo mass of $\sim 10^{12}\ M_{\odot}$ by $z=8$. 
    The black and red curves show the total star formation rate $\dot{M}_{\star}$ and the star cluster formation rate $\dot{M}_{\rm SC}$, respectively. 
    The thick curves show the smoothed value over a redshift window $\Delta z=0.25$, and the underlying thin curves show the unsmoothed values.
    The bottom-right inset panel plots their ratio $\dot{M}_{\rm SC}/\dot{M}_{\star}$, i.e., the cluster formation efficiency, within this halo. 
    }\label{fig:SFRD_Msc}
\end{figure}

\begin{figure*}[!t]
    \centering
\includegraphics[width=1.0\linewidth]{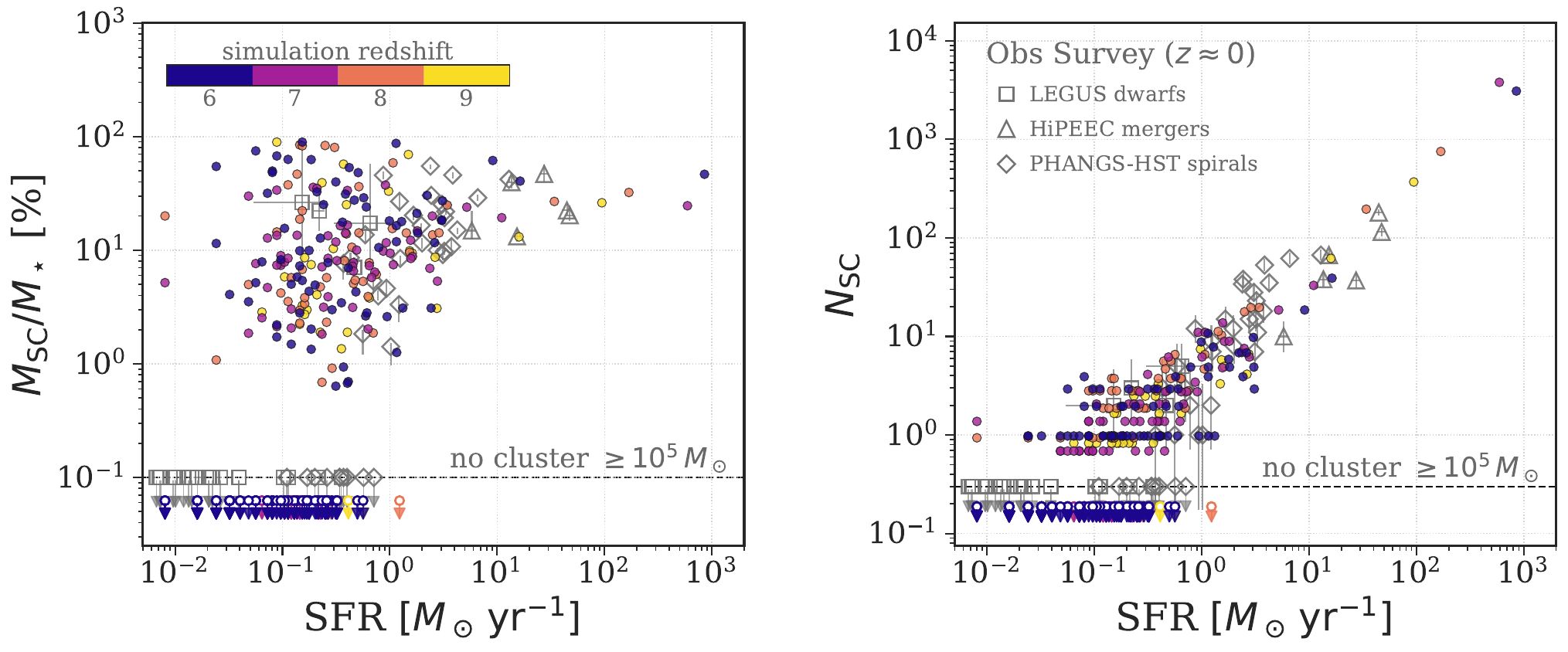}
    \caption{The young massive star cluster populations of individual galaxies as a function of galaxy SFR.
    Filled circles show galaxies in the \SCnR\ simulation at $6\leqslant z\leqslant 9$ colored by redshift. 
    Gray open markers show observations from local surveys: LEGUS \citep{Cook2019, Cook2023}, the HiPEEC \citep{Adamo2020},  and the PHANGS-HST \citep{Lee2022}.
    For both the simulation and observation samples, we select clusters with mass $m_{\rm SC}\geqslant 10^{5}\ M_{\odot}$ and age $\leq 10\ {\rm Myr}$.
    \textit{Left:} Mass ratio of the total mass in the selected clusters to the total stellar mass formed over the past $10$~Myr, i.e., the fraction of the star formation that occurs in these clusters.
    \textit{Right:} Number of these star clusters $N_{\rm SC}$. 
    Galaxies with no clusters are plotted by downward arrows at the horizontal dashed lines.
    }\label{fig:SC_obs}
\end{figure*}

\begin{figure*}[htbp]
    \centering
    \includegraphics[width=1.0\linewidth]{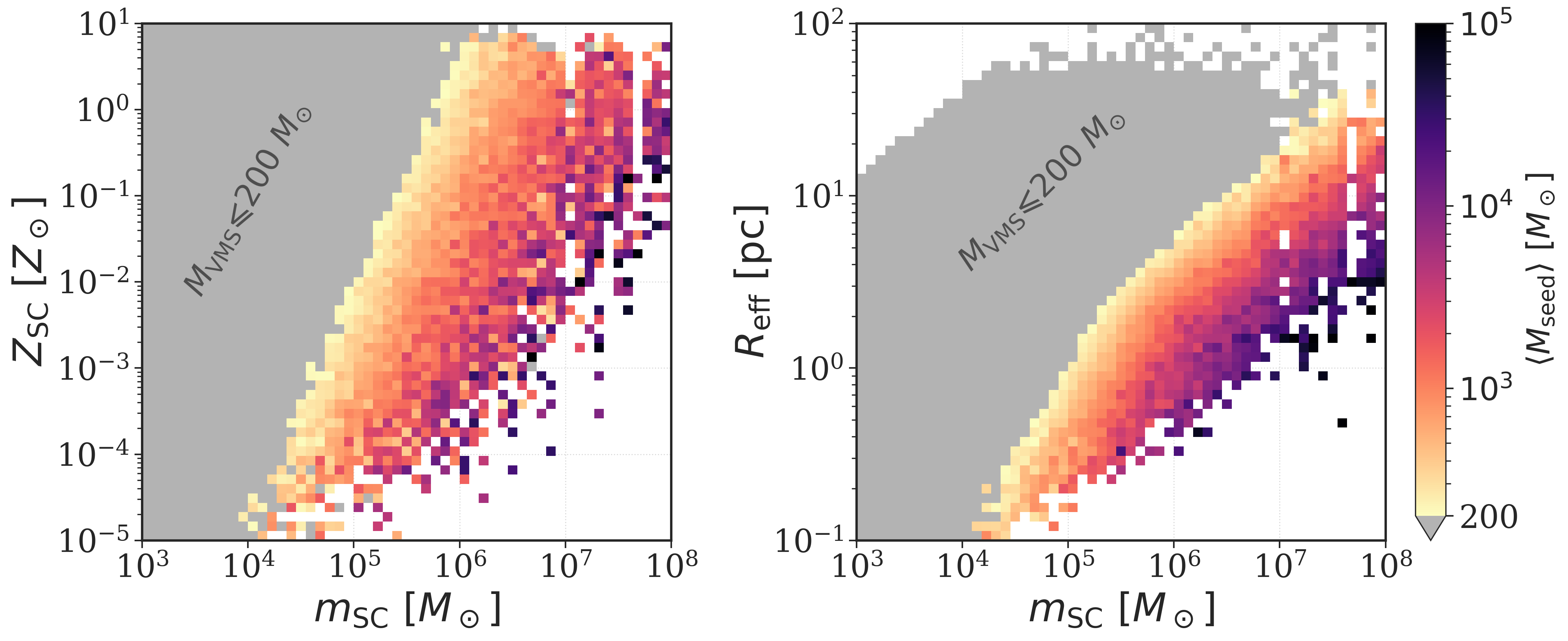}
    \caption{The initial properties of star clusters at formation in \SCnR\ at $z\geqslant 6$.
    \textit{Left:} The distribution of star cluster metallicity $Z_{\rm SC}$ versus star cluster mass $m_{\rm SC}$. 
    \textit{Right:} The star cluster  effective radius $R_{\rm eff}$ versus $m_{\rm SC}$. Both panels are colored by the averaged MBH seed mass $\left<M_{\rm seed}\right>$ in each pixel. The parameter spaces where the VMS cannot collapse to an MBH seed (i.e., hosting $M_{\rm VMS}\leqslant 200\ M_{\odot}$) are colored in gray. 
    }\label{fig:SC_population}
\end{figure*}

\section{Results}\label{sec:results}

In this section we present our results. Fig.~\ref{fig:large_vis} provides a multi-scale visualization of the most massive halo in the \SCnR\ run at $z=6$, from the cosmological environment to the central galaxy. 
MBHs are shown as crosses, with marker size scaling with MBH mass. We change the color of MBH markers in each panel for visual clarity.

\subsection{StarFOF Catalog}

The StarFOFs identified in our simulations represent stellar structures likely to host subgrid populations of star clusters. 
In Fig.~\ref{fig:StarFOF_vs_obs}, we compare the mass and size distribution of the StarFOFs in \SCnR\ at $z=6$ with the observed high-redshift ($z\gtrsim 6$) star cluster complexes and star-forming clumps. 
The observational sample includes Cosmic Gems \citep{Adamo2024_cosmicGems, Messa2026_CosmicGems_new}, 
Firefly Sparkle \citep{Mowla2024_firefly}, Sunrise Arc \citep{Vanzella2023}, Bullet Arc \citep{Bradac2025_BulletArc}, Cosmic Grapes  \citep{Fujimoto2025_CosmicGrapes}, and Cosmic Archipelago \citep{Messa2025_CosmicArchipelago}.
For each observed system, we define the system mass as the summed stellar mass of its cluster members. 
We characterize the system size by twice the effective radius, $2R_{\rm e}$, when it is available. Otherwise, we use the maximum projected separation among the compact members in the source plane.
For the Cosmic Gems, the lower bound on the size includes only the five confirmed clusters, and the upper bound includes the additional candidate clusters reported by \citet{Messa2026_CosmicGems_new}.
Because this observational sample combines different size estimators and includes both confirmed star cluster complexes and star-forming clumps (i.e., Bullet Arc and Cosmic Grapes), Fig.~\ref{fig:StarFOF_vs_obs} should be interpreted as a qualitative consistency check rather than a strict one-to-one comparison. 
The simulated StarFOFs cover the mass range of the observed systems and overlap with the upper end of their size distribution. 
However, the most compact systems, such as the Cosmic Gems, remain unresolved at our limited resolution.

We show the StarFOF population within the most massive DM-FOF halo in \SCnR\ from $z=12$ to $z=8$ in Fig.~\ref{fig:Sfof_evolve}. 
At $z=12$, this DM-FOF hosts three StarFOFs (blue circles).
They are all located in the central regions of separate stellar clumps, with some stars at the outskirts not included in the StarFOFs. 
They are all massive enough to produce MBH seeds (purple crosses), and each of them hosts one MBH seed. 
We only plot the newly formed MBHs in each panel: those seeded within $10$ Myr before the snapshot redshift, so the marked MBHs are those formed \textit{in situ} within the DM-FOF.
As the DM-FOF halo grows, more and more StarFOFs form within the DM-FOF halo. 
Small StarFOFs appear at the outskirts of the DM-FOF halo, while most of them do not host MBH seeds because they are not massive enough.
At the same time, the star formation in the central region of the DM-FOF halo is still active. As we mentioned in Section~\ref{sec:method_seed}, the star cluster-based seeding model allows BH seeds to form in StarFOFs that already host MBHs, while the newly formed star clusters are solely based on the star particles that have not previously contributed to MBH seeding. 
This means that when star formation is still active, the central nuclear star cluster can continue to form new star clusters and produce new MBH seeds: at $z=8$, the central StarFOF has produced three MBH seeds.  
This DM-FOF halo is also eligible for seeding in the halo-based and gas-based seeding models, and would host one MBH seed with $M_{\rm seed}\approx 10^{5}\,M_{\odot}/h$ if either of these two seeding models is applied.
This indicates that the star cluster-based seeding model can capture small-scale star-forming structures within individual DM-FOF halos. 
Moreover, without the restriction of one MBH seed per DM-FOF, the star cluster-based seeding model can produce MBHs as star formation continues. Both of these features make the MBH seeds form in a highly clustered way. As shown later, this can lead to a high MBH merger rate and rapid growth of MBHs in the early Universe.

\subsection{Star Cluster Populations}

Within each StarFOF, we generated star cluster populations according to the prescription described in Section~\ref{sec:method_starcluster}. 
In Fig.~\ref{fig:SFRD_Msc} we show the redshift evolution of the total star formation rate $\dot{M}_{\star}$ and the star cluster formation rate $\dot{M}_{\rm SC}$ within the most massive DM-FOF halo in \SCnR.
The cluster formation rate $\dot{M}_{\rm SC}$ closely traces the overall star formation history of this halo. The star cluster formation efficiency $\dot{M}_{\rm SC}/\dot{M}_{\star}$ (the inset panel) remains $\gtrsim 40\%$. 
This halo forms within a constrained $5\sigma$ density peak and corresponds to an extreme overdense region of the early Universe, so its cluster formation history should not be interpreted as representative of the average cluster formation rate. 

\begin{figure*}[htbp]
    \centering
    \includegraphics[width=1\linewidth]{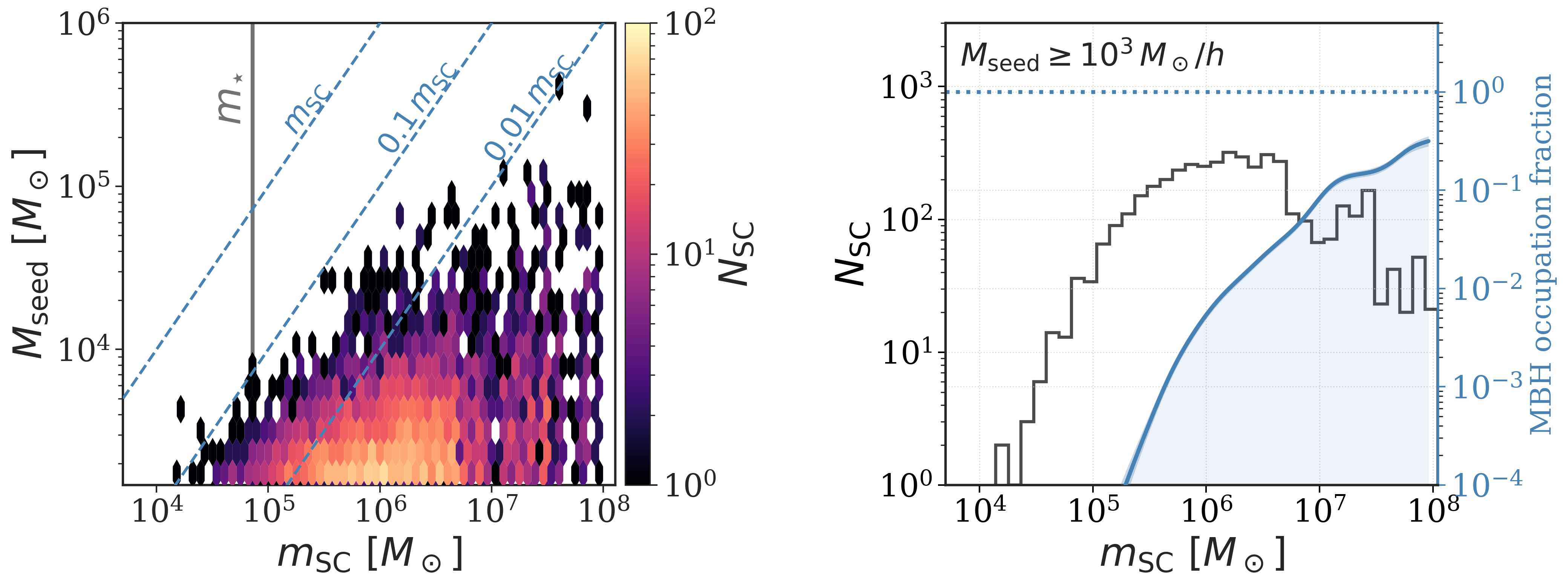}
    \caption{The MBH seeds and their host star clusters in the \texttt{SC-noR} run. 
    \textit{Left:} The distribution of MBH seed mass $M_{\rm seed}$ and their host star cluster mass $m_{\rm SC}$ colored by the star cluster number $N_{\rm SC}$ in each bin. 
    The gray vertical line marks the star particle mass adopted in the simulation, $m_{\star}=5\times 10^{4}\ M_{\odot}$, while the blue dashed lines indicate $M_{\rm seed}/m_{\rm SC}=1$, $0.1$, and $0.01$. 
    \textit{Right:} The distribution of star cluster masses $m_{\rm SC}$ that produce MBH seeds with $M_{\rm seed}\geqslant 10^{3}\ M_{\odot}\,h^{-1}$ (black histogram; left axis) and the fraction of star clusters that host MBH seeds above the same mass limit (blue curves; right axis).
    The horizontal dotted line marks the 100\% occupation fraction.
    Note that we only track the evolution of star clusters hosting MBH seeds, which are typically massive and compact, while most low-mass clusters are not followed. 
    }\label{fig:CWmodel_method}
\end{figure*}

To further validate our star cluster formation model, in Fig.~\ref{fig:SC_obs} we compare the star cluster population in individual galaxies with the local surveys: LEGUS \citep{Cook2019, Cook2023}, the HiPEEC \citep{Adamo2020}, and PHANGS-HST \citep{Lee2022}.
Current observations of star clusters at high redshift are limited to strong-lensing systems, which have a small sample size and lack information of host galaxies. 
This makes the statistical comparison challenging at high redshift, and we therefore compare our simulations with observations of nearby galaxies. 
To match the observational sample, we consider only massive young star clusters with ages $\leq 10\ {\rm Myr}$ and masses $m_{\rm SC}\geqslant 10^{5}\ M_{\odot}$.
The left panel shows the fraction of the stellar mass formed over the past $10$~Myr that is contained in these clusters, while the right panel shows their number $N_{\rm SC}$. Both quantities are plotted as a function of the galaxy SFR.

The star cluster number $N_{\rm SC}$ increases rapidly with the larger galaxy SFR, while the mass fraction of star formation in clusters shows a weak correlation with SFR.
Galaxies hosting no target star clusters are plotted with downward arrows at the horizontal dashed lines. In both simulation and observational samples, these galaxies are concentrated toward the low-SFR end: ${\rm SFR} \lesssim 1\ M_{\odot}\,{\rm yr}^{-1}$.
Even though our simulation is for a redshift range different from those of these local surveys, the simulated star clusters agree well with the observations in both panels. 
This agreement across widely separated epochs supports the idea that star cluster formation follows broadly similar physical processes across cosmic time \citep[e.g.,][]{Adamo2020, Krumholz2019}.

We then present the properties of individual star clusters. Fig.~\ref{fig:SC_population} shows the distribution of cluster metallicity ($Z_{\rm SC}$, left panel) and effective radius ($R_{\rm eff}$, right panel) as functions of star cluster mass. 
Our star cluster population covers a wide range of metallicity, mass, and size. 
We color both $Z_{\rm SC}-m_{\rm SC}$ and $R_{\rm eff}-m_{\rm SC}$ panels by the average mass of MBH seeds $\left<M_{\rm seed}\right>$ in each pixel. 
The parameter space where the VMS cannot collapse to an MBH seed, i.e., host $M_{\rm VMS}\leqslant 200\ M_{\odot}$, is colored gray.
The resultant MBH seed mass strongly depends on star cluster properties: massive seeds form preferentially in massive, compact, and metal-poor star clusters.

\begin{figure*}[!t]
    \centering
    \includegraphics[width=0.9
    \linewidth]{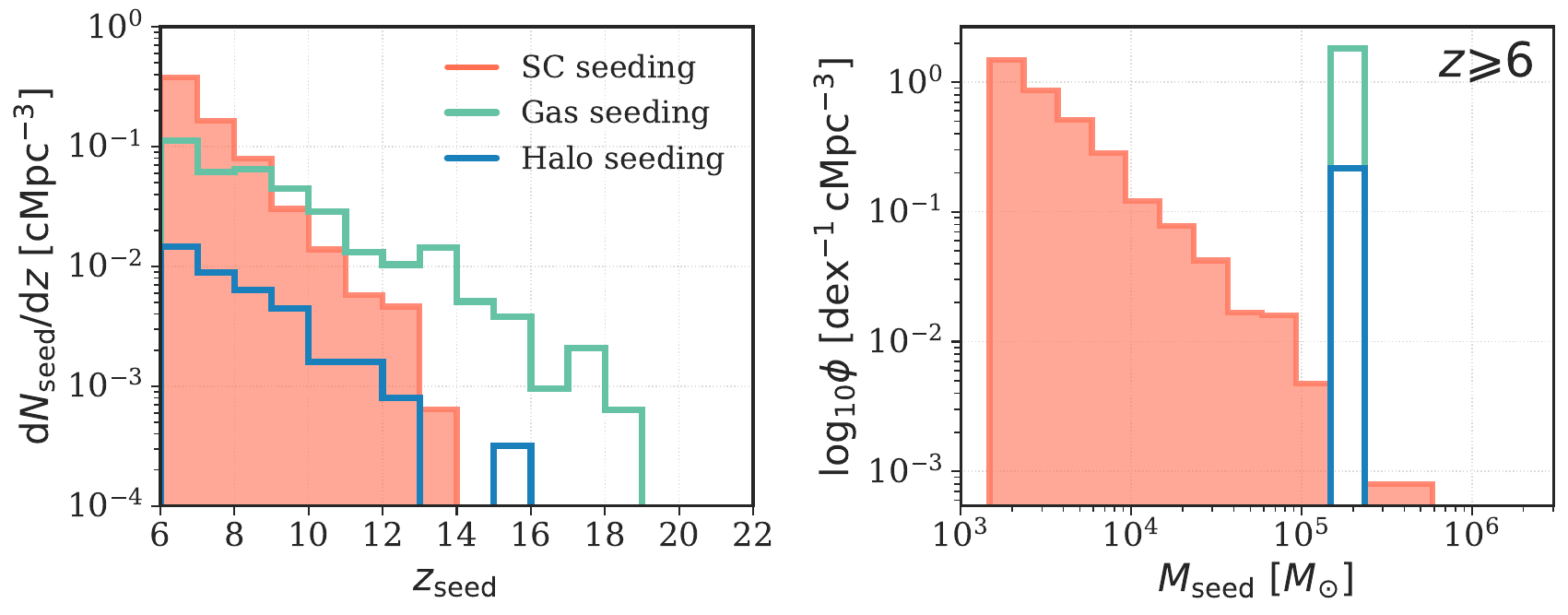}
    \caption{The one-dimensional distribution of the MBH formation redshift $z_{\rm seed}$ (left) and mass $M_{\rm seed}$ (right) for $z\geqslant 6$ with  different seeding methods: star-cluster-based seeding (\SCnR, red), gas-based seeding (\Gas, green), and halo-based seeding (\Halo, blue).
    }\label{fig:vis_number_seed}
\end{figure*}

\begin{figure}[htbp]
    \centering
    \includegraphics[width=1\linewidth]{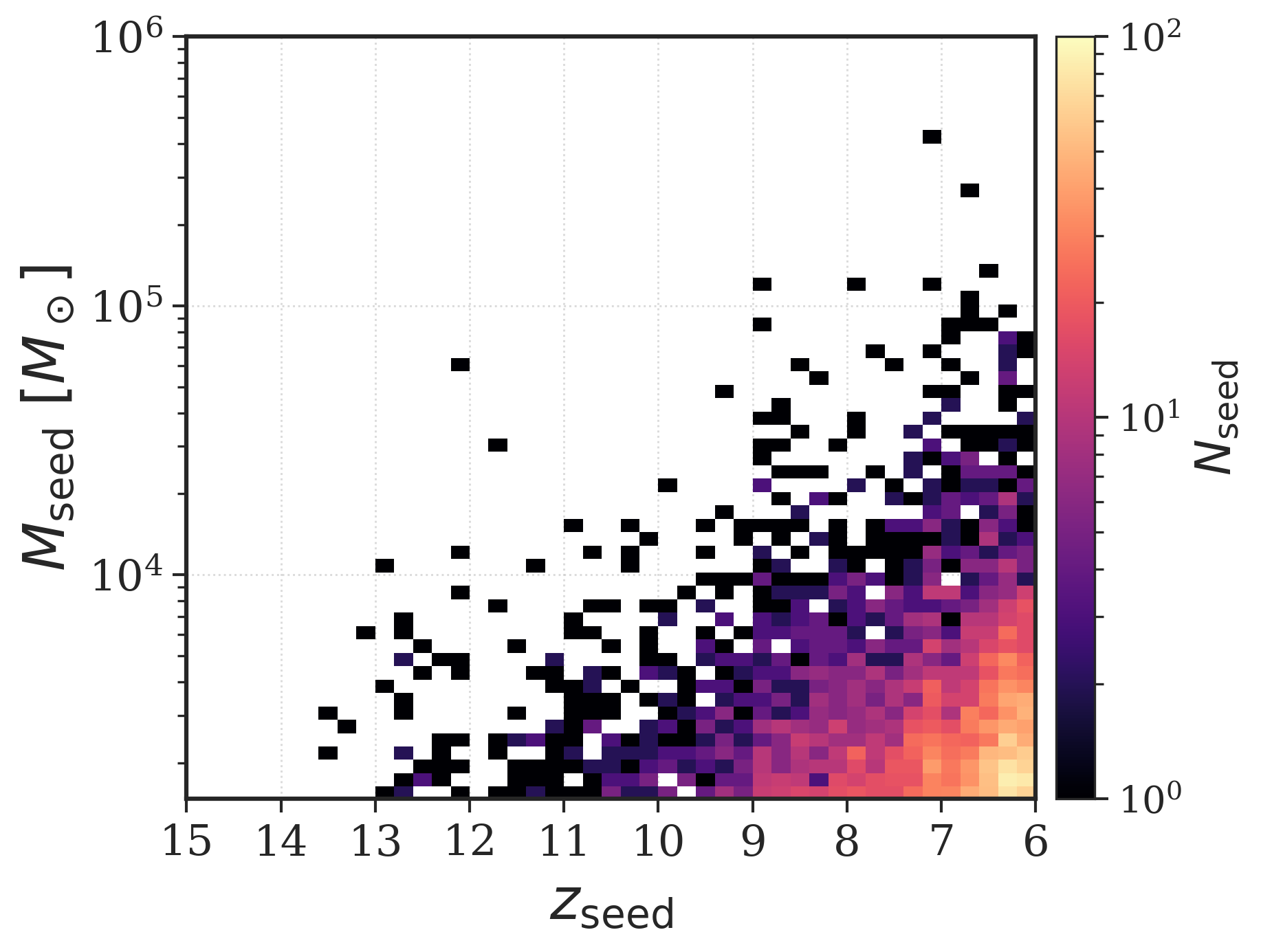}
    \caption{The distribution of MBH seed mass $M_{\rm seed}$ and seeding redshift $z_{\rm seed}$ for \SCnR. We color the plane according to the seed number $N_{\rm seed}$ in each pixel.  
    }\label{fig:seed_mass_vs_z}
\end{figure}

\begin{figure*}[!t]
    \centering
    \includegraphics[width=1\linewidth]{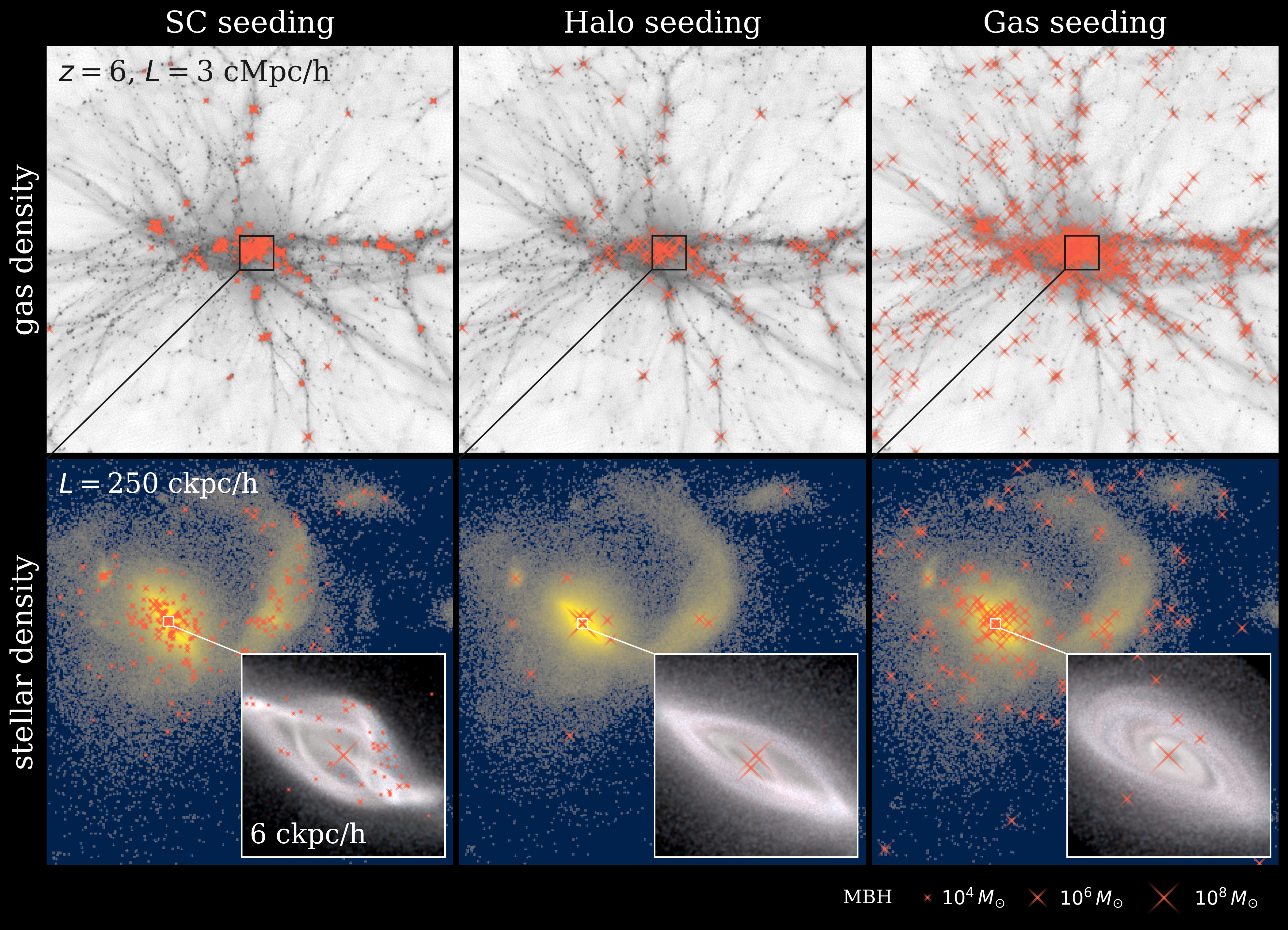}
    \caption{An illustration of the central halo at $z=6$ in the run \SCnR\ (left), \Halo\ (middle), and \Gas\ (right). 
    In each column, we show the gas density field within an $L=3\ {\rm Mpc/h}$ region around the most massive DM-FOF halo (upper), and the stellar density within a zoom-in region of $L=250\ {\rm ckpc/h}$ (lower). 
    The lower-inset panels show the central galaxy of this halo within a region of $L=6\ {\rm ckpc/h}$. 
    We overlay MBH positions (red markers) in each panel, and their marker size is proportional to the MBH mass, as indicated by the legends. 
    }\label{fig:vis_compare}
\end{figure*}

\subsection{MBH Seeds Formed in Star Clusters}
With the star cluster population generated in StarFOFs, MBH seeds are formed following the prescription described in Section~\ref{sec:method_seed}. 
In this section we present the MBH seed populations formed in the \SCnR\ run. 
The left panel of Fig.~\ref{fig:CWmodel_method} shows the distribution of MBH seed mass $M_{\rm seed}$ and their host star cluster mass $m_{\rm SC}$ colored by the sample number $N_{\rm SC}$ in each bin. 
For a given $m_{\rm SC}$, the seed mass peaks at low $M_{\rm seed}$. The upper bound of $M_{\rm seed}$ increases with $m_{\rm SC}$ for $m_{\rm SC}\lesssim 10^{6}\ M_{\odot}$ and flattens at approximately $M_{\rm seed}\sim 10^{5}\ M_{\odot}$. 
The rare massive seeds with $M_{\rm seed}\gtrsim 10^{5}\,M_{\odot}/h$ are exclusively formed in the most massive star clusters with $m_{\rm SC}\gtrsim 10^{6}\,M_{\odot}/h$.

The black histogram on the right panel shows the one-dimensional distribution of the mass of the star cluster $m_{\rm SC}$ that produces MBH seeds with $M_{\rm seed}\geqslant 10^{3}\,M_{\odot}/h$.
These clusters span a mass range of $10^{4}\lesssim m_{\rm SC}\lesssim 10^{8}\,M_{\odot}$, and peak around $10^{6}\ M_{\odot}$, similar to the clusters observed by JWST (e.g., Cosmic Gem \citep{Adamo2024_cosmicGems} and Firefly \citep{Mowla2024_firefly}).
The blue curve in the right panel shows the MBH occupation fraction as a function of $m_{\rm SC}$, i.e., the fraction of star clusters within the given mass bin that produce a seed with $M_{\rm seed}\geqslant 10^{3}\, M_{\odot}/h$. 
This occupation fraction increases rapidly with $m_{\rm SC}$, and reaches $\approx 30\%$ in the highest mass bin.

Fig.~\ref{fig:vis_number_seed} compares the seeds formation redshift distribution $z_{\rm seed}$ and the seed mass distribution $M_{\rm seed}$ in \SCnR\ with those in the gas-based \Gas\ (green curves) and the halo-based seeding run \Halo\ (blue curve). 
Among these three models, \Gas\ produces the earliest seeds at $z\approx 20$, while both \SCnR\ and \Halo\ produce the first seed at $z\approx 15$. 
The later start of seeding in \SCnR\ and \Halo\ is expected because both models depend on sufficient star formation: in \SCnR\ massive star clusters are required to form MBH seeds, and in \Halo\ the DM-FOF halo must host a sufficient stellar mass. 
By contrast, \Gas\ requires only the presence of star-forming gas. 
Once cluster-based seeding begins, its seeding rate increases rapidly toward lower redshift. At $z=6$, 
the seeding rate in
\SCnR\ is about half an order of magnitude higher than that in \Gas.
By $z=6$, within this $L=12.5\ {\rm Mpc}$ box, \SCnR\ produces 4294 seeds in total, compared to 2279 in \Gas.

Two features of the star-cluster-based seeding model contribute to its larger abundance of MBH seeds. On the one hand, as illustrated in Fig.~\ref{fig:Sfof_evolve}, \SCnR\ identifies several stellar structures within a single DM-FOF halo, and each of these structures can host one or more MBH seeds. By contrast, both \Gas\ and \Halo\ allow only one MBH seed per DM-FOF halo. 
On the other hand, \SCnR\ extends the seed population down to lower mass: $M_{\rm seed} = 10^{3}\,M_{\odot}/h$, two orders of magnitude lower than the seed mass adopted in \Gas\ and \Halo\ $M_{\rm seed} = 10^{5}\ M_{\odot}\, h^{-1}$. In the right panel of Fig.~\ref{fig:vis_number_seed}, we can see that the MBH seeds in \SCnR\ are dominated by low-mass seeds. 
When we consider only seeds with $M_{\rm seed} \geqslant  10^{5}\ M_{\odot}\, h^{-1}$, the seed number of \SCnR\ is much lower than that in \Halo\ and \Gas. 

The overall seed mass distribution in \SCnR\ appears as a power-law distribution with a slope of $-1$. 
Rather than treating the MBH seed mass as a free parameter and fixing it at $M_{\rm seed}=10^{5}\ M_{\odot}\,h^{-1}$, the seed mass distribution in \SCnR\ emerges naturally from the properties of the host star clusters, which is physically motivated and correlated with local environments.

We show the correlation between seed mass $M_{\rm seed}$ and seeding redshift $z_{\rm seed}$ in Fig.~\ref{fig:seed_mass_vs_z} for the star-cluster-based model.
The formation of massive MBH seeds requires both a sufficiently massive and dense star cluster to drive runaway stellar collisions and a low-metallicity environment to limit stellar winds. 
The first seeds produced by star clusters at $z\sim 13.5$ have small masses $3\times 10^{3}\ M_{\odot}$. Most seeds in $z\gtrsim 9$ have masses $M_{\rm seed}\leqslant 10^{4}\ M_{\odot}$, while massive seeds appear at later times. This is because the StarFOFs at the highest redshift do not have sufficient mass budget to produce massive star clusters and, therefore, massive seeds. This can be seen from Fig.~\ref{fig:Sfof_evolve}: at $z=12$, the StarFOFs are in general small and have lower $\Gamma$. 
The most massive seeds in \SCnR\ formed around $z=7$. Toward lower redshifts, the enriched gas environment with increasing metallicity would prevent the formation of very massive seeds. 
However, these low-mass seeds can subsequently grow through gas accretion and MBH mergers.

\subsection{ MBHs Populations from Different Seeding Models} 
\label{sec:results_MBH_population}

In this section we make a comparison of the overall MBH population produced by different seeding prescriptions.
We first present illustrations of the central DM-FOF halo at $z=6$ in the runs \SCnR, \Gas, and \Halo\ in Fig.~\ref{fig:vis_compare}. 
Each panel shows the stellar density field within a region of $L_{\rm box}=250\ {\rm ckpc/h}$ with the MBH marked by red crosses. 
The size of the red marker is proportional to the MBH mass, as indicated by the legend. The bottom-right insets show the central galaxy on a smaller scale of $L=6\ {\rm ckpc/h}$.
Because these runs share the ICs, their large-scale structures are similar, while their MBH populations are substantially different. 
The \Halo\ run produces the fewest MBHs. 
With the same fixed seed mass, \Gas\ contains a more abundant MBH population that is more spatially extended. 
Because \Gas\ traces the star-forming gas rather than stars, it can produce MBHs at the outskirts of the stellar structures.  
The \SCnR\ run also hosts a large number of MBHs, which closely trace the stellar structures. The inset panel shows that many low-mass MBHs reside in the stellar disk of the galaxy. 
A few MBHs are outside the disks, and they all have small masses ($M_{\rm BH}\lesssim 10^{4}\ M_{\odot}$).
We show an edge-on view of this galaxy in the lower-right panel of Fig.~\ref{fig:large_vis}, distinguishing the newly seeded MBHs (cyan crosses) from the others (dark blue crosses). 
Most newly seeded MBHs (those formed within the previous 20 Myr) are near the stellar disk, where the cluster formation rate is high. This indicates they are the seeds formed \textit{in situ} within this galaxy. The old MBHs outside the disk are likely accreted from other galaxies.

\begin{figure}[!t]
    \centering
    \includegraphics[width=1\linewidth]{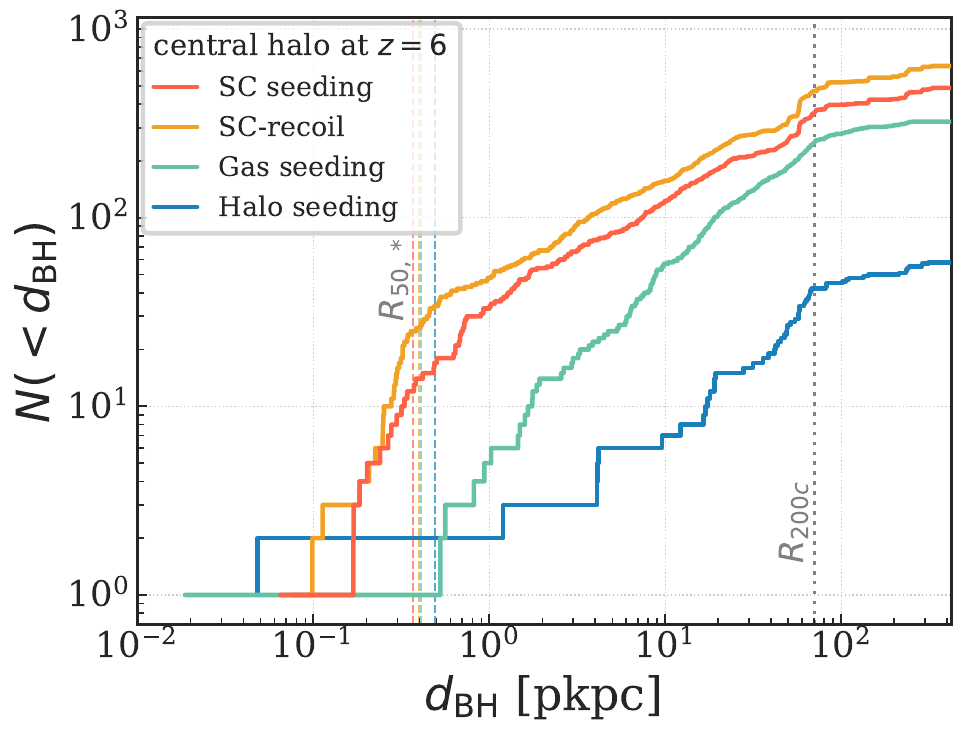}
    \caption{The cumulative distribution of the separation between MBHs and the halo center at $z=6$ for the run \SCnR\ (red), \SCR\ (yellow),
    \Halo\ (blue), and \Gas\ (green). We use the MBH population within the most massive DM-FOF halo in each simulation. 
    The vertical dashed line marks the stellar half-mass radius in each simulation, and the vertical dotted line marks the $R_{\rm 200c}$ for the DM-FOF halo in \SCnR.
    }\label{fig:BH_separation}
\end{figure}

The zoom-in galaxy images in Fig.~\ref{fig:vis_compare} illustrate a fundamental difference between these three seeding models. 
Because \Halo\ and \Gas\ allow at most one MBH to form in each DM-FOF halo, multiple MBHs can only appear in the same halo through hierarchical assembly. 
By contrast, in the star-cluster-based model, multiple MBHs can form \textit{in situ} within single halos. As illustrated in Fig.~\ref{fig:Sfof_evolve}, the central halo in \SCnR\ hosts multiple StarFOFs, each of which can produce one or more MBH seeds. 
This leads to a spatially clustered MBH population, which further increases the MBH merger rate and accelerates the early growth of MBHs.

To give a more quantitative comparison, Fig.~\ref{fig:BH_separation} shows the cumulative number of MBHs $N\left(<d_{\rm BH}\right)$, as a function of their physical separation from the halo center at $z=6$. 
We include all MBHs in the most massive DM-FOF halo of each simulation.
Within the stellar half-mass radius $R_{\rm 50,\star}$, there is only one MBH for \Gas\ and two for \Halo, while $\approx 20$ MBHs for \SCnR.
Beyond $R_{\rm 50,\star}$, the cumulative MBH number in \SCnR\ increases slowly with distance, indicating most of its MBHs are located around the central regions. 
While in \Gas, the MBH counts rise rapidly outside $R_{\rm 50,\star}$. This reflects a more spatially extended population, which is expected for the MBHs formed \textit{ex-situ} and gradually merging into the halo.

\begin{figure}[!t]
    \centering
    \includegraphics[width=1\linewidth]{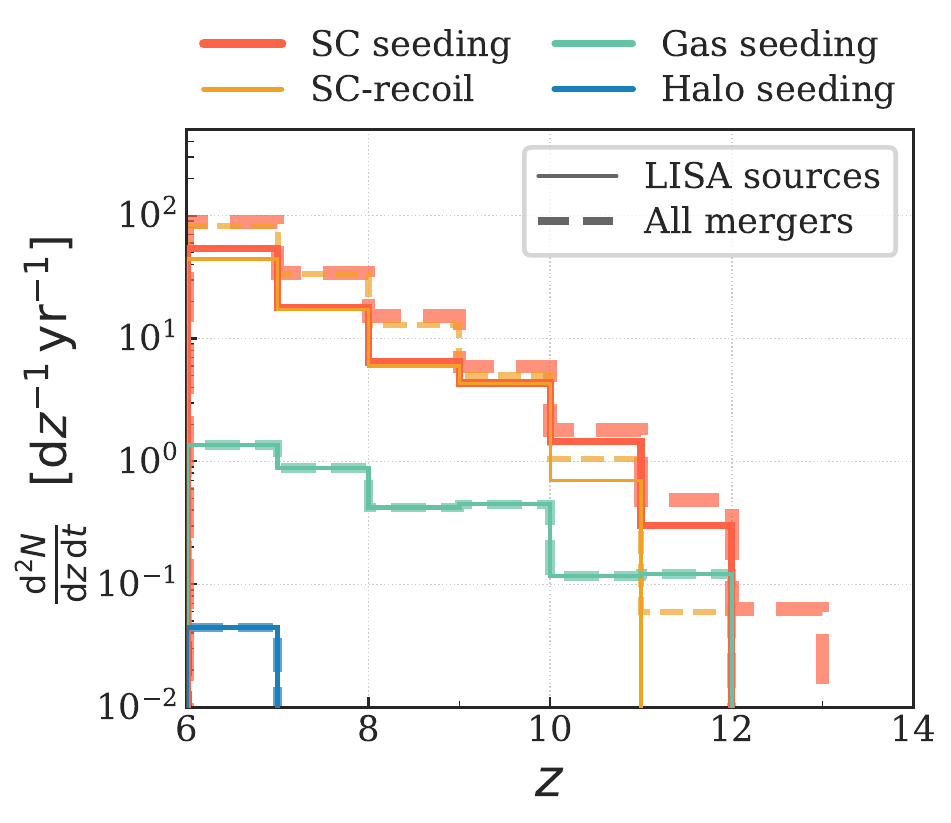}
    \caption{The MBH merger rate in the run \SCnR\ (red), \SCR\ (yellow),
    \Gas\ (green), and \Halo\ (blue) as a function of redshift. The dashed lines show all the mergers, and the solid curves are the mergers that can be detected by LISA.  
    }\label{fig:bh_mergers}
\end{figure}

\begin{figure*}[!t]
    \centering
    \includegraphics[width=0.9\linewidth]{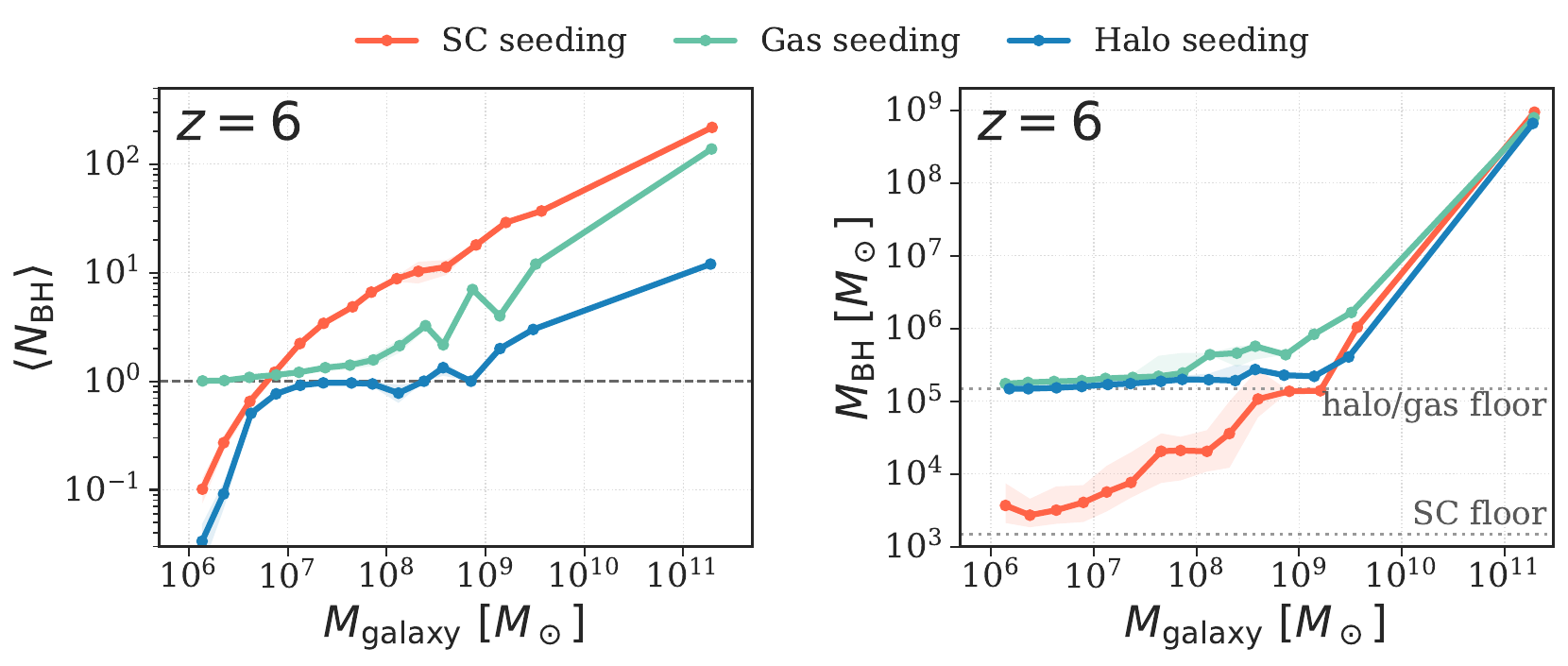}
    \caption{The number of MBHs (left panel) and the mass of the central MBH (right panel) as a function of galaxy mass at $z=6$ in the run \SCnR\ (red), \Gas\ (green), and \Halo\ (blue).  
    The curves show the averaged value among galaxies within the given mass bin. 
    In the left panel, we mark the $N_{\rm BH}=1$ level with a dashed line. In the right panel, we mark the smallest seed mass for Halo-seeding and Gas-seeding, $M_{\rm BH}=10^{5}\ M_{\odot}\,h^{-1}$, and for SC-seeding $M_{\rm BH}=10^{3}\ M_{\odot}\,h^{-1}$, with dotted lines. 
    }\label{fig:bh_gal_relation}
\end{figure*}

The enhanced central abundance of MBHs in \SCnR\ provides a pool of nearby merger partners, which leads to a higher MBH merger rate compared with \Gas\ and \Halo.
In Fig.~\ref{fig:bh_mergers}, we show the MBH merger rate as a function of redshift for the three seeding prescriptions. 
At $z=6$, the merger rate in \SCnR\ exceeds that in \Gas\ by two orders of magnitude.
This is not only caused by the larger number of MBH seeds, because \SCnR\ produces only twice the number of seeds as \Gas\ by $z=6$, which cannot explain the large difference in the merger rate. The spatial clustering of MBHs is therefore the main reason.  

These high-redshift seed mergers are targets for the Laser Interferometer Space Antenna (LISA). In Fig.~\ref{fig:bh_mergers}, the solid curves show the mergers that are detectable by LISA. 
We calculate the LISA signal-to-noise ratio (SNR) following \citet{Wang2025} with the assumption of circular orbits and classify events with ${\rm SNR}\geqslant 10$ as detectable.
We generate $10^{4}$ realizations and randomly place an observational window of $ 4$ years within the last 100 years prior to coalescence. 
The plotted LISA detection rate is scaled to the whole-sky volume. 
Because these rates are extrapolated from a rare $5\sigma$ overdense region, the actual LISA observation is expected to be lower. 
Here we focus only on the comparison between different seeding models. 
Our predicted LISA rate only includes merger between the MBHs tracked in our simulations. Additional GW sources involving stellar-mass BHs, which are not modeled here, could further increase the total LISA detection rate. 
The fixed seed mass of $M_{\rm seed}=10^{5}\ M_{\odot}h^{-1}$ adopted in \Halo\ and \Gas\ produces sufficiently large chirp masses that almost all mergers in these runs are detectable by LISA. 
In contrast, most seeds in \SCnR\ have masses of $M_{\rm BH}\sim 10^{3}\ M_{\odot}h^{-1}$, producing weaker GW signals at frequencies at the edge of the LISA band, so only a fraction of the mergers are detectable. 
However, because of the substantially enhanced merger rate, \SCnR\ still predicts a much larger LISA event rate. 
This suggests that the high-redshift MBH merger population that will be revealed by LISA could provide a direct probe of MBH seeding.

\subsection{MBH-Galaxy Relation and Occupation Fraction}

We next investigate how MBHs depend on their host galaxies. 
In Fig.~\ref{fig:bh_gal_relation}, we show the mean number of MBHs ($\langle N_{\rm BH}\rangle$; left column) and the mass of the most massive MBH (right column) as functions of galaxy mass at $z=6$. 
The different seeding models produce distinct trends in $\langle N_{\rm BH}\rangle$.
In \Gas, $\langle N_{\rm BH}\rangle$ reaches approximately unity at $M_{\rm galaxy}\sim 10^{6}\,M_{\odot}$. 
While at the same $M_{\rm galaxy}$ in \SCnR\ and \Halo\, only $\lesssim 10\%$ of galaxies host MBHs. 
By $M_{\rm galaxy}= 10^{7}\, M_{\odot}$, the mean number of MBHs becomes $\approx 1$ in all three runs.
At higher masses, however, the models diverge significantly. 
In \Halo, $\left<N_{\rm BH}\right>$ remains close to unity over a wide galaxy mass range, and begins to increase only at $M_{\rm galaxy}\gtrsim 10^{9}\, M_{\odot}$.
The more permissive gas-based seeding in \Gas\ allows multiple MBHs to appear at $M_{\rm galaxy}=10^{7-8}\ M_{\odot}$ by hierarchical assembly. 
In \SCnR, the MBH number increases more rapidly with galaxy mass. 
Multiple MBHs are common in galaxies with $M_{\rm galaxy}\gtrsim10^{7}\, M_{\odot}$, and the mean number of MBHs stays about half an order of magnitude higher than \Gas\ at similar $M_{\rm galaxy}$. 
The stronger dependence reflects the direct connection between MBH seeding and host environments: more massive galaxies host larger stellar populations (see Fig.~\ref{fig:SC_obs}), allowing multiple MBH seeds to form within individual galaxies. Hierarchical assembly can further enhance this trend by bringing MBHs formed in progenitor systems. 
Therefore, in the star-cluster-based seeding model, the abundance of MBHs is more directly correlated with the properties of their host galaxies. 
\begin{figure*}[!t]
    \centering
    \includegraphics[width=1\linewidth]{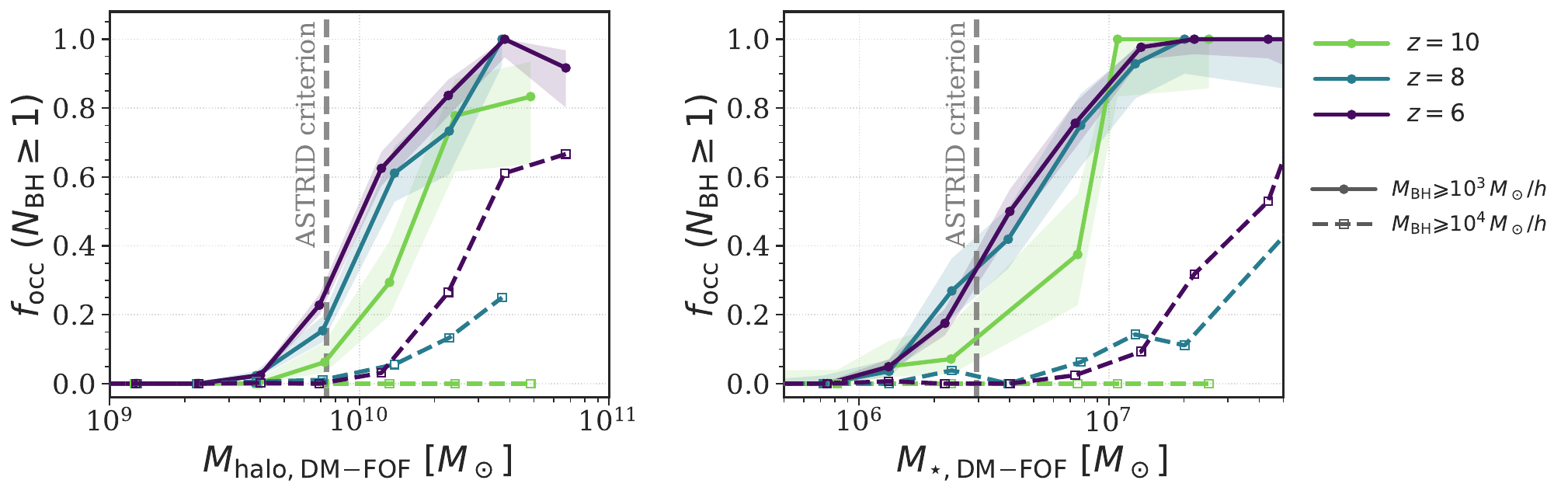}
    \caption{BH occupation fraction in \SCnR\ as a function of DM-FOF total mass (left) and stellar mass (right) at different redshifts: $z=10$ (green), $z=8$ (blue), and $z=6$ (purple). 
    In both panels, we plot the mass bin with at least 3 halos, and the error bars show the 1$\sigma$ uncertainty.
    The solid curves show the halo fraction that hosts MBHs with $M_{\rm BH}\geqslant 10^{3}\ M_{\odot}\,h^{-1}$, and the dashed curves are those hosting MBHs with $M_{\rm BH}\geqslant 10^{4}\ M_{\odot}\,h^{-1}$.
    The vertical dashed lines mark the halo total mass and stellar mass thresholds for the halo-based seeding model, respectively.  
    }\label{fig:bh_occ}
\end{figure*}

The $M_{\rm BH}\!-\!M_{\rm galaxy}$ scaling relation in the right panel of Fig.~\ref{fig:bh_gal_relation} shows a similar environmental dependence in the MBH masses. 
Although the central MBHs in the most massive galaxies converge to similar masses across these three runs, which we will investigate in more detail in Section~\ref{sec:mass_evolution}, the $M_{\rm BH}$ in low-mass galaxies shows significant differences.
With the relatively large $M_{\rm seed}$ in \Gas\ and \Halo, the central MBHs in galaxies with $M_{\rm galaxy}\sim 10^{6-9}\,M_{\odot}$ are bounded by this imposed mass floor, hosting a $10^{5}\, M_{\odot}$ MBH. 
The limited gas reservoirs in these small hosts are generally insufficient to support efficient MBH growth, so many MBHs remain close to their initial seed mass, which produces a nearly flat $M_{\rm BH}\!-\! M_{\rm galaxy}$ relation at the low-mass end. 
Consequently, in gas- and halo-based seeding models, the MBH mass in dwarf galaxies is largely imposed by the seeding prescription rather than by their host environments. 
\citet{Haidar2022} found similar results: in cosmological simulations that adopt halo-based seeding, the MBHs in low-mass galaxies are too massive relative to current observational constraints.

The star-cluster-seeding model, in contrast, extends the seed population down to $M_{\rm seed}=10^{3}\,M_{\odot}h^{-1}$ and produces a much stronger dependence of $M_{\rm BH}$ on host galaxy mass.
The central MBH mass increases from a few $\times 10^{3}\ M_{\odot}$ in galaxies with $M_{\rm galaxy}=10^{6}\ M_{\odot}$ to $10^{5}\ M_{\odot}$ at $M_{\rm galaxy}=10^{9}\ M_{\odot}$
In low-mass galaxies, the overall low gas density produces fewer stars and smaller $\Gamma$, 
making the formation of massive star clusters unlikely and therefore producing lower-mass MBH seeds. 
With $M_{\rm galaxy}$ increasing, the larger stellar population and star-forming environments allow more numerous and more massive star clusters to form, producing both more abundant
and more massive MBH seeds.
Therefore, the star-cluster-based model naturally correlates both the abundance and mass of MBH with their local environments, providing a more physically motivated framework for predicting the MBH populations in dwarf galaxies.

\begin{figure*}[htbp]
    \centering
    \includegraphics[width=0.8\linewidth]{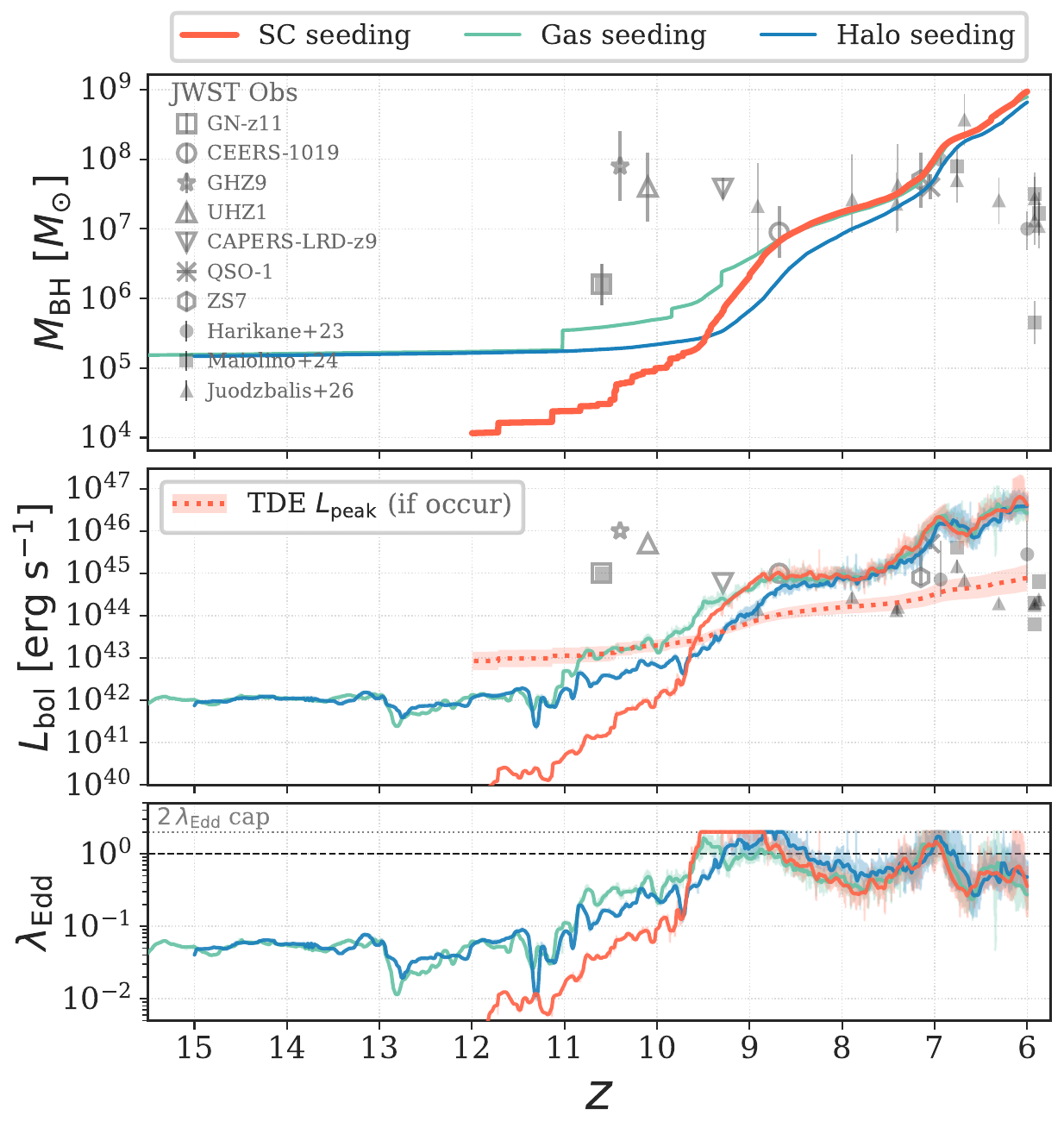}
    \caption{Evolution of the most massive MBH by $z=6$ with different seeding prescriptions: star-cluster-based seeding (\SCnR, red), gas-based seeding (\Gas, green), and halo-based seeding (\Halo, blue). 
    We follow the main branch of the MBH, i.e., the primary progenitor in each merger. 
    From top to bottom, we show the mass evolution (top), bolometric luminosity (middle), and Eddington fraction $\lambda_{\rm Edd}$ (bottom). In the middle panel, the solid curves are the luminosity from gas accretion. 
    For the star-cluster-based model, the red dotted curve shows the expected peak luminosity if a TDE occurs at a given epoch, rather than a continuous contribution to MBH luminosity. 
    Gray symbols with error bars show the high-redshift MBHs detected by JWST, as indicated in the legend. 
    }\label{fig:MBH_mass_evolution}
\end{figure*}

\begin{figure*}[htbp]
    \centering
    \includegraphics[width=1.0\linewidth]{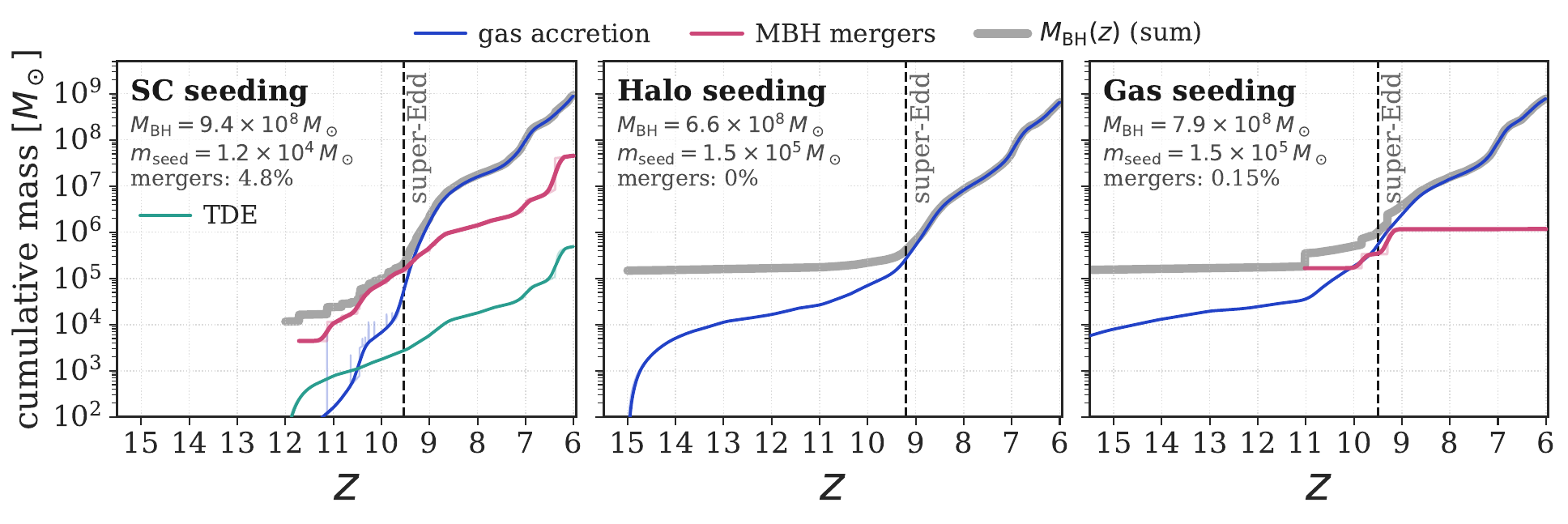}
    \caption{Cumulative mass contributed from different growth channels for the most massive MBHs by $z=6$ in \SCnR\ (left), \Halo\ (middle), and \Gas\ (right). 
    In each panel, the thick gray curve is the MBH mass evolution $M_{\rm BH}$; the blue, red, and green curves show the cumulative mass from gas accretion, MBH mergers, and TDEs, respectively.  
    Only the \SCnR\ has the TDE channel. 
    Curves are smoothed over $\Delta z=0.1$, and the unsmoothed data are plotted underneath with fainter colors. 
    The vertical dashed line marks the redshift when the MBH enters super-Eddington accretion. 
    The final MBH mass at $z=6$, the original seed mass, and the fraction of the final mass contributed by mergers are labeled in the upper left of each panel. 
    }\label{fig:mmbh_massbudget}
\end{figure*}

Finally, we examine the MBH occupation fraction, $f_{\rm occ}$, in \SCnR\ as a function of DM-FOF total mass and stellar mass at $z=10$, $z=8$, and $z=6$ in Fig.~\ref{fig:bh_occ}. 
We use this to assess how well the fixed halo and stellar mass threshold adopted in the halo-based seeding model approximates the MBH occupation fraction predicted by the star-cluster-based model. 
In both panels, the vertical dashed lines indicate the mass threshold adopted in \Halo\ and ASTRID.
Rather than changing steeply at a specific halo mass as assumed by the halo-based seeding, $f_{\rm occ}$ in \SCnR\ gradually increases with halo mass and evolves with redshift.  
At the ASTRID halo-mass threshold $M_{\rm halo}=5\times 10^{9}\,M_{\odot}h^{-1}$, the MBH occupation fraction with $M_{\rm BH}\geqslant 10^{3}\ M_{\odot}$ remains below unity, increasing from $\sim 10\%$ at $z=10$ to $\sim 30\%$ at $z=6$. 
The occupation fraction of more massive MBHs, $M_{\rm BH}\geqslant 10^{4}\ M_{\odot}$, is significantly lower. At $z=6$, these MBHs only appear at $M_{\rm halo}> 10^{10}\ M_{\odot}$. 
A similar smooth transition is also seen in the right panel, which shows $f_{\rm occ}$ versus halo stellar mass. 
Rather than selecting a unique stellar-mass threshold, the star-cluster model produces a $f_{\rm occ}$ that increases continuously over $M_{\star,{\rm DM-FOF}}\sim 10^{6}\ M_{\odot}$ to $10^{7}\ M_{\odot}$. The transition also depends on both redshift and the adopted MBH mass.   
These results suggest that a fixed halo-mass or stellar-mass threshold can capture the broad tendency of $f_{\rm occ}$, but cannot reproduce the gradual and redshift-dependent occupation predicted by the star-cluster seeding model. 


\subsection{Mass Evolution of the Central MBH}
\label{sec:mass_evolution}

\begin{figure}[htbp]
    \centering
    \includegraphics[width=1\linewidth]{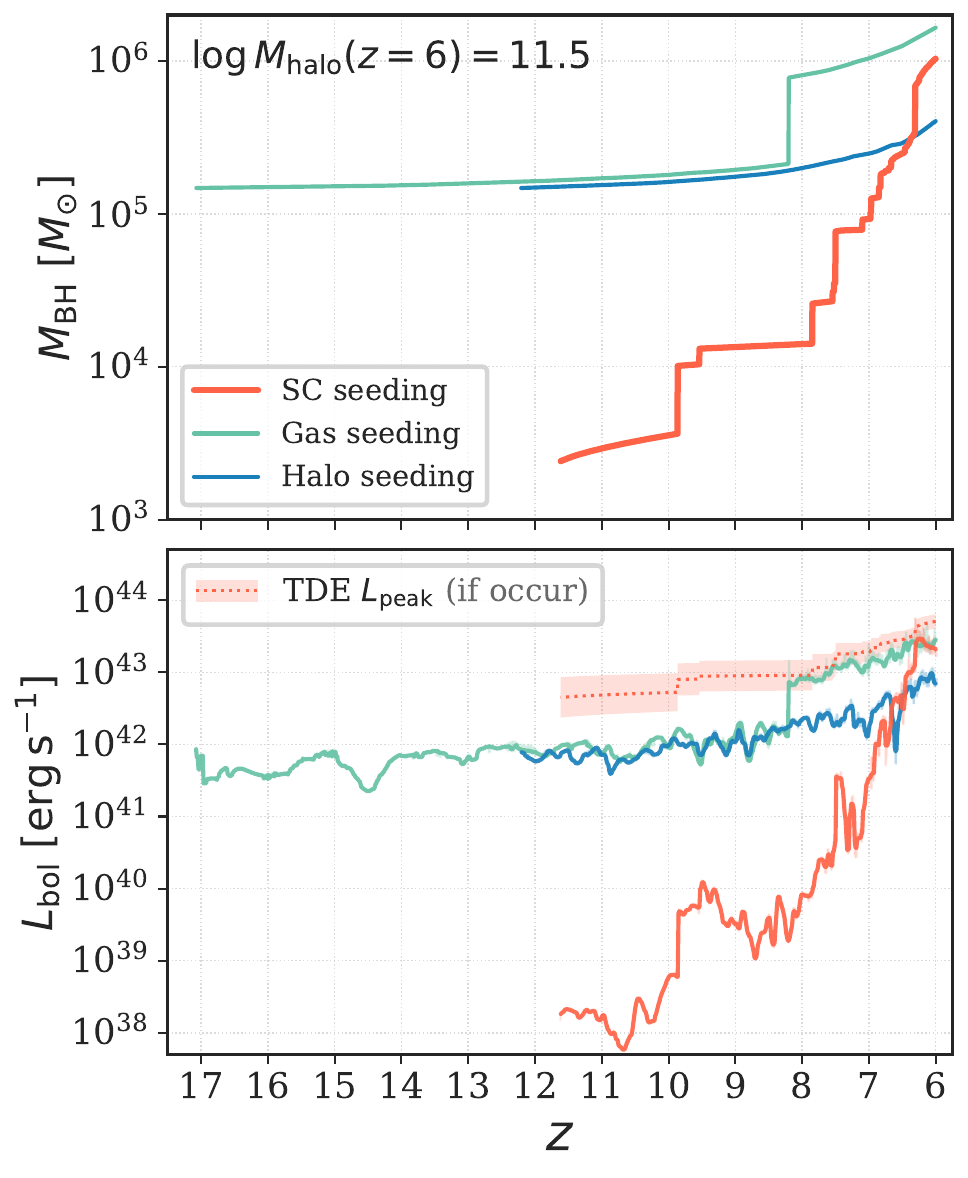}
    \caption{Same as the upper and middle panels in Fig.~\ref{fig:MBH_mass_evolution}, but for the central MBH in the second most massive halo at $z=6$.
    }\label{fig:sec_mmbh_evo}
\end{figure}

As described in Section~\ref{sec:setup}, we use constrained simulations of a $5\sigma$ density peak to study early MBH growth in an extremely overdense region. 
The peak collapses into a halo of $\sim 2\times 10^{11}\ M_{\odot}$ at $z=12$, which grows to $10^{12}\ M_{\odot}$ by $z=8$ and $5\times 10^{12}\ M_{\odot}$ by $z=6$.
In this section, we investigate how the choice of seeding prescription affects the evolution of the central MBH in this massive halo.

For each run, we identify the most massive MBH at $z=6$ and trace its evolution backward in time. 
In Fig.~\ref{fig:MBH_mass_evolution}, we show the mass evolution of the main branch of the MBH merger tree, i.e., the primary progenitor in each merger. 
Gray markers are high-redshift MBH populations observed by JWST.  
Across the runs, the MBH mass converges at $z\lesssim 7$ and reaches $M_{\rm BH}\approx 10^{8.6}\ M_{\odot}$ by $z=6$. However, the various seeding prescriptions lead to different early growth.
The MBHs seeded by the gas-seeding model and star-cluster-seeding model experience rapid growth at $z\gtrsim 8$, producing a more massive MBH than the halo-seeding model at high redshift.
This early growth is more consistent with the JWST observations. 
The rapid growth in \SCnR\ is particularly interesting because, compared to the \Gas\ run, it starts from a much less massive seed ($\log M_{\rm seed}= 4.18$ versus $\log M_{\rm seed}=5.32$) at a much later redshift ($z_{\rm seed}=12$ versus $z_{\rm seed}=18.6$), while it manages to produce a similar MBH mass $\log M_{\rm BH}=6.5$ as that in \Gas\ by $z=9$.  
At $z\approx 9$, the simulated luminosity from \SCnR\ and \Gas\ overlaps with several JWST sources, while the MBH in \Halo\ does not reach the observed luminosity until $z=8$.

To better understand their growth history, the middle and bottom panels in Fig.~\ref{fig:MBH_mass_evolution} plot their bolometric luminosity $L_{\rm bol}$ and Eddington ratio $\lambda_{\rm Edd}$, respectively. 
In all these runs, the MBH experiences super-Eddington accretion with the cap $\lambda_{\rm Edd}=2$ adopted in our simulations. 
The first super-Eddington episode begins at $z\approx 9.6$. It is relatively short-lived, with $\lambda_{\rm Edd}$ dropping to unity by $z\approx 8.7$. There is a second short super-Eddington episode at $z\approx 7$ for all runs. 
This similar self-regulated gas accretion indicates that the MBH growth within these massive halos is mainly determined by the halo environments.

In Fig.~\ref{fig:mmbh_massbudget} we show the cumulative mass contribution from gas accretion, MBH merger, and TDE (only for \SCnR) of the most massive MBH at $z=6$ with different seeding models.  
Although the three MBHs reach similar masses at $z=6$ and gas accretion supplies most of their final masses, the dominant growth channels during the earliest phases are different. 
For \SCnR, frequent mergers grow the light seed $m_{\rm seed}=1.2\times 10^{4}\ M_{\odot}$ rapidly to $\approx 10^{5}\ M_{\odot}$, after which the gas accretion takes over.
The vertical dashed line marks the redshift at which the MBH first enters super-Eddington accretion. 
Across different seeding models, efficient accretion begins at similar MBH masses:
$\log M_{\rm BH}=5.3$, $6.0$, and $5.6$ in \SCnR, \Gas, and \Halo, respectively. 
This suggests that the seeding prescriptions primarily determine how quickly the MBH reaches the mass regime where efficient gas accretion is triggered, while the subsequent growth is insensitive to the seed. 
A similar conclusion was also found in the AMBRA simulation \citep{Zhou2026_ambra}, which showed that high-redshift MBH mergers boost the subsequent gas accretion by increasing the MBH mass. 
Once sustained accretion begins, the gas supply and the subsequent feedback progressively erase the memory of the initial MBH seeds. 
Hence, the seeding prescriptions strongly affect the early assembly history of MBHs at $z\gtrsim 8$, while the final MBH mass by $z=6$ is primarily controlled by the gas accretion. 
This conclusion applies specifically to this massive halo in a $5\sigma$ overdense region, and should not be interpreted as a general result for the MBH population in the early Universe. 
A statistical study of MBH evolution across a wide range of halo environments is therefore needed to fully understand the impact of seeding prescriptions on the MBH population, which we defer to future work (ACORN II).

\begin{figure}[htbp]
    \centering
    \includegraphics[width=1\linewidth]{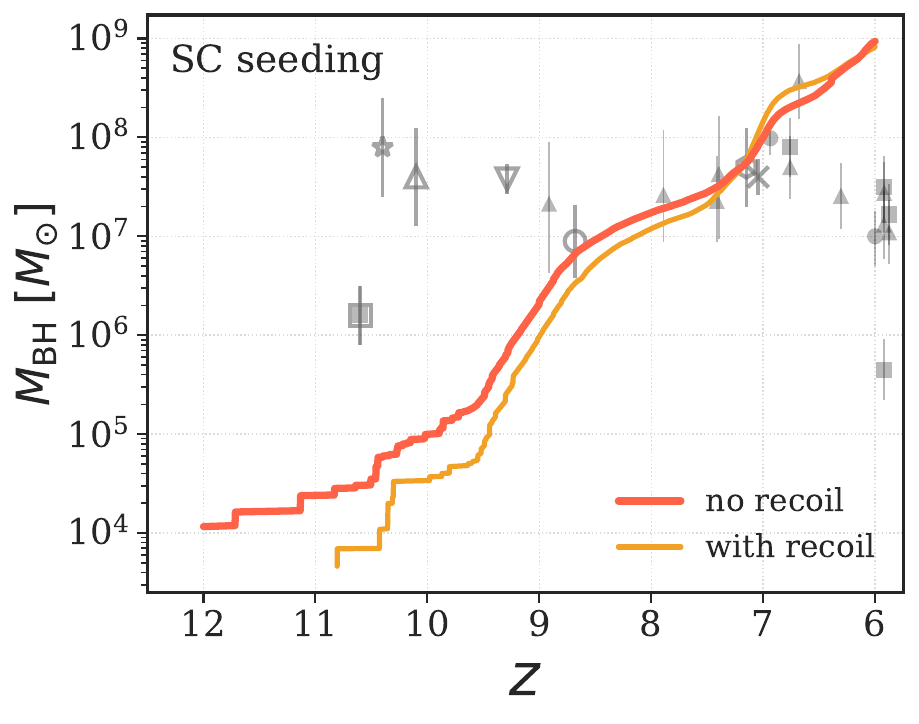}
    \caption{The mass evolution for the most massive MBH by $z=6$ for star-cluster-seeding models with and without the GW recoils, i.e., \SCR\ (yellow) and \SCnR\ (red). Gray symbols with error bars are the same observational population plotted in Fig.~\ref{fig:MBH_mass_evolution}.
    }\label{fig:recoil_effects}
\end{figure}

To show the MBH growth in a more typical halo, where gas accretion is less efficient, in Fig.~\ref{fig:sec_mmbh_evo}, we trace the evolution of the central MBH in the second most massive halo by $z=6$.
\SCnR\ seeds at a similar redshift to \Halo, while with a much smaller seed mass $\approx 10^{3}\ M_{\odot}$ compared to $10^{5}\ M_{\odot}$ in \Halo\ and \Gas. 
After the seeding, the MBH in \Halo\ and \Gas\ experiences little growth. 
In contrast, the MBH seed in \SCnR\ undergoes several mergers, which play an important role in the early growth of MBHs.  
These mergers rely on the large clustered MBH population produced by the star-cluster-seeding model, as we discussed in Section~\ref{sec:results_MBH_population}. 
We include the mass of the host star cluster when we calculate the dynamical friction experienced by each MBH. Hence, the star clusters surrounding MBHs can facilitate the MBH seed mergers by increasing their dynamical friction. 
Beyond that, based on the high-resolution simulations from the MAGICS suite \citep{Chen2024_magicI, Zhou2025_magcisII,Mukherjee2025_magicsIII}, the interaction between the two star clusters within the MBH binary systems could significantly accelerate the MBH orbital decay and boost their merger. 
Although these cluster interactions are unresolved in our simulations, the high-resolution results suggest that real coalescence will occur after the numerical mergers in our simulation, which are implemented on $\sim 1\,\rm kpc$ scales.
Besides the merger, TDEs also make a contribution to the mass growth, especially when the MBHs are newly seeded. We will discuss this in more detail in Section~\ref{sec:results_TDE}.

\begin{figure}[htbp]
    \centering
    \includegraphics[width=1\linewidth]{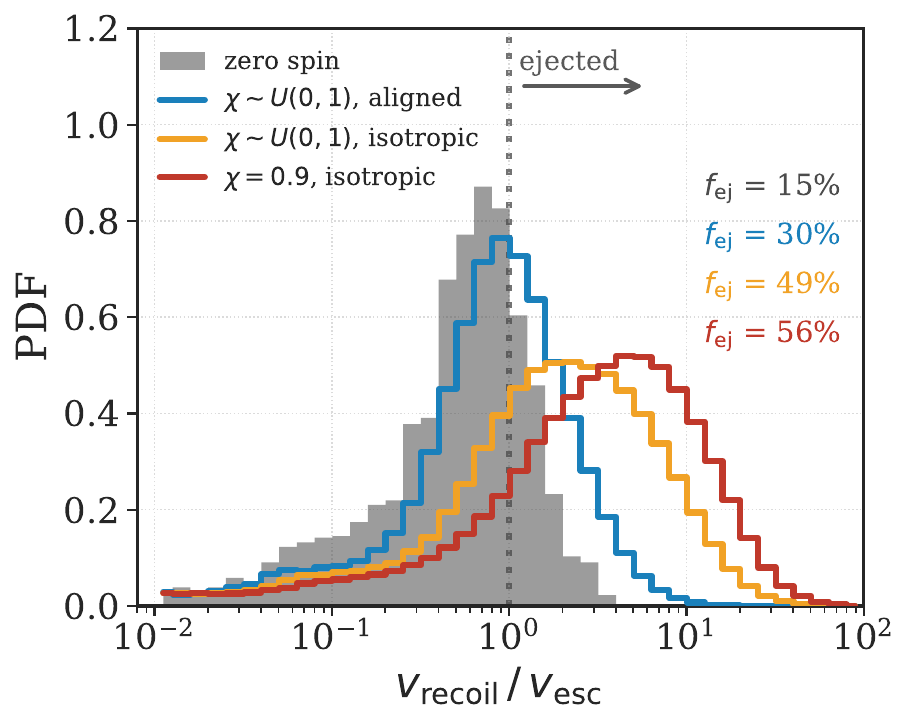}
    \caption{Probability density function (PDF) of the ratio of GW recoil velocity to the escape velocity of the host star clusters, $v_{\rm recoil}/v_{\rm esc}$, for the MBH merger catalog in \SCnR. 
    We consider four spin prescriptions: non-spinning MBHs (gray);
    a uniform magnitude distribution $\mathcal{U}(0,1)$ with directions aligned within $10^{\circ}$ of the orbital angular momentum (blue);  
    a uniform magnitude $\mathcal{U}(0,1)$ with isotropic directions (yellow); 
    isotropic direction with fixed magnitude $\chi=0.9$ (red). 
    The vertical line marks $v_{\rm recoil} = v_{\rm esc}$, above which the MBH merger remnants are ejected from their host clusters. 
    The ejected fraction $f_{\rm ej}$ of each case is shown with corresponding colors. 
    }\label{fig:v_recoil_hist}
\end{figure}


\begin{figure*}[htbp]
    \centering
    \includegraphics[width=0.65\linewidth]{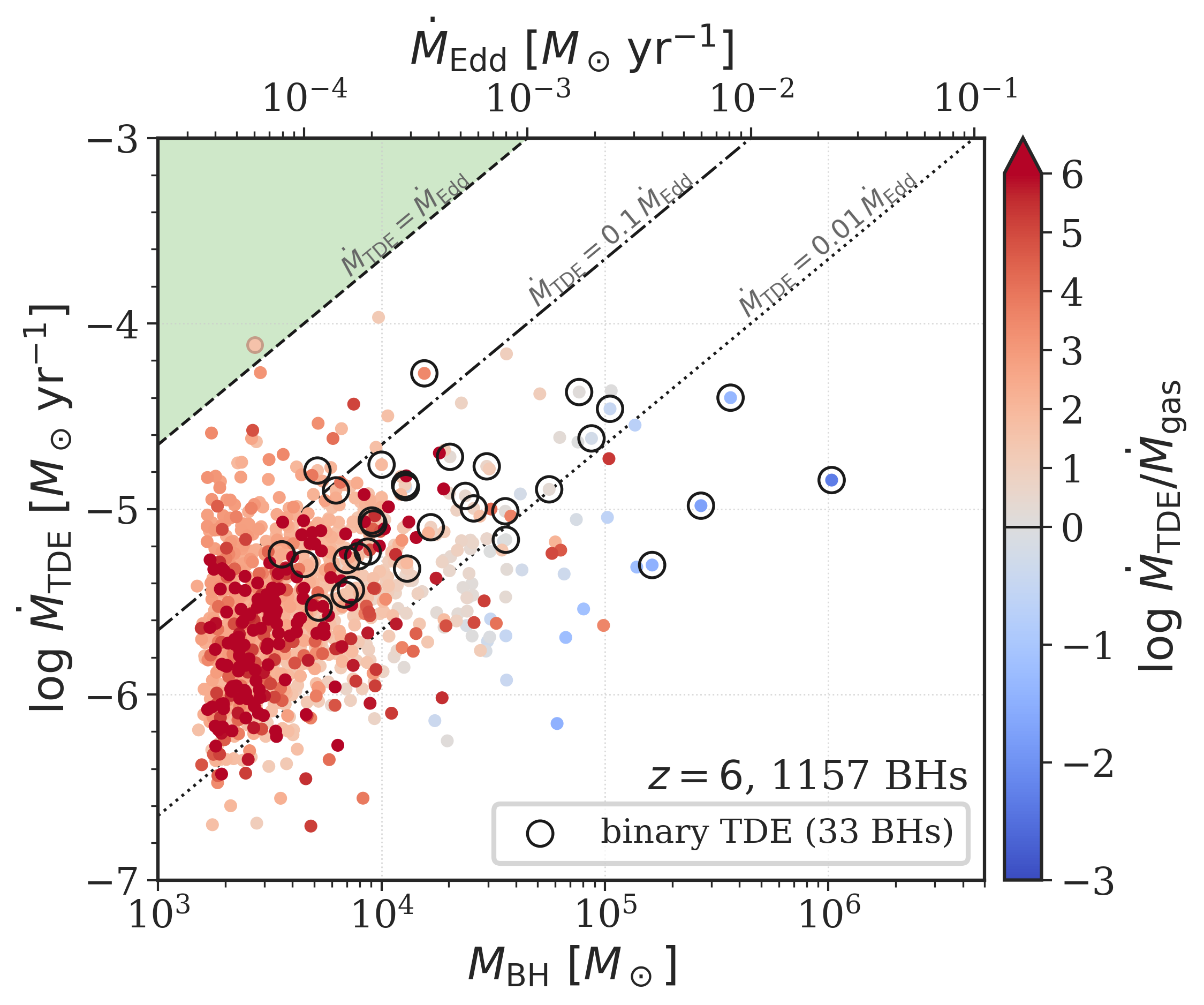}
    \caption{The TDE-driven mass growth rate $\dot{M}_{\rm TDE}$ versus the MBH mass $M_{\rm BH}$ at $z=6$ for the run \SCnR.
    Points are colored by the ratio between $\dot{M}_{\rm TDE}$ and the gas accretion rate $\dot{M}_{\rm gas}$. 
    Black circles mark MBHs that experience binary-enhanced TDEs. For these systems, $M_{\rm BH}$ is the mass of the merger remnant. 
    The dashed, dash-dotted, and dotted lines indicate $\dot{M}_{\rm TDE}=\dot{M}_{\rm Edd}$, $0.1\,\dot{M}_{\rm Edd}$, and $0.01\,\dot{M}_{\rm Edd}$, respectively. $\dot{M}_{\rm Edd}$ is shown on the upper axis. The green shaded region corresponds to where $\dot{M}_{\rm TDE}> \dot{M}_{\rm Edd}$, i.e., super-Eddington growth driven by TDEs. 
    }\label{fig:TDE_rate}
\end{figure*}

\begin{figure*}[htbp]
    \centering
    \includegraphics[width=0.75\linewidth]{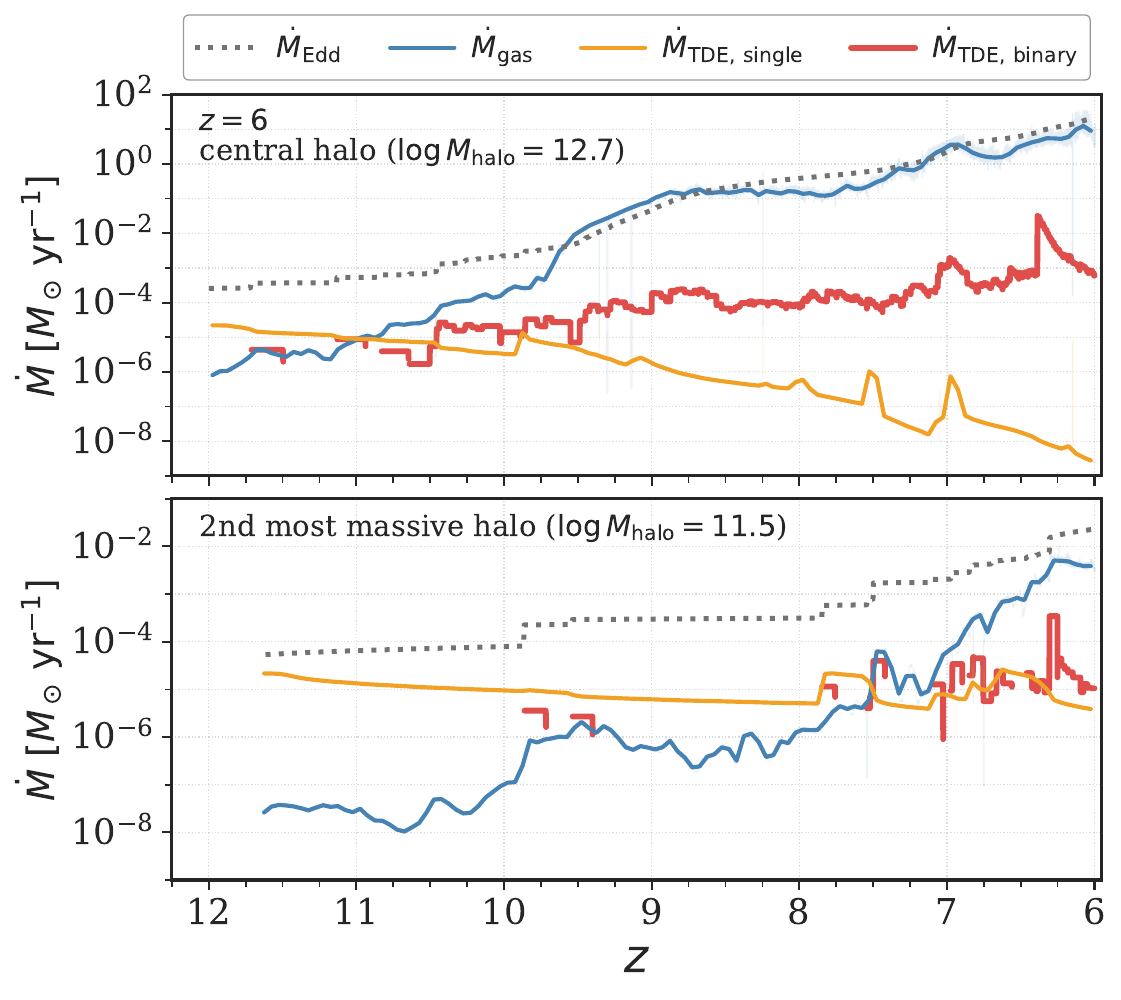}
    \caption{ Evolution of the different mass growth channels for the central MBHs for the two most massive halos at $z=6$. Blue curves show the gas accretion rate $\dot{M}_{\rm gas}$, while yellow and red curves show the 
    isolated- and binary-enhanced-TDE growth rates, respectively.
    The dotted gray curve shows the Eddington accretion rate $\dot{M}_{\rm Edd}$.
    The upper and lower panels correspond to the central MBH in the most massive and second most massive halos, respectively. 
    }\label{fig:mdot_channels}
\end{figure*}

\subsection{Influence of GW Recoil}
\label{sec:results_GWrecoil}
We plot the mass evolution of the central MBH for \SCnR\ and \SCR\ in Fig.~\ref{fig:recoil_effects}.
The GW recoil in \SCR\ delays the early assembly of the MBH at $z\gtrsim 8$ in the star-cluster-based seeding model, but the MBH evolution gradually converges with \SCnR\ with increasing MBH mass ($M_{\rm BH}\sim 10^{7}\ M_{\odot}$) and finally reaches the same mass by $z=6$. 
The similar final $M_{\rm BH}$ in \SCR\ and \SCnR\ is expected because when the MBH becomes massive, the mergers are increasingly dominated by minor mergers, for which the non-spinning recoil velocity is small. 
Moreover, the gravitational potential around the halo center deepens over time, making it harder for the MBH to escape the host system.
We compare the merger rate for \SCnR\ and \SCR\ in Fig.~\ref{fig:bh_mergers}. 
It can be seen that mergers in \SCR\ are delayed at high redshifts compared to \SCnR, while converging toward lower redshifts.  
The kicked MBHs also create more wandering MBHs in \SCR (see Fig.~\ref{fig:BH_separation}).

In our implementation, the MBH binary mass ratio used to calculate the GW recoil is measured at the separation of $\Delta r\approx 2\epsilon_{g}$, the numerical merger scale set by the resolution of our simulations. 
We do not resolve the evolution of the binary mass ratio on smaller scales.
In gas-rich environments, the preferential accretion onto the secondary MBH from a circumbinary disk can drive the binary toward equal masses \citep{Farris2014, Duffell2020, Liao2023, Siwek2023, Zhou2025_magcisII}.
For non-spinning MBHs, the recoil velocity reaches its maximum at $q\approx 0.36$ and decreases for smaller mass ratios \citep{Lousto2013}.
Because cosmological MBH merger populations typically enter the circumbinary phase with mass ratio $q$ below $q\sim0.36$; this unresolved small-scale evolution is expected to increase their recoil velocity on average \citep{Siwek2020, Sayeb2021}. 
Moreover, our assumption of non-spinning MBHs is a conservative choice, as the spin of the progenitor MBHs could produce much larger recoil velocities, $v_{\rm recoil}\gtrsim 1000\ {\rm km}\,{\rm s}^{-1}$. 
Taking into account these effects, our prescription for GW recoil here should be interpreted as a lower limit of the actual influence of GW recoil.

To show the influence of GW recoil with different spin prescriptions, we run a post-processing analysis on the MBH merger catalog from the \SCnR. 
We consider four spin distributions: (1) zero spins; (2) a uniform $\mathcal{U}(0,1)$ magnitude distribution with the direction aligned with the orbital angular momentum; (3) the same uniform magnitude distribution but with isotropic directions; (4) fixed magnitude $\chi=0.9$ with isotropic directions. 
For each merger, we assign the progenitor spins according to these distributions and calculate the GW recoil velocity using the fitted formula from \citet{Lousto2013}. 
We then compare $v_{\rm recoil}$ with the host star cluster's escape velocity, $v_{\rm esc}$, to determine whether the merger remnant is kicked out. 
In this test, we only investigate the MBH retention, while not tracing the subsequent orbital evolution of the kicked MBHs or their effects on future mergers. 
In Fig.~\ref{fig:v_recoil_hist}, we plot the distribution of the ratio of $v_{\rm recoil}/v_{\rm esc}$ for each spin distribution, and mark their ejection fraction, i.e., the fraction of merger remnants with $v_{\rm recoil}>v_{\rm esc}$. 
The non-spinning case gives the lowest ejection fraction: $f_{\rm ej}=15\%$. 
The aligned spins are expected for mergers in gas-rich environments, where torques from circumbinary discs can align both spins with the orbital angular momentum $\mathbf{L}$ \citep{Bogdanovic2007, Dotti2010}.
We model this `aligned' case by placing the spin within $10^{\circ}$ of $\mathbf{L}$, consistent with \citet{Gerosa2015} and \citet{Blecha2016}. 
This will suppress the in-plane spin components and hence avoid the largest recoil velocities (`super-kick') \citep{Campanelli2007_maxRecoil,  Gonzalez2007, Lousto2011}. 
The aligned spins give a relatively modest ejection fraction of 30\%. 
For mergers in gas-poor environments, the spins are randomly oriented. Assuming an isotropic direction distribution increases the ejection fraction to $f_{\rm ej}=49\%$. 
We further consider an optimistic high-recoil scenario with a fixed large magnitude $\chi=0.9$ and isotropic directions, following \citet{Blecha2016}. 
This model predicts the highest ejection fraction among these spin models: $f_{\rm ej}=56\%$. 
These results suggest that non-zero spins can significantly enhance the recoil velocity, increasing the ejection fraction by a factor of $ \approx 2-4$. 
Besides affecting MBH retention in clusters, these strong kicks can also displace merger remnants from the galactic center to the outskirts, where gas accretion is less efficient and subsequent mergers may be delayed, which will suppress MBH mass growth.
As discussed above, frequent mergers are important to the early growth of MBHs. 
In the $z=6$ population of \SCnR, 37\% of the MBHs with $M_{\rm BH}\geq 10^{4}\ M_{\odot}$ went through more than one merger.
Zero-spin recoil in \SCR\ already decreases MBH mergers: the fraction of MBHs above $10^4\ M_{\odot}$ with more than one merger drops to $21\%$. 
With the larger $v_{\rm recoil}$ expected for the spinning MBHs, the mass growth from these repeat mergers would be further suppressed. 


\subsection{TDE Contribution to MBH Early Growth}
\label{sec:results_TDE}

Our cluster-MBH model provides a natural framework for studying the contribution of TDEs to early MBH growth. We first compare the TDE-driven MBH growth rate, $\dot{M}_{\rm TDE}$, with the gas accretion rate $\dot{M}_{\rm gas}$ for the full MBH population at $z=6$. 
Fig.~\ref{fig:TDE_rate} shows the instantaneous $\dot{M}_{\rm TDE}$ as a function of MBH mass $M_{\rm BH}$. 
The points are colored according to the ratio $\dot{M}_{\rm TDE}/\dot{M}_{\rm gas}$. 
The TDE-driven growth rate $\dot{M}_{\rm TDE}$ depends only weakly on $M_{\rm BH}$ over most of the population and typically spans $\approx 10^{-6}-10^{-4}\ M_{\odot}/{\rm yr}$. 
Systems experiencing binary-enhanced TDEs, marked by black circles, are typically at the high-mass end  $M\geqslant 10^{4}\ M_{\odot}$, and have $\dot{M}_{\rm TDE} \approx 10^{-5}\ M_{\odot}/{\rm yr}$.   
For most low-mass MBHs with $M_{\rm BH}\leqslant 10^{4}\ M_{\odot}$, the TDE-driven growth frequently exceeds
$\dot{M}_{\rm gas}$. 
These MBHs generally reside in small halos where gas accretion is inefficient, making stellar feeding from the surrounding cluster the dominant channel. 
For low-mass MBHs, the $\dot{M}_{\rm TDE}$ can have large Eddington ratios. Among this MBH population, one system reaches $\dot{M}_{\rm TDE}> \dot{M}_{\rm Edd}$.  At the high-mass end $M_{\rm BH}\geqslant 10^{5}\ M_{\odot}$, gas accretion typically dominates over $\dot{M}_{\rm TDE}$. 

We then investigate the redshift evolution of the different growth channels for the central MBHs in the two most massive halos at $z=6$ in Fig.~\ref{fig:mdot_channels}. 
At early times, shortly after MBH seeding, $\dot{M}_{\rm TDE, single}$ (yellow curve) makes an important contribution. In the second most massive halo (lower panel), at $z\geq 11$, $\dot{M}_{\rm TDE, single}$ is three orders of magnitude larger than $\dot{M}_{\rm gas}$ and close to $\dot{M}_{\rm Edd}$. 
This is consistent with the behavior of the low-mass MBH population in Fig.~\ref{fig:TDE_rate}. 
As the host star cluster evolves, stellar mass loss and cluster expansion reduce the stellar central densities; this leads to a decrease in the TDE rate toward lower redshift.
Meanwhile, MBH mergers become more frequent (see Fig.~\ref{fig:bh_mergers}), and the binary-enhanced TDE channel $\dot{M}_{\rm TDE, binary}$ (red curve) becomes correspondingly more important.
For the central MBH of the most massive halo (upper panel), $\dot{M}_{\rm TDE, binary}$ exceeds the isolated-cluster contribution 
$\dot{M}_{\rm TDE, single}$ by several orders of magnitude at $z\leq 9$. 
Binary enhancement therefore partly compensates for the decrease in $\dot{M}_{\rm TDE, single}$, making the total TDE contribution remain significant. 
However, as the MBH grows, gas accretion eventually dominates the mass growth. In our model, the Bondi-like gas accretion rate scales as $\dot{M}_{\rm gas}\propto M_{\rm BH}^{2}$ before being capped at $2\times \dot{M}_{\rm Edd}$. 
Consequently, the rapid increase in $M_{\rm BH}$ allows gas accretion to overtake TDE-driven growth at later times. The subsequent MBH evolution is increasingly regulated by the gas supply in its galactic environment, as shown in Section~\ref{sec:mass_evolution}.
This would gradually wash away the memory of initial seeding conditions.

Besides the contribution to the mass growth, TDEs also increase the observability of MBH seeds. 
To estimate the peak luminosity of TDE flares, we use the empirical scaling of \citet{Mummery2026}: 
\begin{equation}
    \log_{10}\left(\frac{L_{\rm peak}}{\rm erg\,s^{-1}}\right) = \alpha + \beta\,\log_{10}\left(\frac{M_{\rm
  BH}}{10^{6}\,M_{\odot}}\right),
    \label{eq:tde_lpeak}
  \end{equation}
where $\alpha = 43.7 \pm 0.1$, and $\beta = 0.4 \pm 0.1$. 
This relation is obtained by fitting a power law to the peak optical/UV blackbody luminosity of the ZTF TDE sample against the $M_{\rm BH}$ inferred from host-galaxy scaling relations \citep{Greene2020}. 
One caveat is that the observed sample in \citet{Mummery2026} spans the mass range $10^{4.5}-10^{9}\ M_{\odot}$, while we will apply this relation down to $10^{3}\ M_{\odot}\,h^{-1}$. 
In Fig.~\ref{fig:MBH_mass_evolution} and Fig.~\ref{fig:sec_mmbh_evo}, the red dotted curve shows the peak luminosity $L_{\rm peak}$ if a TDE occurs. 
Because TDEs are transient events, the curve does not represent continuous emission; instead, it shows the maximum luminosity of individual flares, which typically last weeks to months. 
For the central MBHs in both the most massive and second-most-massive halos, TDEs can temporarily outshine the persistent emission from gas accretion (solid curves) by orders of magnitude at the early stages of MBH growth. 
For MBHs with $M_{\rm BH}\sim 10^{3}-10^{5}\ M_{\odot}$, the empirical relation predicts TDE $L_{\rm peak}\sim 3\times 10^{42}-2\times 10^{43}\ {\rm erg}\,s^{-1}$, where the gas accretion for these MBH seeds remains as low as $10^{38}-10^{42}\ {\rm erg}\,s^{-1}$. 
This indicates that TDE flares substantially enhance the instantaneous 
detectability of MBH seeds, which are otherwise faint. 


\section{Conclusion} 
\label{sec:conclusion}
    This is the first paper of the ACORN project.
    We develop a subgrid MBH seeding model physically motivated by MBH formation through runaway stellar collisions in dense star clusters. 
    We first identify stellar structures (StarFOFs) in the simulations using the FoF algorithm, which represent the star-cluster complexes observed at high redshift, such as Cosmic Gems \citep{Adamo2024_cosmicGems} and Firefly \citep{Mowla2024_firefly}.
    Within each StarFOF, we populate a subgrid star cluster population based on the cluster formation efficiency $\Gamma$ derived through the prescription of \citep{Kruijssen2012} and the properties of the star particles hosted by the StarFOF.
    The MBH seed mass is then determined by the properties of the star cluster following \citet{Williams2026}. 
    We further implement a star cluster evolution model to follow the mass and size evolution of clusters hosting MBHs. 
    Based on the evolving cluster population, we estimate the TDE rates from both isolated MBHs and the enhanced TDEs associated with MBH binaries. 
    Assuming that half of the disrupted stellar mass is accreted by the MBH, we include TDEs as an additional channel for MBH growth. 
    Beyond that, we also model the GW recoil assuming zero MBH spin. 
    We apply this model to constrained ICs representing a $5\sigma$ density peak in a $12.5~{\rm Mpc}/h$-per-side box, targeting the extreme environments where high-redshift MBHs detected by JWST are likely to form. 
    The halo properties are measured from the most massive halo in the ASTRID simulation at $z=8$. 
    Our main results are summarized as follows.

\begin{itemize} 
    \item In our model, the overall star cluster formation closely traces the star formation history of the host galaxy. 
    The resulting young, massive star cluster populations at $z=6-9$ are in good agreement with observations of nearby galaxies ($z\simeq 0$) in terms of both the cluster mass fraction and the number of clusters as a function of galaxy SFR. 
    \item The subgrid star cluster population spans a wide range of metallicity, mass, and size. The resultant MBH seed mass strongly depends on the properties of the host cluster: massive seeds form preferentially in massive, compact, and metal-poor star clusters. With the adopted minimum seed mass $M_{\rm seed}=10^{3}\ M_{\odot}\,h^{-1}$, the host clusters cover the mass range of $10^{4}\lesssim m_{\rm SC}\lesssim 10^{8}\,M_{\odot}$, and peak around $10^{6}\ M_{\odot}$, comparable to the masses of compact clusters observed by JWST at high redshift. 
    \item Compared to the gas-based seeding model, the star-cluster-seeding model begins forming MBHs at lower redshift ($z\approx 15$ versus $z\approx 20$).
     Once seeding begins, however, its seeding rate increases rapidly toward lower redshift. By $z=6$, the seeding rate is about $0.5$ dex higher than that of the gas-based model.
    \item Instead of using a fixed seed mass or stochastically drawing from a $M_{\rm seed}$ distribution, the MBH seed mass in the star-cluster-based seeding model is determined by the properties of the host star clusters, which is physically motivated and correlated with the local environments. The resultant $M_{\rm seed}$ distribution appears as a power-law distribution with a slope of $-1$, similar to the seed mass distribution adopted in AMBRA and ASTRID simulations, while shifted to lower mass. The $M_{\rm seed}$ span a range of $10^{3}\lesssim M_{\rm seed}\lesssim 10^{5}\,M_{\odot}$. 
    \item The star-cluster-seeding model produces a strong connection between MBHs and their host environments. Both the number of MBHs and central MBH masses increase systematically with galaxy masses.  
    In contrast, the fixed seed masses adopted in gas- and halo-based models impose artificially massive MBHs in galaxies with $M_{\rm galaxy}\approx 10^{6-9}\ M_{\odot}$ and produce a much flatter $M_{\rm BH}\!-\!M_{\rm galaxy}$ relation at the low-mass end. 
    This environmental dependence in the star-cluster-seeding model therefore provides a more physically motivated framework for predicting MBH populations in dwarf galaxies. 
    \item For the most massive MBH in the constrained $5\sigma$ overdense region, seeding prescriptions affect its early growth at $z\gtrsim 8$, while the final MBH masses converge across different runs by $z=6$ because it is primarily controlled by self-regulated gas accretion. 
    The rapid growth in the star-cluster-seeding model is particularly interesting because it starts from a much less massive seed ($\log M_{\rm seed}= 4.18$ versus $5.32$) at a much later redshift ($z_{\rm seed}=12$ versus $18.6$), while it manages to produce a similar MBH mass $\log M_{\rm BH}=6.5$ as that in the gas-seeding model by $z=9$.
    \item With the GW recoil implemented, the early growth of the MBHs ($z\gtrsim 9$) is delayed, while the mass evolution gradually converges when the MBHs become massive ($M_{\rm BH}\gtrsim 10^{7}\ M_{\odot}$). This is because, at the early stage, many mergers between MBH seeds are major mergers, which can produce substantial kick velocity. 
    
    \item Compared with halo-based and gas-based seeding prescriptions, star-cluster-based seeding produces more abundant MBHs within the same DM-FOF halo because multiple MBHs are allowed to form \textit{in situ} within a single halo. This spatially clustered MBH population provides a pool of nearby merger partners, increasing early MBH mergers that are detectable by LISA.
    
    \item Rather than a sharp transition assumed by the halo-based seeding, the star-cluster-based seeding model predicts a gradual increase in the MBH occupation fraction $f_{\rm occ}$ with total halo mass and halo stellar mass. 
    At the ASTRID halo mass threshold $M_{\rm halo}=5\times 10^{9}\,M_{\odot}h^{-1}$, the occupation fraction $f_{\rm occ}$ for $M_{\rm BH}\geqslant 10^{3}\ M_{\odot}$ increases from $\sim 10\%$ at $z=10$ to $\sim 30\%$ at $z=6$, while more massive MBHs remain rare. The transition also depends on the redshift. 
    \item TDEs provide an important growth channel for newly seeded, low-mass MBHs. 
    For some of these systems, the TDE-driven accretion rate $\dot{M}_{\rm TDE}$ can exceed the gas accretion rate and even the Eddington limit. As host star clusters evolve, the isolated-cluster contribution $\dot{M}_{\rm TDE, single}$ declines, while the binary-enhanced TDE channel $\dot{M}_{\rm TDE, binary}$ becomes increasingly important. However, as MBHs grow, gas accretion eventually dominates their growth, gradually erasing the imprint of the initial seeding conditions.
    
    \item TDE flares can significantly enhance the instantaneous detectability of MBH seeds. According to the empirically calibrated relation $L_{\rm peak}\!-\!M_{\rm BH}$ from observed TDEs, individual events can reach a peak luminosity of $L_{\rm peak}\approx 10^{42-45}\ {\rm erg}\,s^{-1}$. For low-mass MBHs with $M_{\rm BH}\lesssim 10^{6}\ M_{\odot}$, these transients outshine the persistent luminosity from gas accretion by orders of magnitude, making these otherwise faint MBHs observable. 
\end{itemize}

\section*{Acknowledgments}
Y.Z. thanks Shalini Kurinchi-Vendhan for helpful advice on the visualization. 
Y.Z. and T.D.M. acknowledge the support from the NASA FINESST grant 80NSSC25K0318. 
This work was supported by the National Science Foundation under Cooperative Agreement 2421782 and the Simons Foundation grant MPS-AI-00010515 awarded to the NSF-Simons AI Institute for Cosmic Origins — CosmicAI, https://www.cosmicai.org.
T.D.M. acknowledges funding from NASA ATP 80NSSC20K0519, NSF PHY-2020295, NASA ATP NNX17AK56G, and NASA ATP 80NSSC18K101, NASA Theory grant 80NSSC22K072. A.K.B. acknowledges support from NSF-AST 2510738. 
C.E.W. acknowledges the support of the National Science Foundation Graduate Research Fellowship, the University of California, Los Angeles (UCLA), the Mani L. Bhaumik Institute for Theoretical Physics, and the UCLA Center for Developing Leadership in Science Fellowship.
C.E.W. thanks Howard and Astrid Preston for their generous support.
C.E.W. thanks the support of NASA grant No. 80NSSC24K0773 (ATP-23-ATP23-0149) and the XSEDE/ACCESS AST180056 allocation.
 M.R.-C. acknowledges the support of a fellowship from the ``la Caixa'' Foundation (ID 100010434), with code LCF/BQ/PI26/12120011. 
 M.R.-C. acknowledges funding from the Global Talent Junior Fellowship Programme at the Instituto Galego de F\'isica de Altas Enerx\'ias (IGFAE), supported by grant CEX2023-001318-M (Mar\'ia de Maeztu Unit of Excellence and Agencia Espa\~{n}ola de Investigaci\'on / 10.13039/501100011033). This work has received financial support from the Xunta de Galicia (CIGUS Network of Research Centres) and the European Union through the Galicia Feder 2021-2027 Program. 
  Support for E.G.P.\ was provided by the National Science Foundation Graduate Research Fellowship Program under grant No. DGE-2234667.
 L.B. acknowledges support from NSF awards AST-2307171 and AST-2509457.
 This material is based upon work supported by the National Science Foundation Graduate Research Fellowship Program under Grant No. DGE-2034835. Any opinions, findings, conclusions, or recommendations expressed in this material are those of the author(s) and do not necessarily reflect the views of the National Science Foundation.

The ACORN project was run on the Frontera facility at the Texas Advanced Computing Center.  

\appendix
\counterwithin{figure}{section}

\section{The linking length $\ell_{\star}$ for StarFOF catalogs}
\label{app:linkstar}

\begin{figure*}[htbp]
    \centering
    \includegraphics[width=0.8\linewidth]{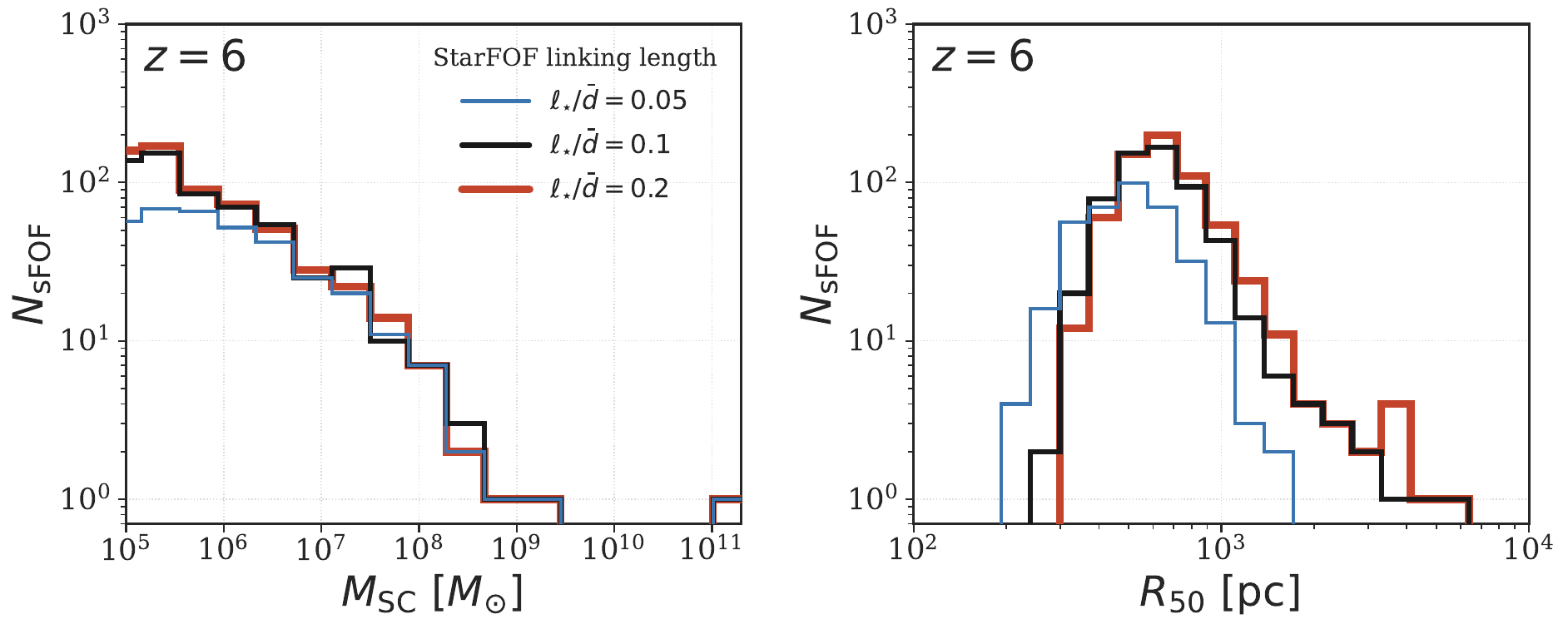}
    \caption{Distribution of the total star cluster mass, $M_{\rm SC}$ (left), and half-mass radius, $R_{\rm 50}$ (right) of the StarFOF population at $z=6$ identified using different linking length: $\ell_{\star}/\bar{d} = 0.05$ (blue), $0.1$ (black), and $0.2$ (red). 
    The catalogs are constructed from the same snapshot of the run \SCnR. 
    }\label{fig:app_linkstar}
\end{figure*}

In this appendix, we examine the sensitivity of the resulting StarFOF population to the adopted linking length $\ell_{\star}$. 
We perform this test using the $z=6$ snapshot of the run \SCnR\ and construct three StarFOF catalogs with $\ell_{\star}/\bar{d} = 0.05$, $0.1$, and $0.2$, where $\bar{d}$ is the mean separation of DM particles in the simulation. 
In our fiducial runs of this work, the linking length for DM-FOF is $\ell_{\rm DM}/\bar{d} = 0.2$, and for the StarFOF is $\ell_{\star}/\bar{d} = 0.1$.
Fig.~\ref{fig:app_linkstar} compares the distributions of the total star cluster mass $M_{\rm SC}$ and half-mass radius $R_{\rm 50}$ of the StarFOF populations identified by different linking lengths.
The mass distributions are broadly insensitive to the choice of $\ell_{\star}$. The main difference occurs at the low-mass end, where the smallest linking length $\ell_{\star}/\bar{d}=0.05$ identifies fewer systems.
The $R_{\rm 50}$ distributions for $\ell_{\star}/\bar{d} = 0.1$ and $0.2$ also agree closely. In contrast, $\ell_{\star}/\bar{d}=0.05$ shifts $R_{\rm 50}$ to smaller radii, as the shorter $\ell_{\star}$ tends to separate spatially extended structures into smaller systems. 
The nearly unchanged $M_{\rm SC}$ distribution indicates that the smaller $\ell_{\star}$ captures most of the star cluster mass budget, probably because the excluded star particles are located at the outskirts where the low density produces small $\Gamma$.
Because our cluster seeding model is primarily determined by massive star clusters ($m_{\rm SC} \gtrsim 10^{5}\ M_{\odot}$), this difference does not significantly affect our results. 
We therefore conclude that our results are robust to the choice of $\ell_{\star}$ over the range tested here.


\section{The $M_{\rm seed,min}$ in star-cluster-based seeding model}
\label{app:M_seed_min}

In this appendix, we examine the sensitivity of the resulting MBH population to the minimum BH seed mass $M_{\rm seed,min}$ in the star-cluster-based seeding model.  
Our fiducial \SCnR\ simulation adopts $M_{\rm seed,min}=10^{3}\ M_{\odot}/h$. 
We run a test simulation with $M_{\rm seed,min}=300\ M_{\odot}/h$ down to $z=9$. This test run uses the same ICs and parameters as \SCnR\ except for $M_{\rm seed,min}$.

\begin{figure*}[htbp]
    \centering
    \includegraphics[width=1\linewidth]{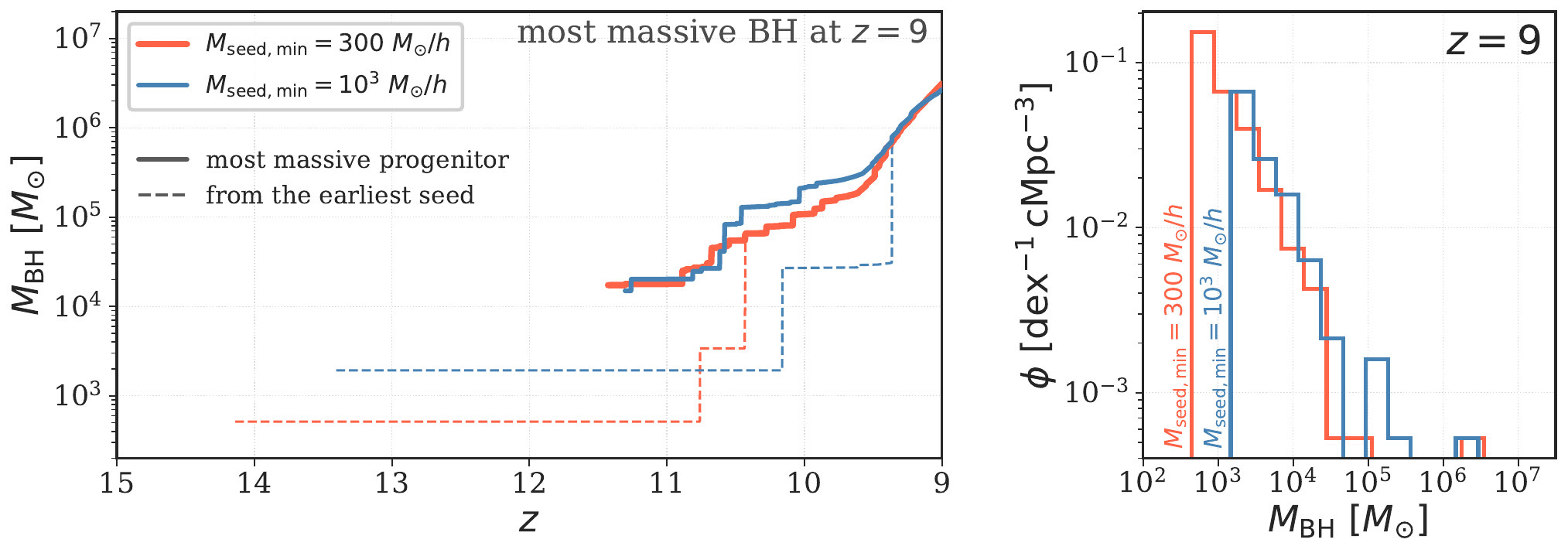}
    \caption{Influence of the minimum seed mass, $M_{\rm seed,min}$, on the MBH population. 
    We compare the fiducial \SCnR\ run with $M_{\rm seed,min}=10^{3}\ M_{\odot}/h$ (blue) to a test run with $M_{\rm seed,min}=300\ M_{\odot}/h$ (red). 
    \textit{Left:} mass growth history of the most massive BH at $z=9$. Solid curves trace its most massive progenitors, and dashed curves trace the earliest seed among its progenitors.  
    \textit{Right:} MBH mass functions at $z=9$.  
    Except for the expected difference in the distribution near $M_{\rm seed,min}$, these two runs produce similar MBH growth histories and mass functions. 
    }\label{fig:app_Mseedmin}
\end{figure*}

Fig.~\ref{fig:app_Mseedmin} compares this test run with the fiducial \SCnR. 
In the left panel, we show the growth history of the most massive BH by $z=9$. The solid curves trace its most massive progenitors, and the dashed curves follow the earliest seed among its progenitors. 
With a smaller $M_{\rm seed,min}$, BH seeds begin to appear at slightly higher redshift.
This is because the earliest population of star clusters is dominated by relatively low-mass systems, which tend to produce low-mass BH seeds. Most of these seeds fall below $10^{3}\ M_{\odot}/h$ but are captured when the threshold is lowered to $300\ M_{\odot}/h$. 
Although a smaller $M_{\rm seed,min}$ changes some details of the early growth of MBH, the MBH evolutions in these two runs converge at $z\lesssim 10$ and reach almost identical masses at $z=9$. 
In the right panel, we show the MBH mass functions at $z=9$. Besides the expected difference near $M_{\rm seed,min}$ as $M_{\rm seed,min}=300\ M_{\odot}/h$ extends the MBH population to smaller masses, these two mass functions agree well in both normalization and shape. 
We therefore conclude that our results are robust to the choice of $M_{\rm seed,min}$ over the range tested here.

\section{Validation of the Constrained Realization}
\label{app:astrid_compare}

\begin{figure*}[htbp]
    \centering
    \includegraphics[width=0.65\linewidth]{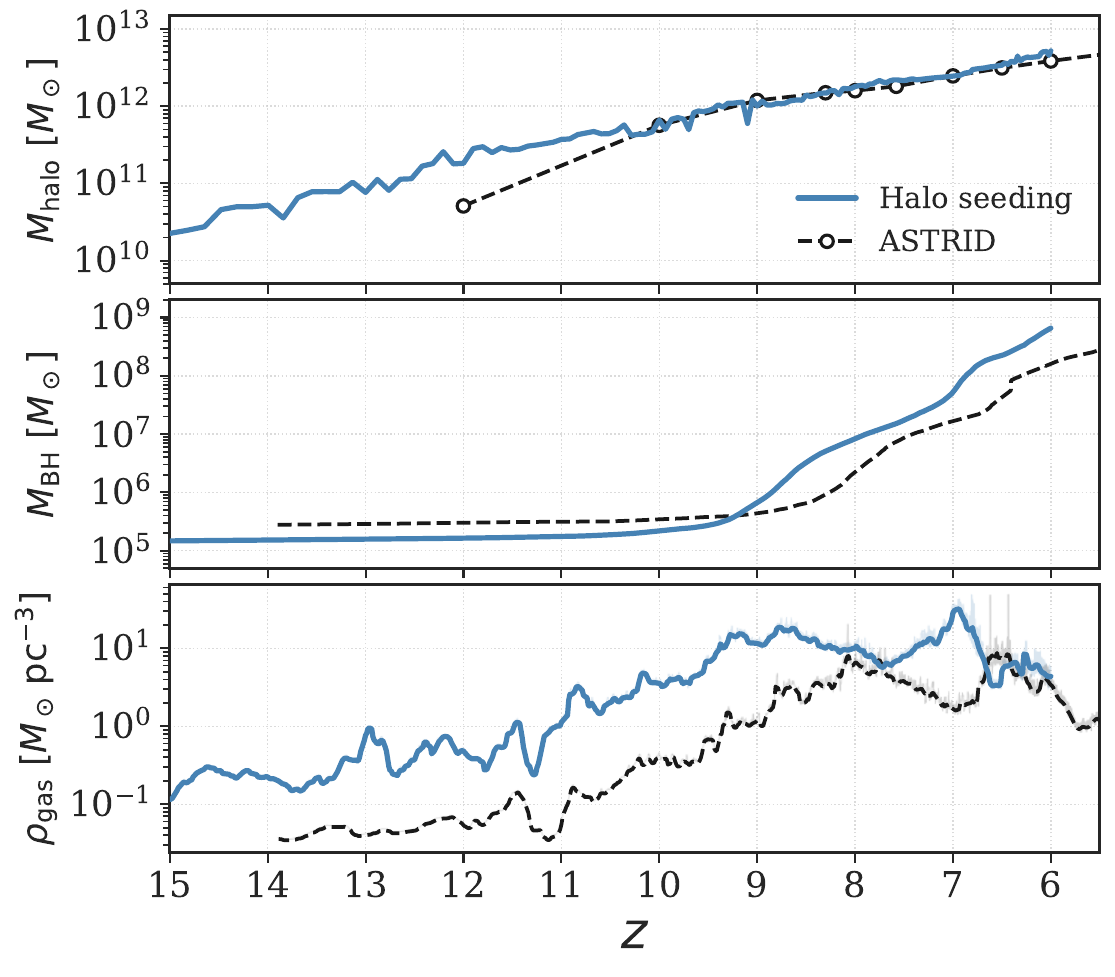}
    \caption{Comparison between the constrained \Halo\ simulation (blue solid curves) and the corresponding halo in the original ASTRID simulation (black dashed curves). 
    From top to bottom, we show the mass of the central DM-FOF halo, the mass of its central MBH, and the local gas density surrounding the MBH. Open circles in the upper panel mark the ASTRID snapshots used for comparison.
    The constrained simulation reproduces the halo mass evolution at $z\lesssim 10$, while its central MBH grows more rapidly because of the denser gas environment, which is likely influenced by the higher mass resolution of the constrained run. 
    }\label{fig:app_astrid_compare}
\end{figure*}

We generate our constrained ICs to reproduce the most massive halo in ASTRID at $z=8$. To verify that our constrained runs reproduce the intended halo assembly history, we compare the \Halo\ run with the original halo in the ASTRID simulation from which we measure the constrained conditions.
The upper panel of Fig.~\ref{fig:app_astrid_compare} shows the evolution of the total halo mass. Although there are differences at earlier times, the halo growth histories converge by $z=10$ and remain in good agreement thereafter. This validates that our constrained simulation successfully recovers the assembly of the targeted massive halo.

However, there is a divergence in the evolution of the central MBH. As shown in the middle panel of Fig.~\ref{fig:app_astrid_compare}, the MBH in \Halo\ grows more rapidly and reaches a higher mass by $z=6$ than its counterpart in the original ASTRID simulation. This difference can be explained by the denser gas surrounding the MBH, as shown in the bottom panel. 
The enhanced gas density likely arises from the higher mass resolution of the constrained simulation compared to ASTRID: $m_{\rm gas}=2\times 10^{5}\ M_{\odot}\,h^{-1}$ versus $m_{\rm gas}=1.26\times 10^{6}\ M_{\odot}\,h^{-1}$. 
The higher resolution allows the dense gas around the MBH to be better resolved and can consequently lead to more rapid gas accretion. 
Therefore, while the CR simulation reproduces the targeted halo assembly, we do not expect the detailed MBH growth history to be identical to that in the original ASTRID simulation.

\bibliography{reference_astridmodel, reference}

@ARTICLE{Bhowmick2024_gasbased_subgrid,
       author = {{Bhowmick}, Aklant K. and {Blecha}, Laura and {Torrey}, Paul and {Weinberger}, Rainer and {Kelley}, Luke Zoltan and {Vogelsberger}, Mark and {Hernquist}, Lars and {Somerville}, Rachel S.},
        title = "{Representing low-mass black hole seeds in cosmological simulations: A new sub-grid stochastic seed model}",
      journal = {\mnras},
         year = 2024,
        month = apr,
       volume = {529},
       number = {4},
        pages = {3768-3792},
          doi = {10.1093/mnras/stae780},
archivePrefix = {arXiv},
       eprint = {2309.15341},
 primaryClass = {astro-ph.GA},
       adsurl = {https://ui.adsabs.harvard.edu/abs/2024MNRAS.529.3768B}
}

@ARTICLE{2021MNRAS.507.2012B,
       author = {{Bhowmick}, Aklant K. and {Blecha}, Laura and {Torrey}, Paul and {Kelley}, Luke Zoltan and {Vogelsberger}, Mark and {Kosciw}, Kaitlyn and {Nelson}, Dylan and {Weinberger}, Rainer and {Hernquist}, Lars},
        title = "{Impact of gas-based seeding on supermassive black hole populations at z {\ensuremath{\geq}} 7}",
      journal = {\mnras},
         year = 2021,
        month = oct,
       volume = {507},
       number = {2},
        pages = {2012-2036},
          doi = {10.1093/mnras/stab2204},
archivePrefix = {arXiv},
       eprint = {2105.08055},
 primaryClass = {astro-ph.GA},
       adsurl = {https://ui.adsabs.harvard.edu/abs/2021MNRAS.507.2012B}
}

@ARTICLE{2026ApJ..1008...30B,
       author = {{Bhowmick}, Aklant K. and {Blecha}, Laura and {Torrey}, Paul and {Kelley}, Luke Zoltan and {Somerville}, Rachel S. and {Weinberger}, Rainer and {Natarajan}, Priyamvada and {Di Matteo}, Tiziana and {Hernquist}, Lars and {Vogelsberger}, Mark and {Garcia}, Alex M.},
        title = "{Supermassive Black Hole Assembly from Heavy Seeds with Dynamical Friction in the BRAHMA Simulations: Implications for JWST, LISA, and the Local Universe}",
      journal = {\apj},
         year = 2026,
        month = sep,
       volume = {1008},
       number = {1},
          eid = {30},
        pages = {30},
          doi = {10.3847/1538-4357/ae90a2},
archivePrefix = {arXiv},
       eprint = {2606.12851},
 primaryClass = {astro-ph.GA},
       adsurl = {https://ui.adsabs.harvard.edu/abs/2026ApJ..1008...30B}
}

@ARTICLE{2025MNRAS.538..518B,
       author = {{Bhowmick}, Aklant K. and {Blecha}, Laura and {Torrey}, Paul and {Somerville}, Rachel S. and {Kelley}, Luke Zoltan and {Weinberger}, Rainer and {Vogelsberger}, Mark and {Hernquist}, Lars and {Natarajan}, Priyamvada and {Kho}, Jonathan and {Di Matteo}, Tiziana},
        title = "{Signatures of black hole seeding in the local Universe: predictions from the BRAHMA cosmological simulations}",
      journal = {\mnras},
         year = 2025,
        month = mar,
       volume = {538},
       number = {1},
        pages = {518-536},
          doi = {10.1093/mnras/staf269},
archivePrefix = {arXiv},
       eprint = {2411.19332},
 primaryClass = {astro-ph.GA},
       adsurl = {https://ui.adsabs.harvard.edu/abs/2025MNRAS.538..518B}
}

@ARTICLE{2024MNRAS.533.1907B,
       author = {{Bhowmick}, Aklant K. and {Blecha}, Laura and {Torrey}, Paul and {Somerville}, Rachel S. and {Kelley}, Luke Zoltan and {Vogelsberger}, Mark and {Weinberger}, Rainer and {Hernquist}, Lars and {Sivasankaran}, Aneesh},
        title = "{Growth of high-redshift supermassive black holes from heavy seeds in the BRAHMA cosmological simulations: implications of overmassive black holes}",
      journal = {\mnras},
         year = 2024,
        month = sep,
       volume = {533},
       number = {2},
        pages = {1907-1926},
          doi = {10.1093/mnras/stae1819},
archivePrefix = {arXiv},
       eprint = {2406.14658},
 primaryClass = {astro-ph.GA},
       adsurl = {https://ui.adsabs.harvard.edu/abs/2024MNRAS.533.1907B}
}

@ARTICLE{2024MNRAS.531.4311B,
       author = {{Bhowmick}, Aklant K. and {Blecha}, Laura and {Torrey}, Paul and {Kelley}, Luke Zoltan and {Weinberger}, Rainer and {Vogelsberger}, Mark and {Hernquist}, Lars and {Somerville}, Rachel S. and {Evans}, Analis Eolyn},
        title = "{Introducing the BRAHMA simulation suite: signatures of low-mass black hole seeding models in cosmological simulations}",
      journal = {\mnras},
         year = 2024,
        month = jul,
       volume = {531},
       number = {4},
        pages = {4311-4335},
          doi = {10.1093/mnras/stae1386},
archivePrefix = {arXiv},
       eprint = {2402.03626},
 primaryClass = {astro-ph.GA},
       adsurl = {https://ui.adsabs.harvard.edu/abs/2024MNRAS.531.4311B}
}

@ARTICLE{Vink2018,
       author = {{Vink}, Jorick S.},
        title = "{Very massive stars: a metallicity-dependent upper-mass limit, slow winds, and the self-enrichment of globular clusters}",
      journal = {\aap},
         year = 2018,
        month = jul,
       volume = {615},
          eid = {A119},
        pages = {A119},
          doi = {10.1051/0004-6361/201832773},
archivePrefix = {arXiv},
       eprint = {1803.08042},
 primaryClass = {astro-ph.SR},
       adsurl = {https://ui.adsabs.harvard.edu/abs/2018A&A...615A.119V}
}

@ARTICLE{Goulding2023_uhz1_obs,
       author = {{Goulding}, Andy D. and {Greene}, Jenny E. and {Setton}, David J. and {Labbe}, Ivo and {Bezanson}, Rachel and {Miller}, Tim B. and {Atek}, Hakim and {Bogd{\'a}n}, {\'A}kos and {Brammer}, Gabriel and {Chemerynska}, Iryna and {Cutler}, Sam E. and {Dayal}, Pratika and {Fudamoto}, Yoshinobu and {Fujimoto}, Seiji and {Furtak}, Lukas J. and {Kokorev}, Vasily and {Khullar}, Gourav and {Leja}, Joel and {Marchesini}, Danilo and {Natarajan}, Priyamvada and {Nelson}, Erica and {Oesch}, Pascal A. and {Pan}, Richard and {Papovich}, Casey and {Price}, Sedona H. and {van Dokkum}, Pieter and {Wang}, Bingjie and {Weaver}, John R. and {Whitaker}, Katherine E. and {Zitrin}, Adi},
        title = "{UNCOVER: The Growth of the First Massive Black Holes from JWST/NIRSpec-Spectroscopic Redshift Confirmation of an X-Ray Luminous AGN at z = 10.1}",
      journal = {\apjl},
         year = 2023,
        month = sep,
       volume = {955},
       number = {1},
          eid = {L24},
        pages = {L24},
          doi = {10.3847/2041-8213/acf7c5},
archivePrefix = {arXiv},
       eprint = {2308.02750},
 primaryClass = {astro-ph.GA},
       adsurl = {https://ui.adsabs.harvard.edu/abs/2023ApJ...955L..24G}
}

@ARTICLE{Inayoshi2020,
       author = {{Inayoshi}, Kohei and {Visbal}, Eli and {Haiman}, Zolt{\'a}n},
        title = "{The Assembly of the First Massive Black Holes}",
      journal = {\araa},
         year = 2020,
        month = aug,
       volume = {58},
        pages = {27-97},
          doi = {10.1146/annurev-astro-120419-014455},
archivePrefix = {arXiv},
       eprint = {1911.05791},
 primaryClass = {astro-ph.GA},
       adsurl = {https://ui.adsabs.harvard.edu/abs/2020ARA&A..58...27I}
}

@ARTICLE{Greene2020,
       author = {{Greene}, Jenny E. and {Strader}, Jay and {Ho}, Luis C.},
        title = "{Intermediate-Mass Black Holes}",
      journal = {\araa},
         year = 2020,
        month = aug,
       volume = {58},
        pages = {257-312},
          doi = {10.1146/annurev-astro-032620-021835},
archivePrefix = {arXiv},
       eprint = {1911.09678},
 primaryClass = {astro-ph.GA},
       adsurl = {https://ui.adsabs.harvard.edu/abs/2020ARA&A..58..257G}
}

@ARTICLE{Gurkan2004,
       author = {{G{\"u}rkan}, M. Atakan and {Freitag}, Marc and {Rasio}, Frederic A.},
        title = "{Formation of Massive Black Holes in Dense Star Clusters. I. Mass Segregation and Core Collapse}",
      journal = {\apj},
         year = 2004,
        month = apr,
       volume = {604},
       number = {2},
        pages = {632-652},
          doi = {10.1086/381968},
archivePrefix = {arXiv},
       eprint = {astro-ph/0308449},
 primaryClass = {astro-ph},
       adsurl = {https://ui.adsabs.harvard.edu/abs/2004ApJ...604..632G}
}

@ARTICLE{Devecchi2009,
       author = {{Devecchi}, B. and {Volonteri}, M.},
        title = "{Formation of the First Nuclear Clusters and Massive Black Holes at High Redshift}",
      journal = {\apj},
         year = 2009,
        month = mar,
       volume = {694},
       number = {1},
        pages = {302-313},
          doi = {10.1088/0004-637X/694/1/302},
archivePrefix = {arXiv},
       eprint = {0810.1057},
 primaryClass = {astro-ph},
       adsurl = {https://ui.adsabs.harvard.edu/abs/2009ApJ...694..302D}
}

@ARTICLE{Freitag2006,
       author = {{Freitag}, Marc and {Rasio}, Frederic A. and {Baumgardt}, Holger},
        title = "{Runaway collisions in young star clusters - I. Methods and tests}",
      journal = {\mnras},
         year = 2006,
        month = may,
       volume = {368},
       number = {1},
        pages = {121-140},
          doi = {10.1111/j.1365-2966.2006.10095.x},
archivePrefix = {arXiv},
       eprint = {astro-ph/0503129},
 primaryClass = {astro-ph},
       adsurl = {https://ui.adsabs.harvard.edu/abs/2006MNRAS.368..121F}
}

@ARTICLE{PortegiesZwart2004,
       author = {{Portegies Zwart}, Simon F. and {Baumgardt}, Holger and {Hut}, Piet and {Makino}, Junichiro and {McMillan}, Stephen L.~W.},
        title = "{Formation of massive black holes through runaway collisions in dense young star clusters}",
      journal = {\nat},
         year = 2004,
        month = apr,
       volume = {428},
       number = {6984},
        pages = {724-726},
          doi = {10.1038/nature02448},
archivePrefix = {arXiv},
       eprint = {astro-ph/0402622},
 primaryClass = {astro-ph},
       adsurl = {https://ui.adsabs.harvard.edu/abs/2004Natur.428..724P}
}

@ARTICLE{Volonteri2012,
       author = {{Volonteri}, Marta and {Bellovary}, Jillian},
        title = "{Black holes in the early Universe}",
      journal = {Reports on Progress in Physics},
         year = 2012,
        month = dec,
       volume = {75},
       number = {12},
          eid = {124901},
        pages = {124901},
          doi = {10.1088/0034-4885/75/12/124901},
archivePrefix = {arXiv},
       eprint = {1209.2243},
 primaryClass = {astro-ph.CO},
       adsurl = {https://ui.adsabs.harvard.edu/abs/2012RPPh...75l4901V}
}

@ARTICLE{Bogdan2024_uhz1_obs,
       author = {{Bogd{\'a}n}, {\'A}kos and {Goulding}, Andy D. and {Natarajan}, Priyamvada and {Kov{\'a}cs}, Orsolya E. and {Tremblay}, Grant R. and {Chadayammuri}, Urmila and {Volonteri}, Marta and {Kraft}, Ralph P. and {Forman}, William R. and {Jones}, Christine and {Churazov}, Eugene and {Zhuravleva}, Irina},
        title = "{Evidence for heavy-seed origin of early supermassive black holes from a z {\ensuremath{\approx}} 10 X-ray quasar}",
      journal = {Nature Astronomy},
         year = 2024,
        month = jan,
       volume = {8},
       number = {1},
        pages = {126-133},
          doi = {10.1038/s41550-023-02111-9},
archivePrefix = {arXiv},
       eprint = {2305.15458},
 primaryClass = {astro-ph.GA},
       adsurl = {https://ui.adsabs.harvard.edu/abs/2024NatAs...8..126B}
}

@ARTICLE{Lahen2025,
       author = {{Lah{\'e}n}, Natalia and {Rantala}, Antti and {Naab}, Thorsten and {Partmann}, Christian and {Johansson}, Peter H. and {Hislop}, Jessica May},
        title = "{The formation, evolution, and disruption of star clusters with improved gravitational dynamics in simulated dwarf galaxies}",
      journal = {\mnras},
         year = 2025,
        month = apr,
       volume = {538},
       number = {3},
        pages = {2129-2148},
          doi = {10.1093/mnras/staf350},
archivePrefix = {arXiv},
       eprint = {2410.01891},
 primaryClass = {astro-ph.GA},
       adsurl = {https://ui.adsabs.harvard.edu/abs/2025MNRAS.538.2129L}
}

@ARTICLE{Lahen2023,
       author = {{Lah{\'e}n}, Natalia and {Naab}, Thorsten and {Kauffmann}, Guinevere and {Sz{\'e}csi}, Dorottya and {Hislop}, Jessica May and {Rantala}, Antti and {Kozyreva}, Alexandra and {Walch}, Stefanie and {Hu}, Chia-Yu},
        title = "{Formation of star clusters and enrichment by massive stars in simulations of low-metallicity galaxies with a fully sampled initial stellar mass function}",
      journal = {\mnras},
         year = 2023,
        month = jun,
       volume = {522},
       number = {2},
        pages = {3092-3116},
          doi = {10.1093/mnras/stad1147},
archivePrefix = {arXiv},
       eprint = {2211.15705},
 primaryClass = {astro-ph.GA},
       adsurl = {https://ui.adsabs.harvard.edu/abs/2023MNRAS.522.3092L}
}

@ARTICLE{Kritos2023,
       author = {{Kritos}, Konstantinos and {Berti}, Emanuele and {Silk}, Joseph},
        title = "{Massive black hole assembly in nuclear star clusters}",
      journal = {\prd},
         year = 2023,
        month = oct,
       volume = {108},
       number = {8},
          eid = {083012},
        pages = {083012},
          doi = {10.1103/PhysRevD.108.083012},
archivePrefix = {arXiv},
       eprint = {2212.06845},
 primaryClass = {astro-ph.HE},
       adsurl = {https://ui.adsabs.harvard.edu/abs/2023PhRvD.108h3012K}
}

@ARTICLE{Kritos2024,
       author = {{Kritos}, Konstantinos and {Berti}, Emanuele and {Silk}, Joseph},
        title = "{Supermassive black holes from runaway mergers and accretion in nuclear star clusters}",
      journal = {\mnras},
         year = 2024,
        month = jun,
       volume = {531},
       number = {1},
        pages = {133-136},
          doi = {10.1093/mnras/stae1145},
archivePrefix = {arXiv},
       eprint = {2404.11676},
 primaryClass = {astro-ph.HE},
       adsurl = {https://ui.adsabs.harvard.edu/abs/2024MNRAS.531..133K}
}

@ARTICLE{Kritos2025_NSCmerger,
       author = {{Kritos}, Konstantinos and {Beckmann}, Ricarda S. and {Silk}, Joseph and {Berti}, Emanuele and {Yi}, Sophia and {Volonteri}, Marta and {Dubois}, Yohan and {Devriendt}, Julien},
        title = "{Supermassive Black Hole Growth in Hierarchically Merging Nuclear Star Clusters}",
      journal = {\apj},
         year = 2025,
        month = sep,
       volume = {991},
       number = {1},
          eid = {58},
        pages = {58},
          doi = {10.3847/1538-4357/adeb44},
archivePrefix = {arXiv},
       eprint = {2412.15334},
 primaryClass = {astro-ph.GA},
       adsurl = {https://ui.adsabs.harvard.edu/abs/2025ApJ...991...58K}
}

@ARTICLE{Mapelli2009,
       author = {{Mapelli}, M. and {Colpi}, M. and {Zampieri}, L.},
        title = "{Low metallicity and ultra-luminous X-ray sources in the Cartwheel galaxy}",
      journal = {\mnras},
         year = 2009,
        month = may,
       volume = {395},
       number = {1},
        pages = {L71-L75},
          doi = {10.1111/j.1745-3933.2009.00645.x},
archivePrefix = {arXiv},
       eprint = {0902.3540},
 primaryClass = {astro-ph.HE},
       adsurl = {https://ui.adsabs.harvard.edu/abs/2009MNRAS.395L..71M}
}

@ARTICLE{Heger2003,
       author = {{Heger}, A. and {Fryer}, C.~L. and {Woosley}, S.~E. and {Langer}, N. and {Hartmann}, D.~H.},
        title = "{How Massive Single Stars End Their Life}",
      journal = {\apj},
         year = 2003,
        month = jul,
       volume = {591},
       number = {1},
        pages = {288-300},
          doi = {10.1086/375341},
archivePrefix = {arXiv},
       eprint = {astro-ph/0212469},
 primaryClass = {astro-ph},
       adsurl = {https://ui.adsabs.harvard.edu/abs/2003ApJ...591..288H}
}

@ARTICLE{Habouzit2017,
       author = {{Habouzit}, M{\'e}lanie and {Volonteri}, Marta and {Dubois}, Yohan},
        title = "{Blossoms from black hole seeds: properties and early growth regulated by supernova feedback}",
      journal = {\mnras},
         year = 2017,
        month = jul,
       volume = {468},
       number = {4},
        pages = {3935-3948},
          doi = {10.1093/mnras/stx666},
archivePrefix = {arXiv},
       eprint = {1605.09394},
 primaryClass = {astro-ph.GA},
       adsurl = {https://ui.adsabs.harvard.edu/abs/2017MNRAS.468.3935H}
}

@ARTICLE{Pacucci2023,
       author = {{Pacucci}, Fabio and {Nguyen}, Bao and {Carniani}, Stefano and {Maiolino}, Roberto and {Fan}, Xiaohui},
        title = "{JWST CEERS and JADES Active Galaxies at z = 4-7 Violate the Local M $_{{\ensuremath{\bullet}}}$-M $_{{\ensuremath{\star}}}$ Relation at >3{\ensuremath{\sigma}}: Implications for Low-mass Black Holes and Seeding Models}",
      journal = {\apjl},
         year = 2023,
        month = nov,
       volume = {957},
       number = {1},
          eid = {L3},
        pages = {L3},
          doi = {10.3847/2041-8213/ad0158},
archivePrefix = {arXiv},
       eprint = {2308.12331},
 primaryClass = {astro-ph.GA},
       adsurl = {https://ui.adsabs.harvard.edu/abs/2023ApJ...957L...3P}
}

@ARTICLE{Kritos2026_NSC_LRD,
       author = {{Kritos}, Konstantinos and {Silk}, Joseph},
        title = "{From Nuclear Star Clusters to Little Red Dots: Black Hole Growth, Mergers, and Tidal Disruptions}",
      journal = {\apj},
         year = 2026,
        month = jun,
       volume = {1004},
       number = {2},
          eid = {226},
        pages = {226},
          doi = {10.3847/1538-4357/ae731c},
archivePrefix = {arXiv},
       eprint = {2510.21709},
 primaryClass = {astro-ph.HE},
       adsurl = {https://ui.adsabs.harvard.edu/abs/2026ApJ..1004..226K}
}

@ARTICLE{Rodriguez2023,
       author = {{Rodriguez}, Carl L. and {Hafen}, Zachary and {Grudi{\'c}}, Michael Y. and {Lamberts}, Astrid and {Sharma}, Kuldeep and {Faucher-Gigu{\`e}re}, Claude-Andr{\'e} and {Wetzel}, Andrew},
        title = "{Great balls of FIRE II: The evolution and destruction of star clusters across cosmic time in a Milky Way-mass galaxy}",
      journal = {\mnras},
         year = 2023,
        month = may,
       volume = {521},
       number = {1},
        pages = {124-147},
          doi = {10.1093/mnras/stad578},
archivePrefix = {arXiv},
       eprint = {2203.16547},
 primaryClass = {astro-ph.GA},
       adsurl = {https://ui.adsabs.harvard.edu/abs/2023MNRAS.521..124R}
}

@ARTICLE{Bruel2024,
       author = {{Bruel}, T. and {Rodriguez}, C.~L. and {Lamberts}, A. and {Grudi{\'c}}, M.~Y. and {Hafen}, Z. and {Feldmann}, R.},
        title = "{Great Balls of FIRE. III. Modelling black hole mergers from massive star clusters in simulations of galaxies}",
      journal = {\aap},
         year = 2024,
        month = jun,
       volume = {686},
          eid = {A106},
        pages = {A106},
          doi = {10.1051/0004-6361/202348716},
archivePrefix = {arXiv},
       eprint = {2311.14855},
 primaryClass = {astro-ph.GA},
       adsurl = {https://ui.adsabs.harvard.edu/abs/2024A&A...686A.106B}
}

@ARTICLE{Rantala2025_Frost_clustersII,
       author = {{Rantala}, Antti and {Lah{\'e}n}, Natalia and {Naab}, Thorsten and {Escobar}, Gast{\'o}n J. and {Iorio}, Giuliano},
        title = "{FROST-CLUSTERS ─ II. Massive stars, binaries, and triples boost supermassive black hole seed formation in assembling star clusters}",
      journal = {\mnras},
         year = 2025,
        month = nov,
       volume = {543},
       number = {3},
        pages = {2130-2158},
          doi = {10.1093/mnras/staf1519},
archivePrefix = {arXiv},
       eprint = {2506.04330},
 primaryClass = {astro-ph.GA},
       adsurl = {https://ui.adsabs.harvard.edu/abs/2025MNRAS.543.2130R}
}

@ARTICLE{Rantala2024_Frost_clusterI,
       author = {{Rantala}, Antti and {Naab}, Thorsten and {Lah{\'e}n}, Natalia},
        title = "{FROST-CLUSTERS - I. Hierarchical star cluster assembly boosts intermediate-mass black hole formation}",
      journal = {\mnras},
         year = 2024,
        month = jul,
       volume = {531},
       number = {3},
        pages = {3770-3799},
          doi = {10.1093/mnras/stae1413},
archivePrefix = {arXiv},
       eprint = {2403.10602},
 primaryClass = {astro-ph.GA},
       adsurl = {https://ui.adsabs.harvard.edu/abs/2024MNRAS.531.3770R}
}

@ARTICLE{Grudic2023,
       author = {{Grudi{\'c}}, Michael Y. and {Hafen}, Zachary and {Rodriguez}, Carl L. and {Guszejnov}, D{\'a}vid and {Lamberts}, Astrid and {Wetzel}, Andrew and {Boylan-Kolchin}, Michael and {Faucher-Gigu{\`e}re}, Claude-Andr{\'e}},
        title = "{Great balls of FIRE - I. The formation of star clusters across cosmic time in a Milky Way-mass galaxy}",
      journal = {\mnras},
         year = 2023,
        month = feb,
       volume = {519},
       number = {1},
        pages = {1366-1380},
          doi = {10.1093/mnras/stac3573},
archivePrefix = {arXiv},
       eprint = {2203.05732},
 primaryClass = {astro-ph.GA},
       adsurl = {https://ui.adsabs.harvard.edu/abs/2023MNRAS.519.1366G}
}

@ARTICLE{Rantala2026_Frost_clusterIII,
       author = {{Rantala}, Antti and {Naab}, Thorsten and {Lah{\'e}n}, Natalia and {Reuter}, Klaus and {Rampp}, Markus and {Chru{\'s}li{\'n}ska}, Martyna and {Reinoso}, Basti{\'a}n},
        title = "{FROST-CLUSTERS ─ III. Metallicity-dependent intermediate-mass black hole formation by runaway collisions in dense star clusters}",
      journal = {\mnras},
         year = 2026,
        month = jul,
       volume = {549},
       number = {4},
          eid = {stag986},
        pages = {stag986},
          doi = {10.1093/mnras/stag986},
archivePrefix = {arXiv},
       eprint = {2601.07917},
 primaryClass = {astro-ph.GA},
       adsurl = {https://ui.adsabs.harvard.edu/abs/2026MNRAS.549ag986R}
}

@ARTICLE{Mowla2024_firefly,
       author = {{Mowla}, Lamiya and {Iyer}, Kartheik and {Asada}, Yoshihisa and {Desprez}, Guillaume and {Tan}, Vivian Yun Yan and {Martis}, Nicholas and {Sarrouh}, Ghassan and {Strait}, Victoria and {Abraham}, Roberto and {Brada{\v{c}}}, Maru{\v{s}}a and {Brammer}, Gabriel and {Muzzin}, Adam and {Pacifici}, Camilla and {Ravindranath}, Swara and {Sawicki}, Marcin and {Willott}, Chris and {Estrada-Carpenter}, Vince and {Jahan}, Nusrath and {Noirot}, Ga{\"e}l and {Matharu}, Jasleen and {Rihtar{\v{s}}i{\v{c}}}, Gregor and {Zabl}, Johannes},
        title = "{Formation of a low-mass galaxy from star clusters in a 600-million-year-old Universe}",
      journal = {\nat},
         year = 2024,
        month = dec,
       volume = {636},
       number = {8042},
        pages = {332-336},
          doi = {10.1038/s41586-024-08293-0},
archivePrefix = {arXiv},
       eprint = {2402.08696},
 primaryClass = {astro-ph.GA},
       adsurl = {https://ui.adsabs.harvard.edu/abs/2024Natur.636..332M}
}

@ARTICLE{Zhou2026_ambra,
       author = {{Zhou}, Yihao and {Bhowmick}, Aklant Kumar and {Di Matteo}, Tiziana and {LaChance}, Patrick and {Croft}, Rupert and {Blecha}, Laura and {Bird}, Simeon and {Torrey}, Paul and {Hernquist}, Lars},
        title = "{First results of AMBRA: Abundant Seeds and Early Mergers as a Pathway to the First Massive Black Holes}",
      journal = {arXiv e-prints},
         year = 2026,
        month = apr,
          eid = {arXiv:2604.01123},
        pages = {arXiv:2604.01123},
          doi = {10.48550/arXiv.2604.01123},
archivePrefix = {arXiv},
       eprint = {2604.01123},
 primaryClass = {astro-ph.GA},
       adsurl = {https://ui.adsabs.harvard.edu/abs/2026arXiv260401123Z}
}

@ARTICLE{Adamo2024_cosmicGems,
       author = {{Adamo}, Angela and {Bradley}, Larry D. and {Vanzella}, Eros and {Claeyssens}, Ad{\'e}la{\"\i}de and {Welch}, Brian and {Diego}, Jose M. and {Mahler}, Guillaume and {Oguri}, Masamune and {Sharon}, Keren and {Abdurro'uf} and {Hsiao}, Tiger Yu-Yang and {Xu}, Xinfeng and {Messa}, Matteo and {Lassen}, Augusto E. and {Zackrisson}, Erik and {Brammer}, Gabriel and {Coe}, Dan and {Kokorev}, Vasily and {Ricotti}, Massimo and {Zitrin}, Adi and {Fujimoto}, Seiji and {Inoue}, Akio K. and {Resseguier}, Tom and {Rigby}, Jane R. and {Jim{\'e}nez-Teja}, Yolanda and {Windhorst}, Rogier A. and {Hashimoto}, Takuya and {Tamura}, Yoichi},
        title = "{Bound star clusters observed in a lensed galaxy 460 Myr after the Big Bang}",
      journal = {\nat},
         year = 2024,
        month = aug,
       volume = {632},
       number = {8025},
        pages = {513-516},
          doi = {10.1038/s41586-024-07703-7},
archivePrefix = {arXiv},
       eprint = {2401.03224},
 primaryClass = {astro-ph.GA},
       adsurl = {https://ui.adsabs.harvard.edu/abs/2024Natur.632..513A}
}

@ARTICLE{Vanzella2023,
       author = {{Vanzella}, Eros and {Claeyssens}, Ad{\'e}la{\"\i}de and {Welch}, Brian and {Adamo}, Angela and {Coe}, Dan and {Diego}, Jose M. and {Mahler}, Guillaume and {Khullar}, Gourav and {Kokorev}, Vasily and {Oguri}, Masamune and {Ravindranath}, Swara and {Furtak}, Lukas J. and {Hsiao}, Tiger Yu-Yang and {Abdurro'uf} and {Mandelker}, Nir and {Brammer}, Gabriel and {Bradley}, Larry D. and {Brada{\v{c}}}, Maru{\v{s}}a and {Conselice}, Christopher J. and {Dayal}, Pratika and {Nonino}, Mario and {Andrade-Santos}, Felipe and {Windhorst}, Rogier A. and {Pirzkal}, Nor and {Sharon}, Keren and {de Mink}, S.~E. and {Fujimoto}, Seiji and {Zitrin}, Adi and {Eldridge}, Jan J. and {Norman}, Colin},
        title = "{JWST/NIRCam Probes Young Star Clusters in the Reionization Era Sunrise Arc}",
      journal = {\apj},
         year = 2023,
        month = mar,
       volume = {945},
       number = {1},
          eid = {53},
        pages = {53},
          doi = {10.3847/1538-4357/acb59a},
archivePrefix = {arXiv},
       eprint = {2211.09839},
 primaryClass = {astro-ph.GA},
       adsurl = {https://ui.adsabs.harvard.edu/abs/2023ApJ...945...53V}
}

@ARTICLE{Lee2023,
       author = {{Lee}, Seungjae and {Kim}, Ji-hoon and {Oh}, Boon Kiat},
        title = "{Growth of a Massive Black Hole in a Dense Star Cluster Via Tidal Disruption Accretion}",
      journal = {\apj},
         year = 2023,
        month = feb,
       volume = {943},
       number = {2},
          eid = {77},
        pages = {77},
          doi = {10.3847/1538-4357/aca75d},
archivePrefix = {arXiv},
       eprint = {2211.02376},
 primaryClass = {astro-ph.GA},
       adsurl = {https://ui.adsabs.harvard.edu/abs/2023ApJ...943...77L}
}

@ARTICLE{Strubbe2009,
       author = {{Strubbe}, Linda E. and {Quataert}, Eliot},
        title = "{Optical flares from the tidal disruption of stars by massive black holes}",
      journal = {\mnras},
         year = 2009,
        month = dec,
       volume = {400},
       number = {4},
        pages = {2070-2084},
          doi = {10.1111/j.1365-2966.2009.15599.x},
archivePrefix = {arXiv},
       eprint = {0905.3735},
 primaryClass = {astro-ph.CO},
       adsurl = {https://ui.adsabs.harvard.edu/abs/2009MNRAS.400.2070S}
}

@ARTICLE{Stone2017,
       author = {{Stone}, Nicholas C. and {K{\"u}pper}, Andreas H.~W. and {Ostriker}, Jeremiah P.},
        title = "{Formation of massive black holes in galactic nuclei: runaway tidal encounters}",
      journal = {\mnras},
         year = 2017,
        month = jun,
       volume = {467},
       number = {4},
        pages = {4180-4199},
          doi = {10.1093/mnras/stx097},
archivePrefix = {arXiv},
       eprint = {1606.01909},
 primaryClass = {astro-ph.GA},
       adsurl = {https://ui.adsabs.harvard.edu/abs/2017MNRAS.467.4180S}
}

@ARTICLE{Rees1988,
       author = {{Rees}, Martin J.},
        title = "{Tidal disruption of stars by black holes of {}10$^{6}$-{}10$^{8}$ solar masses in nearby galaxies}",
      journal = {\nat},
         year = 1988,
        month = jun,
       volume = {333},
       number = {6173},
        pages = {523-528},
          doi = {10.1038/333523a0},
       adsurl = {https://ui.adsabs.harvard.edu/abs/1988Natur.333..523R}
}

@ARTICLE{Chung2026,
       author = {{Chung}, Eunwoo and {Jo}, Yongseok and {Kim}, Ji-hoon and {Jung}, Minyong and {Kwon}, Oh-kyoung},
        title = "{From Dense Gas Clouds to Supermassive Black Hole Seeds: Hybrid Hydro/Direct $N$-body Simulations of Runaway Collision-driven Intermediate-mass Black Hole Formation}",
      journal = {arXiv e-prints},
         year = 2026,
        month = jun,
          eid = {arXiv:2606.10000},
        pages = {arXiv:2606.10000},
          doi = {10.48550/arXiv.2606.10000},
archivePrefix = {arXiv},
       eprint = {2606.10000},
 primaryClass = {astro-ph.GA},
       adsurl = {https://ui.adsabs.harvard.edu/abs/2026arXiv260610000C}
}

@ARTICLE{Marques-Chaves2026,
       author = {{Marques-Chaves}, R. and {Martins}, F. and {Schaerer}, D. and {Dessauges-Zavadsky}, M. and {Palacios}, A.},
        title = "{Signatures of very massive stars in the epoch of reionization}",
      journal = {\aap},
         year = 2026,
        month = jul,
       volume = {711},
          eid = {L3},
        pages = {L3},
          doi = {10.1051/0004-6361/202660493},
archivePrefix = {arXiv},
       eprint = {2604.21493},
 primaryClass = {astro-ph.GA},
       adsurl = {https://ui.adsabs.harvard.edu/abs/2026A&A...711L...3M}
}

@ARTICLE{Mukherjee2025_magicsIII,
       author = {{Mukherjee}, Diptajyoti and {Zhou}, Yihao and {Chen}, Nianyi and {Di Carlo}, Ugo Niccol{\`o} and {Di Matteo}, Tiziana},
        title = "{MAGICS. III. Seeds Sink Swiftly: Nuclear Star Clusters Dramatically Accelerate Seed Black Hole Mergers}",
      journal = {\apj},
         year = 2025,
        month = mar,
       volume = {981},
       number = {2},
          eid = {203},
        pages = {203},
          doi = {10.3847/1538-4357/adb1b0},
archivePrefix = {arXiv},
       eprint = {2409.19095},
 primaryClass = {astro-ph.GA},
       adsurl = {https://ui.adsabs.harvard.edu/abs/2025ApJ...981..203M}
}

@ARTICLE{Zhang1999,
       author = {{Zhang}, Qing and {Fall}, S. Michael},
        title = "{The Mass Function of Young Star Clusters in the ``Antennae'' Galaxies}",
      journal = {\apjl},
         year = 1999,
        month = dec,
       volume = {527},
       number = {2},
        pages = {L81-L84},
          doi = {10.1086/312412},
archivePrefix = {arXiv},
       eprint = {astro-ph/9911229},
 primaryClass = {astro-ph},
       adsurl = {https://ui.adsabs.harvard.edu/abs/1999ApJ...527L..81Z}
}

@ARTICLE{Messa2018,
       author = {{Messa}, M. and {Adamo}, A. and {{\"O}stlin}, G. and {Calzetti}, D. and {Grasha}, K. and {Grebel}, E.~K. and {Shabani}, F. and {Chandar}, R. and {Dale}, D.~A. and {Dobbs}, C.~L. and {Elmegreen}, B.~G. and {Fumagalli}, M. and {Gouliermis}, D.~A. and {Kim}, H. and {Smith}, L.~J. and {Thilker}, D.~A. and {Tosi}, M. and {Ubeda}, L. and {Walterbos}, R. and {Whitmore}, B.~C. and {Fedorenko}, K. and {Mahadevan}, S. and {Andrews}, J.~E. and {Bright}, S.~N. and {Cook}, D.~O. and {Kahre}, L. and {Nair}, P. and {Pellerin}, A. and {Ryon}, J.~E. and {Ahmad}, S.~D. and {Beale}, L.~P. and {Brown}, K. and {Clarkson}, D.~A. and {Guidarelli}, G.~C. and {Parziale}, R. and {Turner}, J. and {Weber}, M.},
        title = "{The young star cluster population of M51 with LEGUS - I. A comprehensive study of cluster formation and evolution}",
      journal = {\mnras},
         year = 2018,
        month = jan,
       volume = {473},
       number = {1},
        pages = {996-1018},
          doi = {10.1093/mnras/stx2403},
archivePrefix = {arXiv},
       eprint = {1709.06101},
 primaryClass = {astro-ph.GA},
       adsurl = {https://ui.adsabs.harvard.edu/abs/2018MNRAS.473..996M}
}

@ARTICLE{Peres1962,
       author = {{Peres}, Asher},
        title = "{Classical Radiation Recoil}",
      journal = {Physical Review},
         year = 1962,
        month = dec,
       volume = {128},
       number = {5},
        pages = {2471-2475},
          doi = {10.1103/PhysRev.128.2471},
       adsurl = {https://ui.adsabs.harvard.edu/abs/1962PhRv..128.2471P}
}

@ARTICLE{Fitchett1983,
       author = {{Fitchett}, M.~J.},
        title = "{The influence of gravitational wave momentum losses on the centre of mass motion of a Newtonian binay system.}",
      journal = {\mnras},
         year = 1983,
        month = jun,
       volume = {203},
        pages = {1049-1062},
          doi = {10.1093/mnras/203.4.1049},
       adsurl = {https://ui.adsabs.harvard.edu/abs/1983MNRAS.203.1049F}
}

@ARTICLE{Lousto2011,
       author = {{Lousto}, Carlos O. and {Zlochower}, Yosef},
        title = "{Hangup Kicks: Still Larger Recoils by Partial Spin-Orbit Alignment of Black-Hole Binaries}",
      journal = {\prl},
         year = 2011,
        month = dec,
       volume = {107},
       number = {23},
          eid = {231102},
        pages = {231102},
          doi = {10.1103/PhysRevLett.107.231102},
archivePrefix = {arXiv},
       eprint = {1108.2009},
 primaryClass = {gr-qc},
       adsurl = {https://ui.adsabs.harvard.edu/abs/2011PhRvL.107w1102L}
}

@ARTICLE{Marks2012,
       author = {{Marks}, M. and {Kroupa}, P.},
        title = "{Inverse dynamical population synthesis. Constraining the initial conditions of young stellar clusters by studying their binary populations}",
      journal = {\aap},
         year = 2012,
        month = jul,
       volume = {543},
          eid = {A8},
        pages = {A8},
          doi = {10.1051/0004-6361/201118231},
archivePrefix = {arXiv},
       eprint = {1205.1508},
 primaryClass = {astro-ph.GA},
       adsurl = {https://ui.adsabs.harvard.edu/abs/2012A&A...543A...8M}
}

@ARTICLE{GonzalezPrieto2025,
       author = {{Gonz{\'a}lez Prieto}, Elena and {Rodriguez}, Carl L. and {Cabrera}, Tom{\'a}s},
        title = "{Growing the Intermediate-mass Black Hole in Omega Centauri}",
      journal = {\apjl},
         year = 2025,
        month = sep,
       volume = {990},
       number = {2},
          eid = {L69},
        pages = {L69},
          doi = {10.3847/2041-8213/adfd4a},
archivePrefix = {arXiv},
       eprint = {2507.06316},
 primaryClass = {astro-ph.HE},
       adsurl = {https://ui.adsabs.harvard.edu/abs/2025ApJ...990L..69G}
}

@ARTICLE{Davies2011,
       author = {{Davies}, Melvyn B. and {Miller}, M. Coleman and {Bellovary}, Jillian M.},
        title = "{Supermassive Black Hole Formation Via Gas Accretion in Nuclear Stellar Clusters}",
      journal = {\apjl},
         year = 2011,
        month = oct,
       volume = {740},
       number = {2},
          eid = {L42},
        pages = {L42},
          doi = {10.1088/2041-8205/740/2/L42},
archivePrefix = {arXiv},
       eprint = {1106.5943},
 primaryClass = {astro-ph.CO},
       adsurl = {https://ui.adsabs.harvard.edu/abs/2011ApJ...740L..42D}
}

@ARTICLE{Campanelli2007,
       author = {{Campanelli}, Manuela and {Lousto}, Carlos and {Zlochower}, Yosef and {Merritt}, David},
        title = "{Large Merger Recoils and Spin Flips from Generic Black Hole Binaries}",
      journal = {\apjl},
         year = 2007,
        month = apr,
       volume = {659},
       number = {1},
        pages = {L5-L8},
          doi = {10.1086/516712},
archivePrefix = {arXiv},
       eprint = {gr-qc/0701164},
 primaryClass = {gr-qc},
       adsurl = {https://ui.adsabs.harvard.edu/abs/2007ApJ...659L...5C}
}

@ARTICLE{Dotti2010,
       author = {{Dotti}, M. and {Volonteri}, M. and {Perego}, A. and {Colpi}, M. and {Ruszkowski}, M. and {Haardt}, F.},
        title = "{Dual black holes in merger remnants - II. Spin evolution and gravitational recoil}",
      journal = {\mnras},
         year = 2010,
        month = feb,
       volume = {402},
       number = {1},
        pages = {682-690},
          doi = {10.1111/j.1365-2966.2009.15922.x},
archivePrefix = {arXiv},
       eprint = {0910.5729},
 primaryClass = {astro-ph.HE},
       adsurl = {https://ui.adsabs.harvard.edu/abs/2010MNRAS.402..682D}
}

@ARTICLE{Bogdanovic2007,
       author = {{Bogdanovi{\'c}}, Tamara and {Reynolds}, Christopher S. and {Miller}, M. Coleman},
        title = "{Alignment of the Spins of Supermassive Black Holes Prior to Coalescence}",
      journal = {\apjl},
         year = 2007,
        month = jun,
       volume = {661},
       number = {2},
        pages = {L147-L150},
          doi = {10.1086/518769},
archivePrefix = {arXiv},
       eprint = {astro-ph/0703054},
 primaryClass = {astro-ph},
       adsurl = {https://ui.adsabs.harvard.edu/abs/2007ApJ...661L.147B}
}

@ARTICLE{Williams2026_PopIII,
       author = {{Williams}, Claire E. and {Naoz}, Smadar and {Yoshida}, Naoki and {Lake}, William and {Burkhart}, Blakesley and {Marinacci}, Federico and {Vogelsberger}, Mark and {Menon}, Shyam H. and {Chen}, Avi and {Kremer}, Kyle},
        title = "{Observing Compact Population III Star Clusters and the Presence of Cosmic Streaming}",
      journal = {\apj},
         year = 2026,
        month = feb,
       volume = {997},
       number = {2},
          eid = {202},
        pages = {202},
          doi = {10.3847/1538-4357/ae27c9},
archivePrefix = {arXiv},
       eprint = {2509.22776},
 primaryClass = {astro-ph.GA},
       adsurl = {https://ui.adsabs.harvard.edu/abs/2026ApJ...997..202W}
}

@ARTICLE{Chisholm2026,
       author = {{Chisholm}, John and {Berg}, Danielle A. and {Boylan-Kolchin}, Michael and {de Graaff}, Anna and {Furtak}, Lukas J. and {Kokorev}, Vasily and {Matthee}, Jorryt and {Mu{\~n}oz}, Julian B. and {Naidu}, Rohan P. and {Sander}, Andreas A.~C.},
        title = "{Little Red Dots as Globular Clusters in Formation}",
      journal = {\apjl},
         year = 2026,
        month = jun,
       volume = {1004},
       number = {1},
          eid = {L4},
        pages = {L4},
          doi = {10.3847/2041-8213/ae6dae},
archivePrefix = {arXiv},
       eprint = {2602.15935},
 primaryClass = {astro-ph.GA},
       adsurl = {https://ui.adsabs.harvard.edu/abs/2026ApJ..1004L...4C}
}

@ARTICLE{Choksi2021,
       author = {{Choksi}, Nick and {Kruijssen}, J.~M. Diederik},
        title = "{On the initial mass-radius relation of stellar clusters}",
      journal = {\mnras},
         year = 2021,
        month = nov,
       volume = {507},
       number = {4},
        pages = {5492-5506},
          doi = {10.1093/mnras/stab2514},
archivePrefix = {arXiv},
       eprint = {1912.05560},
 primaryClass = {astro-ph.GA},
       adsurl = {https://ui.adsabs.harvard.edu/abs/2021MNRAS.507.5492C}
}

@ARTICLE{Bekenstein1973,
       author = {{Bekenstein}, Jacob D.},
        title = "{Gravitational-Radiation Recoil and Runaway Black Holes}",
      journal = {\apj},
         year = 1973,
        month = jul,
       volume = {183},
        pages = {657-664},
          doi = {10.1086/152255},
       adsurl = {https://ui.adsabs.harvard.edu/abs/1973ApJ...183..657B}
}

@ARTICLE{Sayeb2021,
       author = {{Sayeb}, Mohammad and {Blecha}, Laura and {Kelley}, Luke Zoltan and {Gerosa}, Davide and {Kesden}, Michael and {Thomas}, July},
        title = "{Massive black hole binary inspiral and spin evolution in a cosmological framework}",
      journal = {\mnras},
         year = 2021,
        month = feb,
       volume = {501},
       number = {2},
        pages = {2531-2546},
          doi = {10.1093/mnras/staa3826},
archivePrefix = {arXiv},
       eprint = {2006.06647},
 primaryClass = {astro-ph.GA},
       adsurl = {https://ui.adsabs.harvard.edu/abs/2021MNRAS.501.2531S}
}

@ARTICLE{Liao2023,
       author = {{Liao}, Shihong and {Johansson}, Peter H. and {Mannerkoski}, Matias and {Irodotou}, Dimitrios and {Rizzuto}, Francesco Paolo and {McAlpine}, Stuart and {Rantala}, Antti and {Rawlings}, Alexander and {Sawala}, Till},
        title = "{Modelling the accretion and feedback of supermassive black hole binaries in gas-rich galaxy mergers}",
      journal = {\mnras},
         year = 2023,
        month = apr,
       volume = {520},
       number = {3},
        pages = {4463-4489},
          doi = {10.1093/mnras/stad412},
archivePrefix = {arXiv},
       eprint = {2211.11788},
 primaryClass = {astro-ph.GA},
       adsurl = {https://ui.adsabs.harvard.edu/abs/2023MNRAS.520.4463L}
}

@ARTICLE{Cook2019,
       author = {{Cook}, D.~O. and {Lee}, J.~C. and {Adamo}, A. and {Kim}, H. and {Chandar}, R. and {Whitmore}, B.~C. and {Mok}, A. and {Ryon}, J.~E. and {Dale}, D.~A. and {Calzetti}, D. and {Andrews}, J.~E. and {Aloisi}, A. and {Ashworth}, G. and {Bright}, S.~N. and {Brown}, T.~M. and {Christian}, C. and {Cignoni}, M. and {Clayton}, G.~C. and {da Silva}, R. and {de Mink}, S.~E. and {Dobbs}, C.~L. and {Elmegreen}, B.~G. and {Elmegreen}, D.~M. and {Evans}, A.~S. and {Fumagalli}, M. and {Gallagher}, J.~S. and {Gouliermis}, D.~A. and {Grasha}, K. and {Grebel}, E.~K. and {Herrero}, A. and {Hunter}, D.~A. and {Jensen}, E.~I. and {Johnson}, K.~E. and {Kahre}, L. and {Kennicutt}, R.~C. and {Krumholz}, M.~R. and {Lee}, N.~J. and {Lennon}, D. and {Linden}, S. and {Martin}, C. and {Messa}, M. and {Nair}, P. and {Nota}, A. and {{\"O}stlin}, G. and {Parziale}, R.~C. and {Pellerin}, A. and {Regan}, M.~W. and {Sabbi}, E. and {Sacchi}, E. and {Schaerer}, D. and {Schiminovich}, D. and {Shabani}, F. and {Slane}, F.~A. and {Small}, J. and {Smith}, C.~L. and {Smith}, L.~J. and {Taibi}, S. and {Thilker}, D.~A. and {de la Torre}, I.~C. and {Tosi}, M. and {Turner}, J.~A. and {Ubeda}, L. and {Van Dyk}, S.~D. and {Walterbos}, R. AM and {Wofford}, A.},
        title = "{Star cluster catalogues for the LEGUS dwarf galaxies}",
      journal = {\mnras},
         year = 2019,
        month = apr,
       volume = {484},
       number = {4},
        pages = {4897-4919},
          doi = {10.1093/mnras/stz331},
archivePrefix = {arXiv},
       eprint = {1902.00082},
 primaryClass = {astro-ph.GA},
       adsurl = {https://ui.adsabs.harvard.edu/abs/2019MNRAS.484.4897C}
}

@ARTICLE{Cook2023,
       author = {{Cook}, D.~O. and {Lee}, J.~C. and {Adamo}, A. and {Calzetti}, D. and {Chandar}, R. and {Whitmore}, B.~C. and {Aloisi}, A. and {Cignoni}, M. and {Dale}, D.~A. and {Elmegreen}, B.~G. and {Fumagalli}, M. and {Grasha}, K. and {Johnson}, K.~E. and {Kennicutt}, R.~C. and {Kim}, H. and {Linden}, S.~T. and {Messa}, M. and {{\"O}stlin}, G. and {Ryon}, J.~E. and {Sacchi}, E. and {Thilker}, D.~A. and {Tosi}, M. and {Wofford}, A.},
        title = "{Fraction of stars in clusters for the LEGUS dwarf galaxies}",
      journal = {\mnras},
         year = 2023,
        month = mar,
       volume = {519},
       number = {3},
        pages = {3749-3775},
          doi = {10.1093/mnras/stac3748},
archivePrefix = {arXiv},
       eprint = {2212.07519},
 primaryClass = {astro-ph.GA},
       adsurl = {https://ui.adsabs.harvard.edu/abs/2023MNRAS.519.3749C}
}

@ARTICLE{Chon2026,
       author = {{Chon}, Sunmyon and {Hirano}, Shingo and {Chen}, Ke-Jung and {Springel}, Volker},
        title = "{Formation of Heavy Seed Black Holes and Little Red Dots-like Compact Clusters in Metal-enriched Star-forming Regions}",
      journal = {arXiv e-prints},
         year = 2026,
        month = aug,
          eid = {arXiv:2609.00437},
        pages = {arXiv:2609.00437},
archivePrefix = {arXiv},
       eprint = {2609.00437},
 primaryClass = {astro-ph.GA},
       adsurl = {https://ui.adsabs.harvard.edu/abs/2026arXiv260900437C}
}

@ARTICLE{Haidar2022,
       author = {{Haidar}, Houda and {Habouzit}, M{\'e}lanie and {Volonteri}, Marta and {Mezcua}, Mar and {Greene}, Jenny and {Neumayer}, Nadine and {Angl{\'e}s-Alc{\'a}zar}, Daniel and {Martin-Navarro}, Ignacio and {Hoyer}, Nils and {Dubois}, Yohan and {Dav{\'e}}, Romeel},
        title = "{The black hole population in low-mass galaxies in large-scale cosmological simulations}",
      journal = {\mnras},
         year = 2022,
        month = aug,
       volume = {514},
       number = {4},
        pages = {4912-4931},
          doi = {10.1093/mnras/stac1659},
archivePrefix = {arXiv},
       eprint = {2201.09888},
 primaryClass = {astro-ph.GA},
       adsurl = {https://ui.adsabs.harvard.edu/abs/2022MNRAS.514.4912H}
}

@ARTICLE{Lousto2013,
       author = {{Lousto}, Carlos O. and {Zlochower}, Yosef},
        title = "{Nonlinear gravitational recoil from the mergers of precessing black-hole binaries}",
      journal = {\prd},
         year = 2013,
        month = apr,
       volume = {87},
       number = {8},
          eid = {084027},
        pages = {084027},
          doi = {10.1103/PhysRevD.87.084027},
archivePrefix = {arXiv},
       eprint = {1211.7099},
 primaryClass = {gr-qc},
       adsurl = {https://ui.adsabs.harvard.edu/abs/2013PhRvD..87h4027L}
}

@ARTICLE{Mummery2026,
       author = {{Mummery}, Andrew},
        title = "{Accretion discs in (repeating) partial tidal disruption events: rapid state transitions, UV plateaus, and flares from disc─remnant collisions}",
      journal = {\mnras},
         year = 2026,
        month = apr,
       volume = {547},
       number = {4},
          eid = {stag387},
        pages = {stag387},
          doi = {10.1093/mnras/stag387},
archivePrefix = {arXiv},
       eprint = {2602.18698},
 primaryClass = {astro-ph.HE},
       adsurl = {https://ui.adsabs.harvard.edu/abs/2026MNRAS.547ag387M}
}

@ARTICLE{Adamo2020,
       author = {{Adamo}, A. and {Hollyhead}, K. and {Messa}, M. and {Ryon}, J.~E. and {Bajaj}, V. and {Runnholm}, A. and {Aalto}, S. and {Calzetti}, D. and {Gallagher}, J.~S. and {Hayes}, M.~J. and {Kruijssen}, J.~M.~D. and {K{\"o}nig}, S. and {Larsen}, S.~S. and {Melinder}, J. and {Sabbi}, E. and {Smith}, L.~J. and {{\"O}stlin}, G.},
        title = "{Star cluster formation in the most extreme environments: insights from the HiPEEC survey}",
      journal = {\mnras},
         year = 2020,
        month = dec,
       volume = {499},
       number = {3},
        pages = {3267-3294},
          doi = {10.1093/mnras/staa2380},
archivePrefix = {arXiv},
       eprint = {2008.12794},
 primaryClass = {astro-ph.GA},
       adsurl = {https://ui.adsabs.harvard.edu/abs/2020MNRAS.499.3267A}
}

@ARTICLE{Lee2022,
       author = {{Lee}, Janice C. and {Whitmore}, Bradley C. and {Thilker}, David A. and {Deger}, Sinan and {Larson}, Kirsten L. and {Ubeda}, Leonardo and {Anand}, Gagandeep S. and {Boquien}, M{\'e}d{\'e}ric and {Chandar}, Rupali and {Dale}, Daniel A. and {Emsellem}, Eric and {Leroy}, Adam K. and {Rosolowsky}, Erik and {Schinnerer}, Eva and {Schmidt}, Judy and {Lilly}, James and {Turner}, Jordan and {Van Dyk}, Schuyler and {White}, Richard L. and {Barnes}, Ashley T. and {Belfiore}, Francesco and {Bigiel}, Frank and {Blanc}, Guillermo A. and {Cao}, Yixian and {Chevance}, Melanie and {Congiu}, Enrico and {Egorov}, Oleg V. and {Glover}, Simon C.~O. and {Grasha}, Kathryn and {Groves}, Brent and {Henshaw}, Jonathan D. and {Hughes}, Annie and {Klessen}, Ralf S. and {Koch}, Eric and {Kreckel}, Kathryn and {Kruijssen}, J.~M. Diederik and {Liu}, Daizhong and {Lopez}, Laura A. and {Mayker}, Ness and {Meidt}, Sharon E. and {Murphy}, Eric J. and {Pan}, Hsi-An and {Pety}, J{\'e}r{\^o}me and {Querejeta}, Miguel and {Razza}, Alessandro and {Saito}, Toshiki and {S{\'a}nchez-Bl{\'a}zquez}, Patricia and {Santoro}, Francesco and {Sardone}, Amy and {Scheuermann}, Fabian and {Schruba}, Andreas and {Sun}, Jiayi and {Usero}, Antonio and {Watkins}, E. and {Williams}, Thomas G.},
        title = "{The PHANGS-HST Survey: Physics at High Angular Resolution in Nearby Galaxies with the Hubble Space Telescope}",
      journal = {\apjs},
         year = 2022,
        month = jan,
       volume = {258},
       number = {1},
          eid = {10},
        pages = {10},
          doi = {10.3847/1538-4365/ac1fe5},
archivePrefix = {arXiv},
       eprint = {2101.02855},
 primaryClass = {astro-ph.GA},
       adsurl = {https://ui.adsabs.harvard.edu/abs/2022ApJS..258...10L}
}

@ARTICLE{Planck,
       author = {{Planck Collaboration} and {Aghanim}, N. and {Akrami}, Y. and {Ashdown}, M. and {Aumont}, J. and {Baccigalupi}, C. and {Ballardini}, M. and {Banday}, A.~J. and {Barreiro}, R.~B. and {Bartolo}, N. and {Basak}, S. and {Battye}, R. and {Benabed}, K. and {Bernard}, J. -P. and {Bersanelli}, M. and {Bielewicz}, P. and {Bock}, J.~J. and {Bond}, J.~R. and {Borrill}, J. and {Bouchet}, F.~R. and {Boulanger}, F. and {Bucher}, M. and {Burigana}, C. and {Butler}, R.~C. and {Calabrese}, E. and {Cardoso}, J. -F. and {Carron}, J. and {Challinor}, A. and {Chiang}, H.~C. and {Chluba}, J. and {Colombo}, L.~P.~L. and {Combet}, C. and {Contreras}, D. and {Crill}, B.~P. and {Cuttaia}, F. and {de Bernardis}, P. and {de Zotti}, G. and {Delabrouille}, J. and {Delouis}, J. -M. and {Di Valentino}, E. and {Diego}, J.~M. and {Dor{\'e}}, O. and {Douspis}, M. and {Ducout}, A. and {Dupac}, X. and {Dusini}, S. and {Efstathiou}, G. and {Elsner}, F. and {En{\ss}lin}, T.~A. and {Eriksen}, H.~K. and {Fantaye}, Y. and {Farhang}, M. and {Fergusson}, J. and {Fernandez-Cobos}, R. and {Finelli}, F. and {Forastieri}, F. and {Frailis}, M. and {Fraisse}, A.~A. and {Franceschi}, E. and {Frolov}, A. and {Galeotta}, S. and {Galli}, S. and {Ganga}, K. and {G{\'e}nova-Santos}, R.~T. and {Gerbino}, M. and {Ghosh}, T. and {Gonz{\'a}lez-Nuevo}, J. and {G{\'o}rski}, K.~M. and {Gratton}, S. and {Gruppuso}, A. and {Gudmundsson}, J.~E. and {Hamann}, J. and {Handley}, W. and {Hansen}, F.~K. and {Herranz}, D. and {Hildebrandt}, S.~R. and {Hivon}, E. and {Huang}, Z. and {Jaffe}, A.~H. and {Jones}, W.~C. and {Karakci}, A. and {Keih{\"a}nen}, E. and {Keskitalo}, R. and {Kiiveri}, K. and {Kim}, J. and {Kisner}, T.~S. and {Knox}, L. and {Krachmalnicoff}, N. and {Kunz}, M. and {Kurki-Suonio}, H. and {Lagache}, G. and {Lamarre}, J. -M. and {Lasenby}, A. and {Lattanzi}, M. and {Lawrence}, C.~R. and {Le Jeune}, M. and {Lemos}, P. and {Lesgourgues}, J. and {Levrier}, F. and {Lewis}, A. and {Liguori}, M. and {Lilje}, P.~B. and {Lilley}, M. and {Lindholm}, V. and {L{\'o}pez-Caniego}, M. and {Lubin}, P.~M. and {Ma}, Y. -Z. and {Mac{\'\i}as-P{\'e}rez}, J.~F. and {Maggio}, G. and {Maino}, D. and {Mandolesi}, N. and {Mangilli}, A. and {Marcos-Caballero}, A. and {Maris}, M. and {Martin}, P.~G. and {Martinelli}, M. and {Mart{\'\i}nez-Gonz{\'a}lez}, E. and {Matarrese}, S. and {Mauri}, N. and {McEwen}, J.~D. and {Meinhold}, P.~R. and {Melchiorri}, A. and {Mennella}, A. and {Migliaccio}, M. and {Millea}, M. and {Mitra}, S. and {Miville-Desch{\^e}nes}, M. -A. and {Molinari}, D. and {Montier}, L. and {Morgante}, G. and {Moss}, A. and {Natoli}, P. and {N{\o}rgaard-Nielsen}, H.~U. and {Pagano}, L. and {Paoletti}, D. and {Partridge}, B. and {Patanchon}, G. and {Peiris}, H.~V. and {Perrotta}, F. and {Pettorino}, V. and {Piacentini}, F. and {Polastri}, L. and {Polenta}, G. and {Puget}, J. -L. and {Rachen}, J.~P. and {Reinecke}, M. and {Remazeilles}, M. and {Renzi}, A. and {Rocha}, G. and {Rosset}, C. and {Roudier}, G. and {Rubi{\~n}o-Mart{\'\i}n}, J.~A. and {Ruiz-Granados}, B. and {Salvati}, L. and {Sandri}, M. and {Savelainen}, M. and {Scott}, D. and {Shellard}, E.~P.~S. and {Sirignano}, C. and {Sirri}, G. and {Spencer}, L.~D. and {Sunyaev}, R. and {Suur-Uski}, A. -S. and {Tauber}, J.~A. and {Tavagnacco}, D. and {Tenti}, M. and {Toffolatti}, L. and {Tomasi}, M. and {Trombetti}, T. and {Valenziano}, L. and {Valiviita}, J. and {Van Tent}, B. and {Vibert}, L. and {Vielva}, P. and {Villa}, F. and {Vittorio}, N. and {Wandelt}, B.~D. and {Wehus}, I.~K. and {White}, M. and {White}, S.~D.~M. and {Zacchei}, A. and {Zonca}, A.},
        title = "{Planck 2018 results. VI. Cosmological parameters}",
      journal = {\aap},
         year = 2020,
        month = sep,
       volume = {641},
          eid = {A6},
        pages = {A6},
          doi = {10.1051/0004-6361/201833910},
archivePrefix = {arXiv},
       eprint = {1807.06209},
 primaryClass = {astro-ph.CO},
       adsurl = {https://ui.adsabs.harvard.edu/abs/2020A&A...641A...6P}
}

@ARTICLE{Siwek2020,
       author = {{Siwek}, Magdalena S. and {Kelley}, Luke Zoltan and {Hernquist}, Lars},
        title = "{The effect of differential accretion on the gravitational wave background and the present-day MBH binary population}",
      journal = {\mnras},
         year = 2020,
        month = oct,
       volume = {498},
       number = {1},
        pages = {537-547},
          doi = {10.1093/mnras/staa2361},
archivePrefix = {arXiv},
       eprint = {2005.09010},
 primaryClass = {astro-ph.GA},
       adsurl = {https://ui.adsabs.harvard.edu/abs/2020MNRAS.498..537S}
}

@ARTICLE{Farris2014,
       author = {{Farris}, Brian D. and {Duffell}, Paul and {MacFadyen}, Andrew I. and {Haiman}, Zoltan},
        title = "{Binary Black Hole Accretion from a Circumbinary Disk: Gas Dynamics inside the Central Cavity}",
      journal = {\apj},
         year = 2014,
        month = mar,
       volume = {783},
       number = {2},
          eid = {134},
        pages = {134},
          doi = {10.1088/0004-637X/783/2/134},
archivePrefix = {arXiv},
       eprint = {1310.0492},
 primaryClass = {astro-ph.HE},
       adsurl = {https://ui.adsabs.harvard.edu/abs/2014ApJ...783..134F}
}

@ARTICLE{Duffell2020,
       author = {{Duffell}, Paul C. and {D'Orazio}, Daniel and {Derdzinski}, Andrea and {Haiman}, Zoltan and {MacFadyen}, Andrew and {Rosen}, Anna L. and {Zrake}, Jonathan},
        title = "{Circumbinary Disks: Accretion and Torque as a Function of Mass Ratio and Disk Viscosity}",
      journal = {\apj},
         year = 2020,
        month = sep,
       volume = {901},
       number = {1},
          eid = {25},
        pages = {25},
          doi = {10.3847/1538-4357/abab95},
archivePrefix = {arXiv},
       eprint = {1911.05506},
 primaryClass = {astro-ph.SR},
       adsurl = {https://ui.adsabs.harvard.edu/abs/2020ApJ...901...25D}
}

@ARTICLE{Siwek2023,
       author = {{Siwek}, Magdalena and {Weinberger}, Rainer and {Mu{\~n}oz}, Diego J. and {Hernquist}, Lars},
        title = "{Preferential accretion and circumbinary disc precession in eccentric binary systems}",
      journal = {\mnras},
         year = 2023,
        month = feb,
       volume = {518},
       number = {4},
        pages = {5059-5071},
          doi = {10.1093/mnras/stac3263},
archivePrefix = {arXiv},
       eprint = {2203.02514},
 primaryClass = {astro-ph.HE},
       adsurl = {https://ui.adsabs.harvard.edu/abs/2023MNRAS.518.5059S}
}

@ARTICLE{Wang2025,
       author = {{Wang}, Bonny Y. and {Zhou}, Yihao and {Chen}, William and {Chen}, Nianyi and {Di Matteo}, Tiziana and {Croft}, Rupert and {Bird}, Simeon and {Ni}, Yueying},
        title = "{Gravitational Waves from Massive Black Hole Mergers in ASTRID: Predictions for LISA}",
      journal = {\apj},
         year = 2025,
        month = nov,
       volume = {993},
       number = {2},
          eid = {199},
        pages = {199},
          doi = {10.3847/1538-4357/ae073e},
archivePrefix = {arXiv},
       eprint = {2503.24304},
 primaryClass = {astro-ph.GA},
       adsurl = {https://ui.adsabs.harvard.edu/abs/2025ApJ...993..199W}
}

@ARTICLE{Campanelli2007_maxRecoil,
       author = {{Campanelli}, Manuela and {Lousto}, Carlos O. and {Zlochower}, Yosef and {Merritt}, David},
        title = "{Maximum Gravitational Recoil}",
      journal = {\prl},
         year = 2007,
        month = jun,
       volume = {98},
       number = {23},
          eid = {231102},
        pages = {231102},
          doi = {10.1103/PhysRevLett.98.231102},
archivePrefix = {arXiv},
       eprint = {gr-qc/0702133},
 primaryClass = {gr-qc},
       adsurl = {https://ui.adsabs.harvard.edu/abs/2007PhRvL..98w1102C}
}

@ARTICLE{Rose2023,
       author = {{Rose}, Sanaea C. and {Naoz}, Smadar and {Sari}, Re'em and {Linial}, Itai},
        title = "{Stellar Collisions in the Galactic Center: Massive Stars, Collision Remnants, and Missing Red Giants}",
      journal = {\apj},
         year = 2023,
        month = sep,
       volume = {955},
       number = {1},
          eid = {30},
        pages = {30},
          doi = {10.3847/1538-4357/acee75},
archivePrefix = {arXiv},
       eprint = {2304.10569},
 primaryClass = {astro-ph.GA},
       adsurl = {https://ui.adsabs.harvard.edu/abs/2023ApJ...955...30R}
}

@ARTICLE{Merritt2006,
       author = {{Merritt}, David},
        title = "{Mass Deficits, Stalling Radii, and the Merger Histories of Elliptical Galaxies}",
      journal = {\apj},
         year = 2006,
        month = sep,
       volume = {648},
       number = {2},
        pages = {976-986},
          doi = {10.1086/506139},
archivePrefix = {arXiv},
       eprint = {astro-ph/0603439},
 primaryClass = {astro-ph},
       adsurl = {https://ui.adsabs.harvard.edu/abs/2006ApJ...648..976M}
}

@ARTICLE{Gnedin2014,
       author = {{Gnedin}, Oleg Y. and {Ostriker}, Jeremiah P. and {Tremaine}, Scott},
        title = "{Co-evolution of Galactic Nuclei and Globular Cluster Systems}",
      journal = {\apj},
         year = 2014,
        month = apr,
       volume = {785},
       number = {1},
          eid = {71},
        pages = {71},
          doi = {10.1088/0004-637X/785/1/71},
archivePrefix = {arXiv},
       eprint = {1308.0021},
 primaryClass = {astro-ph.CO},
       adsurl = {https://ui.adsabs.harvard.edu/abs/2014ApJ...785...71G}
}

@ARTICLE{Antonini2013,
       author = {{Antonini}, Fabio},
        title = "{Origin and Growth of Nuclear Star Clusters around Massive Black Holes}",
      journal = {\apj},
         year = 2013,
        month = jan,
       volume = {763},
       number = {1},
          eid = {62},
        pages = {62},
          doi = {10.1088/0004-637X/763/1/62},
archivePrefix = {arXiv},
       eprint = {1207.6589},
 primaryClass = {astro-ph.GA},
       adsurl = {https://ui.adsabs.harvard.edu/abs/2013ApJ...763...62A}
}

@ARTICLE{Gerosa2021,
       author = {{Gerosa}, Davide and {Fishbach}, Maya},
        title = "{Hierarchical mergers of stellar-mass black holes and their gravitational-wave signatures}",
      journal = {Nature Astronomy},
         year = 2021,
        month = jul,
       volume = {5},
        pages = {749-760},
          doi = {10.1038/s41550-021-01398-w},
archivePrefix = {arXiv},
       eprint = {2105.03439},
 primaryClass = {astro-ph.HE},
       adsurl = {https://ui.adsabs.harvard.edu/abs/2021NatAs...5..749G}
}

@ARTICLE{Antonini2019,
       author = {{Antonini}, Fabio and {Gieles}, Mark and {Gualandris}, Alessia},
        title = "{Black hole growth through hierarchical black hole mergers in dense star clusters: implications for gravitational wave detections}",
      journal = {\mnras},
         year = 2019,
        month = jul,
       volume = {486},
       number = {4},
        pages = {5008-5021},
          doi = {10.1093/mnras/stz1149},
archivePrefix = {arXiv},
       eprint = {1811.03640},
 primaryClass = {astro-ph.HE},
       adsurl = {https://ui.adsabs.harvard.edu/abs/2019MNRAS.486.5008A}
}

@BOOK{Spitzer1987,
       author = {{Spitzer}, Lyman},
        title = "{Dynamical evolution of globular clusters}",
         year = 1987,
       adsurl = {https://ui.adsabs.harvard.edu/abs/1987degc.book.....S}
}

@ARTICLE{Gieles2014,
       author = {{Gieles}, Mark and {Alexander}, Poul E.~R. and {Lamers}, Henny J.~G.~L.~M. and {Baumgardt}, Holger},
        title = "{A prescription and fast code for the long-term evolution of star clusters - II. Unbalanced and core evolution}",
      journal = {\mnras},
         year = 2014,
        month = jan,
       volume = {437},
       number = {1},
        pages = {916-929},
          doi = {10.1093/mnras/stt1980},
archivePrefix = {arXiv},
       eprint = {1310.3631},
 primaryClass = {astro-ph.GA},
       adsurl = {https://ui.adsabs.harvard.edu/abs/2014MNRAS.437..916G}
}

@ARTICLE{Gieles2023,
       author = {{Gieles}, Mark and {Gnedin}, Oleg Y.},
        title = "{The mass-loss rates of star clusters with stellar-mass black holes: implications for the globular cluster mass function}",
      journal = {\mnras},
         year = 2023,
        month = jul,
       volume = {522},
       number = {4},
        pages = {5340-5357},
          doi = {10.1093/mnras/stad1287},
archivePrefix = {arXiv},
       eprint = {2303.03791},
 primaryClass = {astro-ph.GA},
       adsurl = {https://ui.adsabs.harvard.edu/abs/2023MNRAS.522.5340G}
}

@ARTICLE{Sakurai2017,
       author = {{Sakurai}, Yuya and {Yoshida}, Naoki and {Fujii}, Michiko S. and {Hirano}, Shingo},
        title = "{Formation of intermediate-mass black holes through runaway collisions in the first star clusters}",
      journal = {\mnras},
         year = 2017,
        month = dec,
       volume = {472},
       number = {2},
        pages = {1677-1684},
          doi = {10.1093/mnras/stx2044},
archivePrefix = {arXiv},
       eprint = {1704.06130},
 primaryClass = {astro-ph.GA},
       adsurl = {https://ui.adsabs.harvard.edu/abs/2017MNRAS.472.1677S}
}

@ARTICLE{Gezari2021,
       author = {{Gezari}, Suvi},
        title = "{Tidal Disruption Events}",
      journal = {\araa},
         year = 2021,
        month = sep,
       volume = {59},
        pages = {21-58},
          doi = {10.1146/annurev-astro-111720-030029},
archivePrefix = {arXiv},
       eprint = {2104.14580},
 primaryClass = {astro-ph.HE},
       adsurl = {https://ui.adsabs.harvard.edu/abs/2021ARA&A..59...21G}
}

@ARTICLE{Fialkov2017,
       author = {{Fialkov}, Anastasia and {Loeb}, Abraham},
        title = "{Jetted tidal disruptions of stars as a flag of intermediate mass black holes at high redshifts}",
      journal = {\mnras},
         year = 2017,
        month = nov,
       volume = {471},
       number = {4},
        pages = {4286-4299},
          doi = {10.1093/mnras/stx1755},
archivePrefix = {arXiv},
       eprint = {1611.01386},
 primaryClass = {astro-ph.HE},
       adsurl = {https://ui.adsabs.harvard.edu/abs/2017MNRAS.471.4286F}
}

@ARTICLE{Gerosa2015,
       author = {{Gerosa}, Davide and {Sesana}, Alberto},
        title = "{Missing black holes in brightest cluster galaxies as evidence for the occurrence of superkicks in nature}",
      journal = {\mnras},
         year = 2015,
        month = jan,
       volume = {446},
       number = {1},
        pages = {38-55},
          doi = {10.1093/mnras/stu2049},
archivePrefix = {arXiv},
       eprint = {1405.2072},
 primaryClass = {astro-ph.GA},
       adsurl = {https://ui.adsabs.harvard.edu/abs/2015MNRAS.446...38G}
}

@ARTICLE{Blecha2016,
       author = {{Blecha}, Laura and {Sijacki}, Debora and {Kelley}, Luke Zoltan and {Torrey}, Paul and {Vogelsberger}, Mark and {Nelson}, Dylan and {Springel}, Volker and {Snyder}, Gregory and {Hernquist}, Lars},
        title = "{Recoiling black holes: prospects for detection and implications of spin alignment}",
      journal = {\mnras},
         year = 2016,
        month = feb,
       volume = {456},
       number = {1},
        pages = {961-989},
          doi = {10.1093/mnras/stv2646},
archivePrefix = {arXiv},
       eprint = {1508.01524},
 primaryClass = {astro-ph.GA},
       adsurl = {https://ui.adsabs.harvard.edu/abs/2016MNRAS.456..961B}
}

@ARTICLE{Gonzalez2007,
       author = {{Gonz{\'a}lez}, Jos{\'e} A. and {Sperhake}, Ulrich and {Br{\"u}gmann}, Bernd and {Hannam}, Mark and {Husa}, Sascha},
        title = "{Maximum Kick from Nonspinning Black-Hole Binary Inspiral}",
      journal = {\prl},
         year = 2007,
        month = mar,
       volume = {98},
       number = {9},
          eid = {091101},
        pages = {091101},
          doi = {10.1103/PhysRevLett.98.091101},
archivePrefix = {arXiv},
       eprint = {gr-qc/0610154},
 primaryClass = {gr-qc},
       adsurl = {https://ui.adsabs.harvard.edu/abs/2007PhRvL..98i1101G}
}

@ARTICLE{Zhou2025_magcisII,
       author = {{Zhou}, Yihao and {Mukherjee}, Diptajyoti and {Chen}, Nianyi and {Di Matteo}, Tiziana and {Johansson}, Peter H. and {Rantala}, Antti and {Partmann}, Christian and {Di Carlo}, Ugo Niccol{\`o} and {Bird}, Simeon and {Ni}, Yueying},
        title = "{MAGICS. II. Seed Black Holes Stripped of Their Surrounding Stars Do Not Sink}",
      journal = {\apj},
         year = 2025,
        month = feb,
       volume = {980},
       number = {1},
          eid = {79},
        pages = {79},
          doi = {10.3847/1538-4357/ada283},
archivePrefix = {arXiv},
       eprint = {2409.19914},
 primaryClass = {astro-ph.GA},
       adsurl = {https://ui.adsabs.harvard.edu/abs/2025ApJ...980...79Z}
}

@ARTICLE{Schaye2015_eagle,
       author = {{Schaye}, Joop and {Crain}, Robert A. and {Bower}, Richard G. and {Furlong}, Michelle and {Schaller}, Matthieu and {Theuns}, Tom and {Dalla Vecchia}, Claudio and {Frenk}, Carlos S. and {McCarthy}, I.~G. and {Helly}, John C. and {Jenkins}, Adrian and {Rosas-Guevara}, Y.~M. and {White}, Simon D.~M. and {Baes}, Maarten and {Booth}, C.~M. and {Camps}, Peter and {Navarro}, Julio F. and {Qu}, Yan and {Rahmati}, Alireza and {Sawala}, Till and {Thomas}, Peter A. and {Trayford}, James},
        title = "{The EAGLE project: simulating the evolution and assembly of galaxies and their environments}",
      journal = {\mnras},
         year = 2015,
        month = jan,
       volume = {446},
       number = {1},
        pages = {521-554},
          doi = {10.1093/mnras/stu2058},
archivePrefix = {arXiv},
       eprint = {1407.7040},
 primaryClass = {astro-ph.GA},
       adsurl = {https://ui.adsabs.harvard.edu/abs/2015MNRAS.446..521S}
}

@ARTICLE{Springel2018_illustrisTNGI,
       author = {{Springel}, Volker and {Pakmor}, R{\"u}diger and {Pillepich}, Annalisa and {Weinberger}, Rainer and {Nelson}, Dylan and {Hernquist}, Lars and {Vogelsberger}, Mark and {Genel}, Shy and {Torrey}, Paul and {Marinacci}, Federico and {Naiman}, Jill},
        title = "{First results from the IllustrisTNG simulations: matter and galaxy clustering}",
      journal = {\mnras},
         year = 2018,
        month = mar,
       volume = {475},
       number = {1},
        pages = {676-698},
          doi = {10.1093/mnras/stx3304},
archivePrefix = {arXiv},
       eprint = {1707.03397},
 primaryClass = {astro-ph.GA},
       adsurl = {https://ui.adsabs.harvard.edu/abs/2018MNRAS.475..676S}
}

@ARTICLE{Naiman2018_illustrisTNGII,
       author = {{Naiman}, Jill P. and {Pillepich}, Annalisa and {Springel}, Volker and {Ramirez-Ruiz}, Enrico and {Torrey}, Paul and {Vogelsberger}, Mark and {Pakmor}, R{\"u}diger and {Nelson}, Dylan and {Marinacci}, Federico and {Hernquist}, Lars and {Weinberger}, Rainer and {Genel}, Shy},
        title = "{First results from the IllustrisTNG simulations: a tale of two elements - chemical evolution of magnesium and europium}",
      journal = {\mnras},
         year = 2018,
        month = jun,
       volume = {477},
       number = {1},
        pages = {1206-1224},
          doi = {10.1093/mnras/sty618},
archivePrefix = {arXiv},
       eprint = {1707.03401},
 primaryClass = {astro-ph.GA},
       adsurl = {https://ui.adsabs.harvard.edu/abs/2018MNRAS.477.1206N}
}

@ARTICLE{Gieles2011,
       author = {{Gieles}, Mark and {Heggie}, Douglas C. and {Zhao}, Hongsheng},
        title = "{The life cycle of star clusters in a tidal field}",
      journal = {\mnras},
         year = 2011,
        month = jun,
       volume = {413},
       number = {4},
        pages = {2509-2524},
          doi = {10.1111/j.1365-2966.2011.18320.x},
archivePrefix = {arXiv},
       eprint = {1101.1821},
 primaryClass = {astro-ph.GA},
       adsurl = {https://ui.adsabs.harvard.edu/abs/2011MNRAS.413.2509G}
}

@ARTICLE{Springel2021,
       author = {{Springel}, Volker and {Pakmor}, R{\"u}diger and {Zier}, Oliver and {Reinecke}, Martin},
        title = "{Simulating cosmic structure formation with the GADGET-4 code}",
      journal = {\mnras},
         year = 2021,
        month = sep,
       volume = {506},
       number = {2},
        pages = {2871-2949},
          doi = {10.1093/mnras/stab1855},
archivePrefix = {arXiv},
       eprint = {2010.03567},
 primaryClass = {astro-ph.IM},
       adsurl = {https://ui.adsabs.harvard.edu/abs/2021MNRAS.506.2871S}
}

@ARTICLE{Chen2024_magicI,
       author = {{Chen}, Nianyi and {Mukherjee}, Diptajyoti and {Di Matteo}, Tiziana and {Ni}, Yueying and {Bird}, Simeon and {Croft}, Rupert},
        title = "{MAGICS I. The First Few Orbits Encode the Fate of Seed Massive Black Hole Pairs}",
      journal = {The Open Journal of Astrophysics},
         year = 2024,
        month = apr,
       volume = {7},
          eid = {28},
        pages = {28},
          doi = {10.33232/001c.116179},
archivePrefix = {arXiv},
       eprint = {2312.09183},
 primaryClass = {astro-ph.GA},
       adsurl = {https://ui.adsabs.harvard.edu/abs/2024OJAp....7E..28C}
}

@ARTICLE{Henon1961,
       author = {{H{\'e}non}, M.},
        title = "{Sur l'{\'e}volution dynamique des amas globulaires}",
      journal = {Annales d'Astrophysique},
         year = 1961,
        month = feb,
       volume = {24},
        pages = {369},
       adsurl = {https://ui.adsabs.harvard.edu/abs/1961AnAp...24..369H}
}

@ARTICLE{Alexander2012,
       author = {{Alexander}, Poul E.~R. and {Gieles}, Mark},
        title = "{A prescription and fast code for the long-term evolution of star clusters}",
      journal = {\mnras},
         year = 2012,
        month = jun,
       volume = {422},
       number = {4},
        pages = {3415-3432},
          doi = {10.1111/j.1365-2966.2012.20867.x},
archivePrefix = {arXiv},
       eprint = {1203.4744},
 primaryClass = {astro-ph.GA},
       adsurl = {https://ui.adsabs.harvard.edu/abs/2012MNRAS.422.3415A}
}

@ARTICLE{Brown2021,
       author = {{Brown}, Gillen and {Gnedin}, Oleg Y.},
        title = "{Radii of young star clusters in nearby galaxies}",
      journal = {\mnras},
         year = 2021,
        month = dec,
       volume = {508},
       number = {4},
        pages = {5935-5953},
          doi = {10.1093/mnras/stab2907},
archivePrefix = {arXiv},
       eprint = {2106.12420},
 primaryClass = {astro-ph.GA},
       adsurl = {https://ui.adsabs.harvard.edu/abs/2021MNRAS.508.5935B}
}

@ARTICLE{Marinacci2018_illustrisTNGIII,
       author = {{Marinacci}, Federico and {Vogelsberger}, Mark and {Pakmor}, R{\"u}diger and {Torrey}, Paul and {Springel}, Volker and {Hernquist}, Lars and {Nelson}, Dylan and {Weinberger}, Rainer and {Pillepich}, Annalisa and {Naiman}, Jill and {Genel}, Shy},
        title = "{First results from the IllustrisTNG simulations: radio haloes and magnetic fields}",
      journal = {\mnras},
         year = 2018,
        month = nov,
       volume = {480},
       number = {4},
        pages = {5113-5139},
          doi = {10.1093/mnras/sty2206},
archivePrefix = {arXiv},
       eprint = {1707.03396},
 primaryClass = {astro-ph.CO},
       adsurl = {https://ui.adsabs.harvard.edu/abs/2018MNRAS.480.5113M}
}

@ARTICLE{Nelson2018_illustrisTNGIV,
       author = {{Nelson}, Dylan and {Pillepich}, Annalisa and {Springel}, Volker and {Weinberger}, Rainer and {Hernquist}, Lars and {Pakmor}, R{\"u}diger and {Genel}, Shy and {Torrey}, Paul and {Vogelsberger}, Mark and {Kauffmann}, Guinevere and {Marinacci}, Federico and {Naiman}, Jill},
        title = "{First results from the IllustrisTNG simulations: the galaxy colour bimodality}",
      journal = {\mnras},
         year = 2018,
        month = mar,
       volume = {475},
       number = {1},
        pages = {624-647},
          doi = {10.1093/mnras/stx3040},
archivePrefix = {arXiv},
       eprint = {1707.03395},
 primaryClass = {astro-ph.GA},
       adsurl = {https://ui.adsabs.harvard.edu/abs/2018MNRAS.475..624N}
}

@ARTICLE{Pillepich2018_illustrisTNGV,
       author = {{Pillepich}, Annalisa and {Nelson}, Dylan and {Hernquist}, Lars and {Springel}, Volker and {Pakmor}, R{\"u}diger and {Torrey}, Paul and {Weinberger}, Rainer and {Genel}, Shy and {Naiman}, Jill P. and {Marinacci}, Federico and {Vogelsberger}, Mark},
        title = "{First results from the IllustrisTNG simulations: the stellar mass content of groups and clusters of galaxies}",
      journal = {\mnras},
         year = 2018,
        month = mar,
       volume = {475},
       number = {1},
        pages = {648-675},
          doi = {10.1093/mnras/stx3112},
archivePrefix = {arXiv},
       eprint = {1707.03406},
 primaryClass = {astro-ph.GA},
       adsurl = {https://ui.adsabs.harvard.edu/abs/2018MNRAS.475..648P}
}

@ARTICLE{Zhou2026_astrid,
       author = {{Zhou}, Yihao and {Di Matteo}, Tiziana and {Bird}, Simeon and {Croft}, Rupert and {Ni}, Yueying and {Yang}, Yanhui and {Chen}, Nianyi and {Lachance}, Patrick and {Zhang}, Xiaowen and {Hafezianzadeh}, Fatemeh},
        title = "{The ASTRID Simulation at z = 0: From Massive Black Holes to Large-scale Structure}",
      journal = {\apj},
         year = 2026,
        month = mar,
       volume = {999},
       number = {1},
          eid = {41},
        pages = {41},
          doi = {10.3847/1538-4357/ae3c08},
archivePrefix = {arXiv},
       eprint = {2510.13976},
 primaryClass = {astro-ph.GA},
       adsurl = {https://ui.adsabs.harvard.edu/abs/2026ApJ...999...41Z}
}

@ARTICLE{Rizzuto2023,
       author = {{Rizzuto}, Francesco Paolo and {Naab}, Thorsten and {Rantala}, Antti and {Johansson}, Peter H. and {Ostriker}, Jeremiah P. and {Stone}, Nicholas C. and {Liao}, Shihong and {Irodotou}, Dimitrios},
        title = "{The growth of intermediate mass black holes through tidal captures and tidal disruption events}",
      journal = {\mnras},
         year = 2023,
        month = may,
       volume = {521},
       number = {2},
        pages = {2930-2948},
          doi = {10.1093/mnras/stad734},
archivePrefix = {arXiv},
       eprint = {2211.13320},
 primaryClass = {astro-ph.GA},
       adsurl = {https://ui.adsabs.harvard.edu/abs/2023MNRAS.521.2930R}
}

@ARTICLE{Bahcall1976,
       author = {{Bahcall}, J.~N. and {Wolf}, R.~A.},
        title = "{Star distribution around a massive black hole in a globular cluster.}",
      journal = {\apj},
         year = 1976,
        month = oct,
       volume = {209},
        pages = {214-232},
          doi = {10.1086/154711},
       adsurl = {https://ui.adsabs.harvard.edu/abs/1976ApJ...209..214B}
}

@ARTICLE{Chen2011,
       author = {{Chen}, Xian and {Sesana}, Alberto and {Madau}, Piero and {Liu}, F.~K.},
        title = "{Tidal Stellar Disruptions by Massive Black Hole Pairs. II. Decaying Binaries}",
      journal = {\apj},
         year = 2011,
        month = mar,
       volume = {729},
       number = {1},
          eid = {13},
        pages = {13},
          doi = {10.1088/0004-637X/729/1/13},
archivePrefix = {arXiv},
       eprint = {1012.4466},
 primaryClass = {astro-ph.GA},
       adsurl = {https://ui.adsabs.harvard.edu/abs/2011ApJ...729...13C}
}

@ARTICLE{Li2015,
       author = {{Li}, Gongjie and {Naoz}, Smadar and {Kocsis}, Bence and {Loeb}, Abraham},
        title = "{Implications of the eccentric Kozai-Lidov mechanism for stars surrounding supermassive black hole binaries}",
      journal = {\mnras},
         year = 2015,
        month = aug,
       volume = {451},
       number = {2},
        pages = {1341-1349},
          doi = {10.1093/mnras/stv1031},
archivePrefix = {arXiv},
       eprint = {1502.03825},
 primaryClass = {astro-ph.GA},
       adsurl = {https://ui.adsabs.harvard.edu/abs/2015MNRAS.451.1341L}
}

@ARTICLE{ContrerasGuerra2026,
       author = {{Contreras Guerra}, Pablo and {Grand}, Robert J.~J. and {Reina-Campos}, Marta and {Dalla Vecchia}, Claudio},
        title = "{Introducing AuriGLOBES: The effects of compressive tides, compact object-induced mass loss, and size evolution on modelling globular clusters}",
      journal = {arXiv e-prints},
         year = 2026,
        month = jun,
          eid = {arXiv:2606.30746},
        pages = {arXiv:2606.30746},
          doi = {10.48550/arXiv.2606.30746},
archivePrefix = {arXiv},
       eprint = {2606.30746},
 primaryClass = {astro-ph.GA},
       adsurl = {https://ui.adsabs.harvard.edu/abs/2026arXiv260630746C}
}

@ARTICLE{Reina-Campos2022,
       author = {{Reina-Campos}, Marta and {Keller}, Benjamin W. and {Kruijssen}, J.~M. Diederik and {Gensior}, Jindra and {Trujillo-Gomez}, Sebastian and {Jeffreson}, Sarah M.~R. and {Pfeffer}, Joel L. and {Sills}, Alison},
        title = "{Introducing EMP-Pathfinder: modelling the simultaneous formation and evolution of stellar clusters in their host galaxies}",
      journal = {\mnras},
         year = 2022,
        month = dec,
       volume = {517},
       number = {3},
        pages = {3144-3180},
          doi = {10.1093/mnras/stac1934},
archivePrefix = {arXiv},
       eprint = {2202.06961},
 primaryClass = {astro-ph.GA},
       adsurl = {https://ui.adsabs.harvard.edu/abs/2022MNRAS.517.3144R}
}

@ARTICLE{Melchor2024,
       author = {{Melchor}, Denyz and {Mockler}, Brenna and {Naoz}, Smadar and {Rose}, Sanaea C. and {Ramirez-Ruiz}, Enrico},
        title = "{Tidal Disruption Events from the Combined Effects of Two-body Relaxation and the Eccentric Kozai-Lidov Mechanism}",
      journal = {\apj},
         year = 2024,
        month = jan,
       volume = {960},
       number = {1},
          eid = {39},
        pages = {39},
          doi = {10.3847/1538-4357/acfee0},
archivePrefix = {arXiv},
       eprint = {2306.05472},
 primaryClass = {astro-ph.GA},
       adsurl = {https://ui.adsabs.harvard.edu/abs/2024ApJ...960...39M}
}

@ARTICLE{Mockler2023,
       author = {{Mockler}, Brenna and {Melchor}, Denyz and {Naoz}, Smadar and {Ramirez-Ruiz}, Enrico},
        title = "{Uncovering Hidden Massive Black Hole Companions with Tidal Disruption Events}",
      journal = {\apj},
         year = 2023,
        month = dec,
       volume = {959},
       number = {1},
          eid = {18},
        pages = {18},
          doi = {10.3847/1538-4357/ad0234},
archivePrefix = {arXiv},
       eprint = {2306.05510},
 primaryClass = {astro-ph.HE},
       adsurl = {https://ui.adsabs.harvard.edu/abs/2023ApJ...959...18M}
}

@ARTICLE{Naoz2016,
       author = {{Naoz}, Smadar},
        title = "{The Eccentric Kozai-Lidov Effect and Its Applications}",
      journal = {\araa},
         year = 2016,
        month = sep,
       volume = {54},
        pages = {441-489},
          doi = {10.1146/annurev-astro-081915-023315},
archivePrefix = {arXiv},
       eprint = {1601.07175},
 primaryClass = {astro-ph.EP},
       adsurl = {https://ui.adsabs.harvard.edu/abs/2016ARA&A..54..441N}
}

@ARTICLE{Pfeffer2018,
       author = {{Pfeffer}, Joel and {Kruijssen}, J.~M. Diederik and {Crain}, Robert A. and {Bastian}, Nate},
        title = "{The E-MOSAICS project: simulating the formation and co-evolution of galaxies and their star cluster populations}",
      journal = {\mnras},
         year = 2018,
        month = apr,
       volume = {475},
       number = {4},
        pages = {4309-4346},
          doi = {10.1093/mnras/stx3124},
archivePrefix = {arXiv},
       eprint = {1712.00019},
 primaryClass = {astro-ph.GA},
       adsurl = {https://ui.adsabs.harvard.edu/abs/2018MNRAS.475.4309P}
}

@ARTICLE{Kruijssen2019,
       author = {{Kruijssen}, J.~M. Diederik and {Pfeffer}, Joel L. and {Crain}, Robert A. and {Bastian}, Nate},
        title = "{The E-MOSAICS project: tracing galaxy formation and assembly with the age-metallicity distribution of globular clusters}",
      journal = {\mnras},
         year = 2019,
        month = jul,
       volume = {486},
       number = {3},
        pages = {3134-3179},
          doi = {10.1093/mnras/stz968},
archivePrefix = {arXiv},
       eprint = {1904.04261},
 primaryClass = {astro-ph.GA},
       adsurl = {https://ui.adsabs.harvard.edu/abs/2019MNRAS.486.3134K}
}

@ARTICLE{Lahen2020_griffin,
       author = {{Lah{\'e}n}, Natalia and {Naab}, Thorsten and {Johansson}, Peter H. and {Elmegreen}, Bruce and {Hu}, Chia-Yu and {Walch}, Stefanie and {Steinwandel}, Ulrich P. and {Moster}, Benjamin P.},
        title = "{The GRIFFIN Project{\textemdash}Formation of Star Clusters with Individual Massive Stars in a Simulated Dwarf Galaxy Starburst}",
      journal = {\apj},
         year = 2020,
        month = mar,
       volume = {891},
       number = {1},
          eid = {2},
        pages = {2},
          doi = {10.3847/1538-4357/ab7190},
archivePrefix = {arXiv},
       eprint = {1911.05093},
 primaryClass = {astro-ph.GA},
       adsurl = {https://ui.adsabs.harvard.edu/abs/2020ApJ...891....2L}
}

@ARTICLE{Deng2026,
       author = {{Deng}, Yunwei and {Li}, Hui and {Yan}, Zhiqiang and {Kong}, Chuizheng and {Zhang}, Zhi-Yu},
        title = "{Individual Star Sampling in Star Formation Simulations: A Semi-Deterministic Model}",
      journal = {arXiv e-prints},
         year = 2026,
        month = jun,
          eid = {arXiv:2606.09999},
        pages = {arXiv:2606.09999},
          doi = {10.48550/arXiv.2606.09999},
archivePrefix = {arXiv},
       eprint = {2606.09999},
 primaryClass = {astro-ph.GA},
       adsurl = {https://ui.adsabs.harvard.edu/abs/2026arXiv260609999D}
}

@ARTICLE{Escala2025,
       author = {{Escala}, Andr{\'e}s and {Zimmermann}, Lucas and {Valdebenito}, Sebasti{\'a}n and {Vergara}, Marcelo C. and {Schleicher}, Dominik R.~G. and {Liempi}, Mat{\'\i}as},
        title = "{On the Fate of Little Red Dots}",
      journal = {\apj},
         year = 2025,
        month = dec,
       volume = {995},
       number = {1},
          eid = {44},
        pages = {44},
          doi = {10.3847/1538-4357/ae200e},
archivePrefix = {arXiv},
       eprint = {2509.20453},
 primaryClass = {astro-ph.GA},
       adsurl = {https://ui.adsabs.harvard.edu/abs/2025ApJ...995...44E}
}

@ARTICLE{GonzalezPrieto2024,
       author = {{Gonz{\'a}lez Prieto}, Elena and {Weatherford}, Newlin C. and {Fragione}, Giacomo and {Kremer}, Kyle and {Rasio}, Frederic A.},
        title = "{Intermediate-mass Black Hole Progenitors from Stellar Collisions in Dense Star Clusters}",
      journal = {\apj},
         year = 2024,
        month = jul,
       volume = {969},
       number = {1},
          eid = {29},
        pages = {29},
          doi = {10.3847/1538-4357/ad43d6},
archivePrefix = {arXiv},
       eprint = {2404.11646},
 primaryClass = {astro-ph.GA},
       adsurl = {https://ui.adsabs.harvard.edu/abs/2024ApJ...969...29G}
}

@ARTICLE{Pacucci2025,
       author = {{Pacucci}, Fabio and {Hernquist}, Lars and {Fujii}, Michiko},
        title = "{Little Red Dots are Nurseries of Massive Black Holes}",
      journal = {\apj},
         year = 2025,
        month = nov,
       volume = {994},
       number = {1},
          eid = {40},
        pages = {40},
          doi = {10.3847/1538-4357/ae1619},
archivePrefix = {arXiv},
       eprint = {2509.02664},
 primaryClass = {astro-ph.GA},
       adsurl = {https://ui.adsabs.harvard.edu/abs/2025ApJ...994...40P}
}

@ARTICLE{Williams2025_simulation,
       author = {{Williams}, Claire E. and {Naoz}, Smadar and {Lake}, William and {Burkhart}, Blakesley and {Marinacci}, Federico and {Vogelsberger}, Mark and {Yoshida}, Naoki and {Menon}, Shyam H. and {Chen}, Avi and {Adamo}, Angela},
        title = "{{\ensuremath{\Lambda}}CDM Star Clusters at Cosmic Dawn: Stellar Densities, Environment, and Equilibrium}",
      journal = {\apj},
         year = 2025,
        month = sep,
       volume = {990},
       number = {2},
          eid = {135},
        pages = {135},
          doi = {10.3847/1538-4357/adf19d},
archivePrefix = {arXiv},
       eprint = {2502.17561},
 primaryClass = {astro-ph.GA},
       adsurl = {https://ui.adsabs.harvard.edu/abs/2025ApJ...990..135W}
}

@ARTICLE{Williams2026,
       author = {{Williams}, Claire E. and {Naoz}, Smadar and {Rose}, Sanaea C. and {Burkhart}, Blakesley and {Yoshida}, Naoki and {Chen}, Avi and {Kremer}, Kyle and {Lake}, William and {Marinacci}, Federico and {Menon}, Shyam H. and {Vogelsberger}, Mark},
        title = "{A cosmological framework for stellar collisions at high redshift in proto-globular clusters, nuclear star clusters, and Little Red Dots}",
      journal = {arXiv e-prints},
         year = 2026,
        month = mar,
          eid = {arXiv:2603.26872},
        pages = {arXiv:2603.26872},
          doi = {10.48550/arXiv.2603.26872},
archivePrefix = {arXiv},
       eprint = {2603.26872},
 primaryClass = {astro-ph.GA},
       adsurl = {https://ui.adsabs.harvard.edu/abs/2026arXiv260326872W}
}

@ARTICLE{Ni2022_astrid,
       author = {{Ni}, Yueying and {Di Matteo}, Tiziana and {Bird}, Simeon and {Croft}, Rupert and {Feng}, Yu and {Chen}, Nianyi and {Tremmel}, Michael and {DeGraf}, Colin and {Li}, Yin},
        title = "{The ASTRID simulation: the evolution of supermassive black holes}",
      journal = {\mnras},
         year = 2022,
        month = jun,
       volume = {513},
       number = {1},
        pages = {670-692},
          doi = {10.1093/mnras/stac351},
archivePrefix = {arXiv},
       eprint = {2110.14154},
 primaryClass = {astro-ph.GA},
       adsurl = {https://ui.adsabs.harvard.edu/abs/2022MNRAS.513..670N}
}

@ARTICLE{Bird2022,
       author = {{Bird}, Simeon and {Ni}, Yueying and {Di Matteo}, Tiziana and {Croft}, Rupert and {Feng}, Yu and {Chen}, Nianyi},
        title = "{The ASTRID simulation: galaxy formation and reionization}",
      journal = {\mnras},
         year = 2022,
        month = may,
       volume = {512},
       number = {3},
        pages = {3703-3716},
          doi = {10.1093/mnras/stac648},
archivePrefix = {arXiv},
       eprint = {2111.01160},
 primaryClass = {astro-ph.GA},
       adsurl = {https://ui.adsabs.harvard.edu/abs/2022MNRAS.512.3703B}
}

@ARTICLE{Ni2025,
       author = {{Ni}, Yueying and {Chen}, Nianyi and {Zhou}, Yihao and {Park}, Minjung and {Yang}, Yanhui and {Di Matteo}, Tiziana and {Bird}, Simeon and {Croft}, Rupert},
        title = "{The Astrid Simulation: Evolution of Black Holes and Galaxies to z = 0.5 and Different Evolution Pathways for Galaxy Quenching}",
      journal = {\apj},
         year = 2025,
        month = sep,
       volume = {990},
       number = {2},
          eid = {120},
        pages = {120},
          doi = {10.3847/1538-4357/adf3a7},
archivePrefix = {arXiv},
       eprint = {2409.10666},
 primaryClass = {astro-ph.GA},
       adsurl = {https://ui.adsabs.harvard.edu/abs/2025ApJ...990..120N}
}

@ARTICLE{Katz2015,
       author = {{Katz}, Harley and {Sijacki}, Debora and {Haehnelt}, Martin G.},
        title = "{Seeding high-redshift QSOs by collisional runaway in primordial star clusters}",
      journal = {\mnras},
         year = 2015,
        month = aug,
       volume = {451},
       number = {3},
        pages = {2352-2369},
          doi = {10.1093/mnras/stv1048},
archivePrefix = {arXiv},
       eprint = {1502.03448},
 primaryClass = {astro-ph.GA},
       adsurl = {https://ui.adsabs.harvard.edu/abs/2015MNRAS.451.2352K}
}

@ARTICLE{Larson2023_ceers_obs,
       author = {{Larson}, Rebecca L. and {Finkelstein}, Steven L. and {Kocevski}, Dale D. and {Hutchison}, Taylor A. and {Trump}, Jonathan R. and {Arrabal Haro}, Pablo and {Bromm}, Volker and {Cleri}, Nikko J. and {Dickinson}, Mark and {Fujimoto}, Seiji and {Kartaltepe}, Jeyhan S. and {Koekemoer}, Anton M. and {Papovich}, Casey and {Pirzkal}, Nor and {Tacchella}, Sandro and {Zavala}, Jorge A. and {Bagley}, Micaela and {Behroozi}, Peter and {Champagne}, Jaclyn B. and {Cole}, Justin W. and {Jung}, Intae and {Morales}, Alexa M. and {Yang}, Guang and {Zhang}, Haowen and {Zitrin}, Adi and {Amor{\'\i}n}, Ricardo O. and {Burgarella}, Denis and {Casey}, Caitlin M. and {Ch{\'a}vez Ortiz}, {\'O}scar A. and {Cox}, Isabella G. and {Chworowsky}, Katherine and {Fontana}, Adriano and {Gawiser}, Eric and {Grazian}, Andrea and {Grogin}, Norman A. and {Harish}, Santosh and {Hathi}, Nimish P. and {Hirschmann}, Michaela and {Holwerda}, Benne W. and {Juneau}, St{\'e}phanie and {Leung}, Gene C.~K. and {Lucas}, Ray A. and {McGrath}, Elizabeth J. and {P{\'e}rez-Gonz{\'a}lez}, Pablo G. and {Rigby}, Jane R. and {Seill{\'e}}, Lise-Marie and {Simons}, Raymond C. and {de La Vega}, Alexander and {Weiner}, Benjamin J. and {Wilkins}, Stephen M. and {Yung}, L.~Y. Aaron and {Ceers Team}},
        title = "{A CEERS Discovery of an Accreting Supermassive Black Hole 570 Myr after the Big Bang: Identifying a Progenitor of Massive z > 6 Quasars}",
      journal = {\apjl},
         year = 2023,
        month = aug,
       volume = {953},
       number = {2},
          eid = {L29},
        pages = {L29},
          doi = {10.3847/2041-8213/ace619},
archivePrefix = {arXiv},
       eprint = {2303.08918},
 primaryClass = {astro-ph.GA},
       adsurl = {https://ui.adsabs.harvard.edu/abs/2023ApJ...953L..29L}
}

@ARTICLE{Maiolino2024_gnz11,
       author = {{Maiolino}, Roberto and {Scholtz}, Jan and {Witstok}, Joris and {Carniani}, Stefano and {D'Eugenio}, Francesco and {de Graaff}, Anna and {{\"U}bler}, Hannah and {Tacchella}, Sandro and {Curtis-Lake}, Emma and {Arribas}, Santiago and {Bunker}, Andrew and {Charlot}, St{\'e}phane and {Chevallard}, Jacopo and {Curti}, Mirko and {Looser}, Tobias J. and {Maseda}, Michael V. and {Rawle}, Timothy D. and {Rodr{\'\i}guez del Pino}, Bruno and {Willott}, Chris J. and {Egami}, Eiichi and {Eisenstein}, Daniel J. and {Hainline}, Kevin N. and {Robertson}, Brant and {Williams}, Christina C. and {Willmer}, Christopher N.~A. and {Baker}, William M. and {Boyett}, Kristan and {DeCoursey}, Christa and {Fabian}, Andrew C. and {Helton}, Jakob M. and {Ji}, Zhiyuan and {Jones}, Gareth C. and {Kumari}, Nimisha and {Laporte}, Nicolas and {Nelson}, Erica J. and {Perna}, Michele and {Sandles}, Lester and {Shivaei}, Irene and {Sun}, Fengwu},
        title = "{A small and vigorous black hole in the early Universe}",
      journal = {\nat},
         year = 2024,
        month = mar,
       volume = {627},
       number = {8002},
        pages = {59-63},
          doi = {10.1038/s41586-024-07052-5},
archivePrefix = {arXiv},
       eprint = {2305.12492},
 primaryClass = {astro-ph.GA},
       adsurl = {https://ui.adsabs.harvard.edu/abs/2024Natur.627...59M}
}

@ARTICLE{Messa2026_CosmicGems_new,
       author = {{Messa}, M. and {Vanzella}, E. and {Loiacono}, F. and {Adamo}, A. and {Oguri}, M. and {Sharon}, K. and {Bradley}, L.~D. and {Christensen}, L. and {Claeyssens}, A. and {Richard}, J. and {Abdurro'uf} and {Bauer}, F.~E. and {Bergamini}, P. and {Bolamperti}, A. and {Brada{\v{c}}}, M. and {Calura}, F. and {Coe}, D. and {Diego}, J.~M. and {Grillo}, C. and {Hsiao}, T.~Y.-Y. and {Inoue}, A.~K. and {Fujimoto}, S. and {Lombardi}, M. and {Meneghetti}, M. and {Resseguier}, T. and {Ricotti}, M. and {Rosati}, P. and {Welch}, B. and {Windhorst}, R.~A. and {Xu}, X. and {Zackrisson}, E. and {Zanella}, A. and {Zitrin}, A.},
        title = "{JWST spectroscopic confirmation of the Cosmic Gems arc at z = 9.625: Insights into the small-scale structure of a post-burst system}",
      journal = {\aap},
         year = 2026,
        month = jan,
       volume = {705},
          eid = {A173},
        pages = {A173},
          doi = {10.1051/0004-6361/202556574},
archivePrefix = {arXiv},
       eprint = {2507.18705},
 primaryClass = {astro-ph.GA},
       adsurl = {https://ui.adsabs.harvard.edu/abs/2026A&A...705A.173M}
}

@article{Fujimoto2025_CosmicGrapes,
  author  = {Fujimoto, Seiji and Ouchi, Masami and Kohno, Kotaro
             and others},
  title   = {Primordial rotating disk composed of at least
             15 dense star-forming clumps at cosmic dawn},
  journal = {Nature Astronomy},
  volume  = {9},
  pages   = {1553--1567},
  year    = {2025},
  doi     = {10.1038/s41550-025-02592-w},
  eprint  = {2402.18543},
  archivePrefix = {arXiv}
}

@ARTICLE{Alexander2017,
       author = {{Alexander}, Tal},
        title = "{Stellar Dynamics and Stellar Phenomena Near a Massive Black Hole}",
      journal = {\araa},
         year = 2017,
        month = aug,
       volume = {55},
       number = {1},
        pages = {17-57},
          doi = {10.1146/annurev-astro-091916-055306},
archivePrefix = {arXiv},
       eprint = {1701.04762},
 primaryClass = {astro-ph.GA},
       adsurl = {https://ui.adsabs.harvard.edu/abs/2017ARA&A..55...17A}
}

@ARTICLE{Krumholz2019,
       author = {{Krumholz}, Mark R. and {McKee}, Christopher F. and {Bland-Hawthorn}, Joss},
        title = "{Star Clusters Across Cosmic Time}",
      journal = {\araa},
         year = 2019,
        month = aug,
       volume = {57},
        pages = {227-303},
          doi = {10.1146/annurev-astro-091918-104430},
archivePrefix = {arXiv},
       eprint = {1812.01615},
 primaryClass = {astro-ph.GA},
       adsurl = {https://ui.adsabs.harvard.edu/abs/2019ARA&A..57..227K}
}

@ARTICLE{Baumgardt2004,
       author = {{Baumgardt}, Holger and {Makino}, Junichiro and {Ebisuzaki}, Toshikazu},
        title = "{Massive Black Holes in Star Clusters. I. Equal-Mass Clusters}",
      journal = {\apj},
         year = 2004,
        month = oct,
       volume = {613},
       number = {2},
        pages = {1133-1142},
          doi = {10.1086/423298},
archivePrefix = {arXiv},
       eprint = {astro-ph/0406227},
 primaryClass = {astro-ph},
       adsurl = {https://ui.adsabs.harvard.edu/abs/2004ApJ...613.1133B}
}

@ARTICLE{Preto2004,
       author = {{Preto}, Miguel and {Merritt}, David and {Spurzem}, Rainer},
        title = "{N-Body Growth of a Bahcall-Wolf Cusp around a Black Hole}",
      journal = {\apjl},
         year = 2004,
        month = oct,
       volume = {613},
       number = {2},
        pages = {L109-L112},
          doi = {10.1086/425139},
archivePrefix = {arXiv},
       eprint = {astro-ph/0406324},
 primaryClass = {astro-ph},
       adsurl = {https://ui.adsabs.harvard.edu/abs/2004ApJ...613L.109P}
}

@ARTICLE{Messa2025_CosmicArchipelago,
       author = {{Messa}, M. and {Vanzella}, E. and {Loiacono}, F. and {Bergamini}, P. and {Castellano}, M. and {Sun}, B. and {Willott}, C. and {Windhorst}, R.~A. and {Yan}, H. and {Angora}, G. and {Rosati}, P. and {Adamo}, A. and {Annibali}, F. and {Bolamperti}, A. and {Brada{\v{c}}}, M. and {Bradley}, L.~D. and {Calura}, F. and {Claeyssens}, A. and {Comastri}, A. and {Conselice}, C.~J. and {D'Silva}, J.~C.~J. and {Dickinson}, M. and {Frye}, B.~L. and {Grillo}, C. and {Grogin}, N.~A. and {Gruppioni}, C. and {Koekemoer}, A.~M. and {Meneghetti}, M. and {Me{\v{s}}tri{\'c}}, U. and {Pascale}, R. and {Ravindranath}, S. and {Ricotti}, M. and {Summers}, J. and {Zanella}, A.},
        title = "{Anatomy of a z = 6 Lyman-{\ensuremath{\alpha}} emitter down to parsec scales: Extreme UV slopes, metal-poor regions, and possibly leaking star clusters}",
      journal = {\aap},
         year = 2025,
        month = feb,
       volume = {694},
          eid = {A59},
        pages = {A59},
          doi = {10.1051/0004-6361/202451695},
archivePrefix = {arXiv},
       eprint = {2407.20331},
 primaryClass = {astro-ph.GA},
       adsurl = {https://ui.adsabs.harvard.edu/abs/2025A&A...694A..59M}
}

@ARTICLE{Bradac2025_BulletArc,
       author = {{Brada{\v{c}}}, Maru{\v{s}}a and {Jude{\v{z}}}, Jon and {Willott}, Chris and {Rihtar{\v{s}}ic}, Gregor and {Martis}, Nicholas S. and {Harshan}, Anishya and {Felicioni}, Giordano and {Asada}, Yoshihisa and {Desprez}, Guillaume and {Clowe}, Douglas and {Gonzalez}, Anthony H. and {Jones}, Christine and {Lemaux}, Brian C. and {Markevitch}, Maxim and {Markov}, Vladan and {Mowla}, Lamiya and {Noirot}, Ga{\"e}l and {Peter}, Annika H.~G. and {Robertson}, Andrew and {Sarrouh}, Ghassan T.~E. and {Sawicki}, Marcin and {Schrabback}, Tim and {Tripodi}, Roberta},
        title = "{Star Formation under a Cosmic Microscope: Highly Magnified z = 11 Galaxy behind the Bullet Cluster}",
      journal = {\apjl},
         year = 2025,
        month = dec,
       volume = {995},
       number = {2},
          eid = {L74},
        pages = {L74},
          doi = {10.3847/2041-8213/ae27d2},
archivePrefix = {arXiv},
       eprint = {2509.20446},
 primaryClass = {astro-ph.GA},
       adsurl = {https://ui.adsabs.harvard.edu/abs/2025ApJ...995L..74B}
}

@ARTICLE{Kruijssen2012,
       author = {{Kruijssen}, J.~M. Diederik},
        title = "{On the fraction of star formation occurring in bound stellar clusters}",
      journal = {\mnras},
         year = 2012,
        month = nov,
       volume = {426},
       number = {4},
        pages = {3008-3040},
          doi = {10.1111/j.1365-2966.2012.21923.x},
archivePrefix = {arXiv},
       eprint = {1208.2963},
 primaryClass = {astro-ph.CO},
       adsurl = {https://ui.adsabs.harvard.edu/abs/2012MNRAS.426.3008K}
}

@software{Feng2018_mpgadget,
       author = {{Feng}, Yu and {Bird}, Simeon and {Anderson}, Lauren and {Font-Ribera}, Andreu and {Pedersen}, Chris},
        title = "{MP-Gadget/MP-Gadget: A tag for getting a DOI}",
         year = 2018,
        month = oct,
          eid = {10.5281/zenodo.1451799},
          doi = {10.5281/zenodo.1451799},
      version = {FirstDOI},
    publisher = {Zenodo},
       adsurl = {https://ui.adsabs.harvard.edu/abs/2018zndo...1451799F}
}

@ARTICLE{Katz1996,
       author = {{Katz}, Neal and {Weinberg}, David H. and {Hernquist}, Lars},
        title = "{Cosmological Simulations with TreeSPH}",
      journal = {\apjs},
         year = 1996,
        month = jul,
       volume = {105},
        pages = {19},
          doi = {10.1086/192305},
archivePrefix = {arXiv},
       eprint = {astro-ph/9509107},
 primaryClass = {astro-ph},
       adsurl = {https://ui.adsabs.harvard.edu/abs/1996ApJS..105...19K}
}

@ARTICLE{DiMatteo2005_BH_model,
       author = {{Di Matteo}, Tiziana and {Springel}, Volker and {Hernquist}, Lars},
        title = "{Energy input from quasars regulates the growth and activity of black holes and their host galaxies}",
      journal = {\nat},
         year = 2005,
        month = feb,
       volume = {433},
       number = {7026},
        pages = {604-607},
          doi = {10.1038/nature03335},
archivePrefix = {arXiv},
       eprint = {astro-ph/0502199},
 primaryClass = {astro-ph},
       adsurl = {https://ui.adsabs.harvard.edu/abs/2005Natur.433..604D}
}

@ARTICLE{Weinberger2017_tng,
       author = {{Weinberger}, Rainer and {Springel}, Volker and {Hernquist}, Lars and {Pillepich}, Annalisa and {Marinacci}, Federico and {Pakmor}, R{\"u}diger and {Nelson}, Dylan and {Genel}, Shy and {Vogelsberger}, Mark and {Naiman}, Jill and {Torrey}, Paul},
        title = "{Simulating galaxy formation with black hole driven thermal and kinetic feedback}",
      journal = {\mnras},
         year = 2017,
        month = mar,
       volume = {465},
       number = {3},
        pages = {3291-3308},
          doi = {10.1093/mnras/stw2944},
archivePrefix = {arXiv},
       eprint = {1607.03486},
 primaryClass = {astro-ph.GA},
       adsurl = {https://ui.adsabs.harvard.edu/abs/2017MNRAS.465.3291W}
}

@ARTICLE{BondiHoyle1944,
   author = {{Bondi}, H. and {Hoyle}, F.},
    title = "{On the mechanism of accretion by stars}",
  journal = {\mnras},
     year = 1944,
   volume = 104,
    pages = {273-+},
   adsurl = {http://adsabs.harvard.edu/abs/1944MNRAS.104..273B}
}

@ARTICLE{Chen2022_DF,
       author = {{Chen}, Nianyi and {Ni}, Yueying and {Tremmel}, Michael and {Di Matteo}, Tiziana and {Bird}, Simeon and {DeGraf}, Colin and {Feng}, Yu},
        title = "{Dynamical friction modelling of massive black holes in cosmological simulations and effects on merger rate predictions}",
      journal = {\mnras},
         year = 2022,
        month = feb,
       volume = {510},
       number = {1},
        pages = {531-550},
          doi = {10.1093/mnras/stab3411},
archivePrefix = {arXiv},
       eprint = {2104.00021},
 primaryClass = {astro-ph.GA},
       adsurl = {https://ui.adsabs.harvard.edu/abs/2022MNRAS.510..531C}
}

@ARTICLE{Tremmel2015,
       author = {{Tremmel}, M. and {Governato}, F. and {Volonteri}, M. and {Quinn}, T.~R.},
        title = "{Off the beaten path: a new approach to realistically model the orbital decay of supermassive black holes in galaxy formation simulations}",
      journal = {\mnras},
         year = 2015,
        month = aug,
       volume = {451},
       number = {2},
        pages = {1868-1874},
          doi = {10.1093/mnras/stv1060},
archivePrefix = {arXiv},
       eprint = {1501.07609},
 primaryClass = {astro-ph.GA},
       adsurl = {https://ui.adsabs.harvard.edu/abs/2015MNRAS.451.1868T}
}

@ARTICLE{Karakas2010,
       author = {{Karakas}, A.~I.},
        title = "{Updated stellar yields from asymptotic giant branch models}",
      journal = {\mnras},
         year = 2010,
        month = apr,
       volume = {403},
       number = {3},
        pages = {1413-1425},
          doi = {10.1111/j.1365-2966.2009.16198.x},
archivePrefix = {arXiv},
       eprint = {0912.2142},
 primaryClass = {astro-ph.SR},
       adsurl = {https://ui.adsabs.harvard.edu/abs/2010MNRAS.403.1413K}
}

@ARTICLE{Doherty2014_metallIII,
       author = {{Doherty}, Carolyn L. and {Gil-Pons}, Pilar and {Lau}, Herbert H.~B. and {Lattanzio}, John C. and {Siess}, Lionel and {Campbell}, Simon W.},
        title = "{Super and massive AGB stars - III. Nucleosynthesis in metal-poor and very metal-poor stars - Z = 0.001 and 0.0001}",
      journal = {\mnras},
         year = 2014,
        month = jun,
       volume = {441},
       number = {1},
        pages = {582-598},
          doi = {10.1093/mnras/stu571},
archivePrefix = {arXiv},
       eprint = {1403.5054},
 primaryClass = {astro-ph.SR},
       adsurl = {https://ui.adsabs.harvard.edu/abs/2014MNRAS.441..582D}
}

@ARTICLE{Doherty2014_metallII,
       author = {{Doherty}, Carolyn L. and {Gil-Pons}, Pilar and {Lau}, Herbert H.~B. and {Lattanzio}, John C. and {Siess}, Lionel},
        title = "{Super and massive AGB stars - II. Nucleosynthesis and yields - Z = 0.02, 0.008 and 0.004}",
      journal = {\mnras},
         year = 2014,
        month = jan,
       volume = {437},
       number = {1},
        pages = {195-214},
          doi = {10.1093/mnras/stt1877},
archivePrefix = {arXiv},
       eprint = {1310.2614},
 primaryClass = {astro-ph.SR},
       adsurl = {https://ui.adsabs.harvard.edu/abs/2014MNRAS.437..195D}
}

@ARTICLE{Kobayashi2006,
       author = {{Kobayashi}, Chiaki and {Umeda}, Hideyuki and {Nomoto}, Ken'ichi and {Tominaga}, Nozomu and {Ohkubo}, Takuya},
        title = "{Galactic Chemical Evolution: Carbon through Zinc}",
      journal = {\apj},
         year = 2006,
        month = dec,
       volume = {653},
       number = {2},
        pages = {1145-1171},
          doi = {10.1086/508914},
archivePrefix = {arXiv},
       eprint = {astro-ph/0608688},
 primaryClass = {astro-ph},
       adsurl = {https://ui.adsabs.harvard.edu/abs/2006ApJ...653.1145K}
}

@ARTICLE{Nomoto1997,
       author = {{Nomoto}, K. and {Iwamoto}, K. and {Nakasato}, N. and {Thielemann}, F.-K. and {Brachwitz}, F. and {Tsujimoto}, T. and {Kubo}, Y. and {Kishimoto}, N.},
        title = "{Nucleosynthesis in type Ia supernovae}",
      journal = {\nphysa},
         year = 1997,
        month = feb,
       volume = {621},
        pages = {467-476},
          doi = {10.1016/S0375-9474(97)00291-1},
archivePrefix = {arXiv},
       eprint = {astro-ph/9706025},
 primaryClass = {astro-ph},
       adsurl = {https://ui.adsabs.harvard.edu/abs/1997NuPhA.621..467N}
}

@ARTICLE{Okamoto2010,
       author = {{Okamoto}, Takashi and {Frenk}, Carlos S. and {Jenkins}, Adrian and {Theuns}, Tom},
        title = "{The properties of satellite galaxies in simulations of galaxy formation}",
      journal = {\mnras},
         year = 2010,
        month = jul,
       volume = {406},
       number = {1},
        pages = {208-222},
          doi = {10.1111/j.1365-2966.2010.16690.x},
archivePrefix = {arXiv},
       eprint = {0909.0265},
 primaryClass = {astro-ph.CO},
       adsurl = {https://ui.adsabs.harvard.edu/abs/2010MNRAS.406..208O}
}

@ARTICLE{Vogelsberger2013,
       author = {{Vogelsberger}, Mark and {Genel}, Shy and {Sijacki}, Debora and {Torrey}, Paul and {Springel}, Volker and {Hernquist}, Lars},
        title = "{A model for cosmological simulations of galaxy formation physics}",
      journal = {\mnras},
         year = 2013,
        month = dec,
       volume = {436},
       number = {4},
        pages = {3031-3067},
          doi = {10.1093/mnras/stt1789},
archivePrefix = {arXiv},
       eprint = {1305.2913},
 primaryClass = {astro-ph.CO},
       adsurl = {https://ui.adsabs.harvard.edu/abs/2013MNRAS.436.3031V}
}

@ARTICLE{Springel2003,
       author = {{Springel}, Volker and {Hernquist}, Lars},
        title = "{Cosmological smoothed particle hydrodynamics simulations: a hybrid multiphase model for star formation}",
      journal = {\mnras},
         year = 2003,
        month = feb,
       volume = {339},
       number = {2},
        pages = {289-311},
          doi = {10.1046/j.1365-8711.2003.06206.x},
archivePrefix = {arXiv},
       eprint = {astro-ph/0206393},
 primaryClass = {astro-ph},
       adsurl = {https://ui.adsabs.harvard.edu/abs/2003MNRAS.339..289S}
}

@ARTICLE{Bellovary2011,
       author = {{Bellovary}, Jillian and {Volonteri}, Marta and {Governato}, Fabio and {Shen}, Sijing and {Quinn}, Thomas and {Wadsley}, James},
        title = "{The First Massive Black Hole Seeds and Their Hosts}",
      journal = {\apj},
         year = 2011,
        month = nov,
       volume = {742},
       number = {1},
          eid = {13},
        pages = {13},
          doi = {10.1088/0004-637X/742/1/13},
archivePrefix = {arXiv},
       eprint = {1104.3858},
 primaryClass = {astro-ph.CO},
       adsurl = {https://ui.adsabs.harvard.edu/abs/2011ApJ...742...13B}
}

@ARTICLE{Chabrier2003,
       author = {{Chabrier}, Gilles},
        title = "{Galactic Stellar and Substellar Initial Mass Function}",
      journal = {\pasp},
         year = 2003,
        month = jul,
       volume = {115},
       number = {809},
        pages = {763-795},
          doi = {10.1086/376392},
archivePrefix = {arXiv},
       eprint = {astro-ph/0304382},
 primaryClass = {astro-ph},
       adsurl = {https://ui.adsabs.harvard.edu/abs/2003PASP..115..763C}
}

@ARTICLE{Krumholz2011_h2cooling,
       author = {{Krumholz}, Mark R. and {Gnedin}, Nickolay Y.},
        title = "{A Comparison of Methods for Determining the Molecular Content of Model Galaxies}",
      journal = {\apj},
         year = 2011,
        month = mar,
       volume = {729},
       number = {1},
          eid = {36},
        pages = {36},
          doi = {10.1088/0004-637X/729/1/36},
archivePrefix = {arXiv},
       eprint = {1011.4065},
 primaryClass = {astro-ph.CO},
       adsurl = {https://ui.adsabs.harvard.edu/abs/2011ApJ...729...36K}
}

@ARTICLE{Wiersma2009,
       author = {{Wiersma}, Robert P.~C. and {Schaye}, Joop and {Theuns}, Tom and {Dalla Vecchia}, Claudio and {Tornatore}, Luca},
        title = "{Chemical enrichment in cosmological, smoothed particle hydrodynamics simulations}",
      journal = {\mnras},
         year = 2009,
        month = oct,
       volume = {399},
       number = {2},
        pages = {574-600},
          doi = {10.1111/j.1365-2966.2009.15331.x},
archivePrefix = {arXiv},
       eprint = {0902.1535},
 primaryClass = {astro-ph.CO},
       adsurl = {https://ui.adsabs.harvard.edu/abs/2009MNRAS.399..574W}
}

@ARTICLE{Pillepich2018,
       author = {{Pillepich}, Annalisa and {Springel}, Volker and {Nelson}, Dylan and {Genel}, Shy and {Naiman}, Jill and {Pakmor}, R{\"u}diger and {Hernquist}, Lars and {Torrey}, Paul and {Vogelsberger}, Mark and {Weinberger}, Rainer and {Marinacci}, Federico},
        title = "{Simulating galaxy formation with the IllustrisTNG model}",
      journal = {\mnras},
         year = 2018,
        month = jan,
       volume = {473},
       number = {3},
        pages = {4077-4106},
          doi = {10.1093/mnras/stx2656},
archivePrefix = {arXiv},
       eprint = {1703.02970},
 primaryClass = {astro-ph.GA},
       adsurl = {https://ui.adsabs.harvard.edu/abs/2018MNRAS.473.4077P}
}

@ARTICLE{Tremmel2017,
       author = {{Tremmel}, M. and {Karcher}, M. and {Governato}, F. and {Volonteri}, M. and {Quinn}, T.~R. and {Pontzen}, A. and {Anderson}, L. and {Bellovary}, J.},
        title = "{The Romulus cosmological simulations: a physical approach to the formation, dynamics and accretion models of SMBHs}",
      journal = {\mnras},
         year = 2017,
        month = sep,
       volume = {470},
       number = {1},
        pages = {1121-1139},
          doi = {10.1093/mnras/stx1160},
archivePrefix = {arXiv},
       eprint = {1607.02151},
 primaryClass = {astro-ph.GA},
       adsurl = {https://ui.adsabs.harvard.edu/abs/2017MNRAS.470.1121T}
}
\bibliographystyle{aasjournal}


\end{CJK*}
\end{document}